\documentclass[10pt,a4paper,twoside]{article} 
\usepackage{latexsym}
\usepackage{amsmath}
\usepackage{amssymb}
\usepackage{a4wide}
\usepackage{amsfonts}
\usepackage{mathrsfs}               
\usepackage{bbm}                    
\usepackage{bbold}                  
\usepackage{textcomp}
\usepackage{theorem} 
\DeclareMathAlphabet{\mathpzc}{OT1}{pzc}{m}{it}     
\newcommand{\CC}{\mathbb{C}}

\newcommand{\NN}{\mathbb{N}}

\newcommand{\RR}{\mathbb{R}}
\newcommand{\TT}{\mathbb{T}}
\newcommand{\ZZ}{\mathbb{Z}}
\newcommand{\sU}{\mathscr{U}}
\newcommand{\sW}{\mathscr{W}}

\newcommand{\ve}{\varepsilon}  
\newcommand{\vp}{\varphi}      
\newcommand{\vk}{\varkappa}
\newcommand{\vr}{\varrho}
\newcommand{\vt}{\vartheta}
\newcommand{\vs}{\varsigma}
\newcommand{\supp}{\mathrm{supp}\,}
\newcommand{\const}{\mathrm{const}}
\newcommand{\dist}{\mathrm{dist}}

\newcommand{\Ran}{\mathrm{Ran}\;}

\newcommand{\sgn}{\mathrm{sgn}}
\newcommand{\diz}{\mathrm{div}\,}
\newcommand{\aaa}{\mathrm{aa}}
\newcommand{\id}{\mathbb{1}}   
\newcommand{\SL}{\langle \,}   
\newcommand{\SR}{\, \rangle}          
\newcommand{\klg}{\leqslant}   
\newcommand{\grg}{\geqslant}          
\newcommand{\LO}{\mathscr{L}}    
\newcommand{\HR}{\mathscr{H}}    
\newcommand{\Borel}{\mathfrak{B}}
\newcommand{\Fdist}{d_{F}}       
\newcommand{\HVF}[1]{X_{#1}}     
\newcommand{\KVF}[1]{K_{#1}}     
\newcommand{\HFA}{\widehat{X}_a} 
\newcommand{\KFA}{\widehat{K}_a} 
\newcommand{\bigO}{\mathcal{O}}  
\newcommand{\SMS}{\mathfrak{S}}  
\newcommand{\SPn}[2]{\langle\,#1\,|\,#2\,\rangle} 
\newcommand{\SPb}[2]{\big\langle\,#1\,\big|\,#2\,\big\rangle} 
\newcommand{\SYC}[2]{\widetilde{\sigma}(\,#1\,,\,#2\,)}     
\newcommand{\matr}[1]{\left(\begin{array}{cc}#1\end{array}\right)}
\newcommand{\Det}[1]{\left|\begin{array}{cc}#1\end{array}\right|}
\newcommand{\Cor}{\mathrm{Cor}}                
\newcommand{\spec}{\mathrm{spec}}              
\newcommand{\Fesh}{\mathcal{F}}                
\newcommand{\xyhalf}{{\textstyle\frac{x+y}{2}}}
\newcommand{\xyh}{{\textstyle\frac{x-y}{h}}}   
\newcommand{\FT}{\mathscr{F}}                  
\newcommand{\Figx}{\mathbb{f}}                 
\newcommand{\Indx}{\mathbb{i}}                 
\newcommand{\Bx}{\mathbb{k}}                   
\newcommand{\FS}{F}                            
\newcommand{\nv}{\mathring{v}}                 
\newcommand{\FM}{\mathbb{F}}                   
\newcommand{\Symbb}[1]{S_b(#1)}                
\newcommand{\Op}{\mathrm{Op}_h^W}              
\newcommand{\bbA}{\mathbb{A}}                   
\newcommand{\bbB}{\mathbb{B}}                   %
\newcommand{\CF}{\omega}                        
\newcommand{\Schwartz}{\mathscr{S}}             
\newcommand{\wtG}{\widetilde{\Gamma}}          
\newcommand{\proof}{{\sc Proof: }}
\newcommand{\qed}{\hfill$\Box$}
\theoremstyle{plain}
\newtheorem{theorem}{Theorem.}[section]
\newtheorem{lemma}[theorem]{Lemma.}
\newtheorem{proposition}[theorem]{Proposition.}
\newtheorem{corollary}[theorem]{Corollary.}

\newtheorem{example}[theorem]{Example.}
\newtheorem{remark}[theorem]{Remark.}
\newtheorem{hypothesis}[theorem]{Hypothesis.}
\begin{document}

\begin{center} 
{\Large{Correlation asymptotics for non-translation invariant\\

\smallskip

lattice spin systems}}
\end{center}

\bigskip

\begin{center}       
\noindent {\large\textsc{Oliver Matte}}  

\smallskip

\noindent Mathematisches Institut der
Universit\"at M\"unchen\\
Theresienstra{\ss}e 39, D-80333 M\"unchen, Germany.

\smallskip

\noindent {\tt matte@mathematik.uni-muenchen.de}

\end{center}

\begin{abstract}
\noindent We obtain asymptotic expressions for the Green kernels of
certain
non-translation invariant transition matrices 
using methods of semiclassical and microlocal analysis.
Combined with a result by Bach and M{\o}ller this yields
asymptotic formulas for the truncated two-point correlation
functions of certain non-translation invariant lattice models of
real-valued spins.

\smallskip

\end{abstract}

\section{Introduction}

This article is a further contribution
to the study of correlation asymptotics
for lattice systems of real-valued spins
via the Witten-Laplacian
approach
\cite{BJS,BaMo1,BaMo2,Ma,Sj1,Sj2}.
Like the authors did in the cited articles, we consider
a lattice system at very
small temperatures
and assume that the on-site potentials have only
one non-degenerated minimum
and that the ferromagnetic pair interaction
is sufficiently weak. 
In this situation one expects the truncated
two-point correlation functions of the
associated Gibbs measure to decay exponentially.
In fact, in \cite{BJS,Sj1} the leading asymptotics, as the temperature
tends to zero and the distance on the lattice tends to infinity,
of the exponential decay were computed under
similar hypotheses. In \cite{Sj2} even a full asymptotic expansion
has been obtained.
The results of \cite{BJS,Sj1,Sj2} hold for 
translation invariant systems
on finite discrete tori and are uniform in the number of spins.
(In \cite{Sj2} the thermodynamic limit of the derived
expansions is discussed, too.)
They provide a precise description of the
Ornstein-Zernike behaviour
which is actually well-known 
for translation invariant systems since a long time;
see, e.g., \cite{P-L}.
The starting point of the analysis of the correlation
in \cite{BJS,BaMo1,BaMo2,Sj1,Sj2} is the Helffer-Sj\"ostrand formula
\cite{HeSj,Sj1}
which represents the truncated
two-point correlation functions as matrix elements
of the resolvent of a certain Witten-Laplacian on one-forms.
If one works directly in the thermodynamic limit, or
more precisely, if one considers a spin system on 
the lattice $\ZZ^d$ whose equilibrium distribution
is given by a translation invariant tempered Gibbs measure, then 
the formula for the leading asymptotics
derived in \cite{BJS,Sj1} is still valid.
This has been proved in \cite{Ma} by
replacing the Witten-Laplacian by a supersymmetric Dirichlet
operator and again using the Helffer-Sj\"ostrand formula.
In \cite{BaMo1,BaMo2} the authors were able to drop the
assumption of translation invariance in the study
of the correlation asymptotics. (At the same time they were
able to relax the conditions on the on-site potentials
of \cite{BJS,Sj1,Sj2}.) They showed that, still for 
very small temperatures and for energy functionals
with only one global minimum,
the correlation asymptotics are determined by the Hessian
of the energy functional at the global minimum.
They derived this result for finite spin systems on graphs.
Since all their results are again uniform in the number
of spins, it is, however, easy to see that their methods
work also in the infinite-dimensional case.

In this article we choose to work in the thermodynamic
limit and our goal is to analyse the formal Hessian
of the energy functional evaluated at the global minimum
more precisely. The latter is an example of a transition
operator in $\LO(\ell^2)$.
So the main result of this article is actually to develop
a method to calculate the asymptotics of the Green kernel
of a certain class of operators in $\LO(\ell^2)$.
Instead of considering the large distance asymptotics
we fix two points in $\RR^d$ and assume that they sit
on lattices with vanishing lattice spacing,
$\ZZ^d_h=\{h\,x:\,x\in\ZZ^d\}$, $h\in(0,1]$, $d\grg2$. 
At the same time we assume that the spatial variation
of the interaction potentials is fixed, too.
In this way we obtain a kind of continuum limit
which permits to interprete the Hessian
of our energy functional as a $h$-pseudodifferential operator
on $\RR^d$. We remark, however, that we introduce the
continuum limit only for methodological reasons and
that, in the limit, one would end up with a trivial theory.
In order to analyse the Green kernel of the
Hessian we therefore construct a parametrix for
the $h$-pseudodifferential operator.
This is essentially done via a geometric optics 
-or WKB- construction for the solution of a corresponding
heat equation.  
Here we are lead to work with complex-valued symbols
by the introduction of exponential weights.
Consequently, we shall deal with
Fourier integral operators with complex-valued
phase functions \cite{MeSj1,MeSj2,MenSj1,Sj0}. 
Since the results of \cite{MeSj2,MenSj1}
do not directly apply to our situation
we show in detail how they can be suitably modified.
In the end one verifies that the values of the
Green kernel are, up to error terms of infinite order in $h$,
given by the Fourier coefficients of the parametrix.
Applying the method of stationary phase to these
coefficients we find the leading asymptotics, as $h\to0$,
of the Green kernel.
We obtain a formula which is similar
to the expression known from the translation invariant
case and which is determined by a certain Finsler
structure naturally associated to the
$h$-pseudodifferential operator.
If we consider the correlation,
for two points in $\RR^d$ which are
very close to each other, then we 
essentially observe again an Ornstein-Zernike behaviour.
We remark that our results are closely related to those obtained
in \cite{KlRo,Ro} where the authors develop a symbol calculus
on rescaled lattices and discuss Finsler structures associated
to pseudodifferential operators in order to study the tunnel
effect for a general class of difference operators.

The article is organized as follows.
In Section~\ref{1} we state our hypotheses
and our main result precisely.
In Section~\ref{sec-psido}
we determine the $h$-pseudodifferential operator
mentioned above and conjugate it with exponential weights.
In Section~\ref{sec-weight-fct} we construct 
the appropriate exponential weight functions.
In Section~\ref{sec-Ham-Jac} we discuss the eikonal,
or, time-dependent Hamilton-Jacobi equation
proceeding along the lines of \cite{MeSj2,MenSj1}. 
The transport equations are considered
in Section~\ref{sec-heat}.
In Section~\ref{sec-inv} we construct the inverse
of the formal Hessian of our energy functional at the global minimum
and, finally, in Section~\ref{sec-7}
we calculate the leading asymptotics.
The text is followed by three appendices:
In Appendix~\ref{app-BaMo} we prove that
the results of \cite{BaMo2} are applicable
in our situation. In Appendix~\ref{app-OZ-ti2}
we present some elementary calculations which are
used to reformulate the formula for the leading
asymptotics in the translation invariant case.
Some basic facts about almost analytic extensions
are collected in Appendix~\ref{subsec-aa}.


\section{Hypotheses and main results}
\label{1}

For every $h\in(0,1]$ and some $d\in\NN$, $d\grg2$,
we consider a classical system of
real-valued spins on the lattice $\ZZ^d_h:=\{h\,x:\,x\in\ZZ^d\}$,
which is endowed with the Euclidean distance.
A spin configuration is given by an element of $\RR^{\ZZ^d_h}$
and denoted by $\sigma=(\sigma_x)_{x\in\ZZ^d_h}$.
Furthermore, we write $\sigma_\Upsilon=(\sigma_x)_{x\in\Upsilon}$,
for subsets $\Upsilon\subset\ZZ^d_h$.
Our lattice spin model is determined by the choice
of a heuristic energy functional of the system.
The latter describes the interaction of the spins 
and is
(formally) given by
$$
E(\sigma)\,=\,\sum_{x\in\ZZ^d_h}D(x,\sigma_x)
\,+\,\frac{J}{2}\sum_{{x,y\in\ZZ^d_h:\atop0<|x-y|\klg h\,R}} 
W\big(\xyhalf,\xyh,\,\sigma_x-\sigma_y\big)\;,\qquad\sigma\in\RR^{\ZZ^d_h}\,.
$$
Here, the coupling constant $J>0$ will eventually
be assumed to be sufficiently small. 
$R\in[1,\infty)$ is some arbitrarily large but finite
interaction radius. We assume that the on-site potentials, $D$,
and the pair interactions, $W$, fulfill the following

\begin{hypothesis}\label{hyp-UV}
(i) It holds $D\in C^\infty(\RR^d\times\RR,\RR)$ and there exist
$C,\widetilde{C},\varkappa>0$ and $L>2$
such that, for all $(x,\theta)\in\RR^d\times\RR$,
\begin{eqnarray}
|\partial_\theta^\nu D(x,\theta)|&\klg&C\,\SL\theta\SR^{L-\nu}\,,\qquad
\nu=0,1,2\,,\label{karla1}
\\
\partial_{\theta}^2D(x,\theta)&\grg&-C\,,
\\
\big|\,\partial_{\theta}^2D(x,\theta)\,
-\,\partial_{\theta}^2D(x,0)\,\big|&\klg&C\,|\partial_{\theta}D(x,\theta)|\,,
\\
\partial_{\theta}D(x,\theta)\,\sgn(\theta)&\grg&\widetilde{C}\,
\max\big\{\,|\theta|^{1+\varkappa}\,,\,|\theta|\,\big\}\,.
\end{eqnarray}
For $\theta=0$, we have $\partial_\theta^2D(\cdot,0)\in C_b^\infty(\RR^d)$
and
\begin{equation}\label{heribert}
2\,\inf_{x\in\RR^d}\partial_\theta^2D(x,0)\,>\,
\sup_{x\in\RR^d}\partial_\theta^2D(x,0)\,.
\end{equation}

\smallskip

\noindent(ii) For all $\ell\in\ZZ^d$, $0<|\ell|\klg R$, 
we have $W(\cdot,\ell,\cdot\cdot)\in C^\infty(\RR^d\times\RR,\RR)$
and, for $(x,\theta)\in\RR^d\times\RR$,
\begin{eqnarray}
W(x,\ell,\theta)&=&W(x,-\ell,\theta)\;=\;W(x,\ell,-\theta)
\,,\label{karla5}
\\
W(x,\ell,\theta)&\grg&0\,,\quad  W(x,\ell,0) \;=\;0\,,\label{karla6}
\\
|\ell|=1&\Rightarrow&
\partial_\theta^2W(x,\ell,0)\,\grg\,1\,,
\\
\sup_{(x,\theta)\in\RR^{d+1}}\big\{\,
|\partial_\theta^2W(x,\ell,\theta)|&
+&|\partial_\theta^3W(x,\ell,\theta)|
\,\big\}\;<\;\infty\,.\label{karla8}
\end{eqnarray}
Finally, 
$\partial_\theta^2W(\cdot,\ell,0)\in C_b^{\infty}(\RR^d)$.
\end{hypothesis}

\begin{remark} 
The estimates (\ref{karla1})-(\ref{karla8}) are required in
order to apply the results of \cite{AKRT,BaMo2,Ma}.
We assume that $\partial_\theta^2D(\cdot,0)$
and $\partial_\theta^2W(\cdot,\ell,0)$ are elements of
$C_b^{\infty}(\RR^d)$, because this will allow us to
interprete the formal Hessian of $E$ at zero as a
$h$-pseudodifferential operator.
We notice that Hypothesis~\ref{hyp-UV} implies that, for every
$x\in\ZZ^d_h$, the on-site potential $D(x,\cdot)$ has only
one local minimum, namely a non-degenerate one at zero, and
tends to infinity, as $\theta\rightarrow\pm\infty$, faster
than any quadratic polynomial. We also point out that
$D(x,\cdot)$ does not have to be convex.
(\ref{karla5}) and (\ref{karla6}) imply that 
$\partial_\theta^2W(x,\ell,0)\grg0$, for all $x\in\RR^d$
and $\ell\in\ZZ^d$, $0<|\ell|\klg R$. 
\end{remark}

\begin{example}
We let $m\in\NN$, $m\grg2$, and suppose that there are functions
$c_0,c_1,\dots,c_{2m}\in C_b^\infty(\RR^d)$ such that
$\inf c_{2m}>0$, $2\inf c_{2}>\sup c_2$, 
$$
D(x,\theta)\,=\,c_{2m}(x)\,\theta^{2m}\,+\,\dots
\,+\,c_1(x)\,\theta\,+\,c_0(x)\,,\qquad x\in\RR^d\,,\;\theta\in\RR\,,
$$
and $\partial_\theta D(x,\theta)=0$ $\Leftrightarrow$ $\theta=0$,
for $x\in\RR^d$.
Furthermore, we suppose that 
$W(\cdot,\ell,\cdot\cdot)\equiv0$, for $|\ell|\not=1$,
and that there is some $w\in C_b^\infty(\RR^d)$,
$w\grg1/2$, such that
$$
W(x,\ell,\theta)\,=\,w(x)\,\theta^2\,,
$$
for $|\ell|=1$, $x\in\RR^d$, and $\theta\in\RR$.
Then Hypothesis~\ref{hyp-UV} is fulfilled. 
If $c_0,\dots,c_{2m}$ and $w$ are constant, 
we obtain examples of translation invariant
models, which
arise in lattice approximations of Euclidean quantum field
theories of polynomial type and are known as
``$P(\vp)_d$-models''.
\end{example}

\bigskip

\noindent Next, we recall the notion of a tempered Gibbs measure
for the spin system under consideration.
To begin with we introduce, 
for every finite subset $\Upsilon\subset\ZZ^d_h$, 
a (well-defined)
local energy functional with boundary condition 
$\omega\in\RR^{\ZZ^d_h}$,
\begin{eqnarray*}
E_\Upsilon(\sigma_\Upsilon|\omega_{\Upsilon^c})
&:=&\sum_{x\in\Upsilon}D(x,\sigma_x)
\,+\,\frac{J}{2}\sum_{{x,y\in\Upsilon:\atop0<|x-y|\klg h\,R}} 
W\big(\xyhalf,\xyh,\,\sigma_x-\sigma_y\big)
\\
& &
\,+\,J\sum_{{x\in\Upsilon,\,y\in\Upsilon^c:\atop|x-y|\klg h\,R}} 
W\big(\xyhalf,\xyh,\,\sigma_x-\omega_y\big)\;,
\qquad\sigma_\Upsilon\in\RR^{\Upsilon}.
\end{eqnarray*}
We further introduce a local specification, which is
a family of stochastic kernels
indexed by all finite subsets $\Upsilon\subset\ZZ^d_h$
and depending on the parameters
$h\in(0,1]$
and the so-called inverse temperature $\beta\in[1,\infty)$.
Given a tempered boundary condition,
$\omega\in\bigcup_{\alpha\in\NN}\ell_{-\alpha}^2$, where
$\ell^2_\alpha=\{\sigma\in\RR^{\ZZ^d_h}:
\,(\SL x\SR^{\alpha}\,\sigma_x)_x\in\ell^2_\RR(\ZZ^d_h)\}$,
their values are defined by
\begin{equation*}\label{def-localG}
\mu_{\beta,\Upsilon,h}(A,\omega)\,:=\,\!
\frac{1}{\mathscr{Z}_{\Upsilon}(\beta,\omega,h)}
\int\limits_{\RR^{\Upsilon}}
\id_{A}(\sigma_{\Upsilon}|\omega_{\Upsilon^c})
\exp\Big(\!\!
-\beta\,E_{\Upsilon}(\sigma_{\Upsilon}|\omega_{\Upsilon^c})
\Big)\,d\sigma_{\Upsilon}\;,
\end{equation*}
for $A\in \Borel(\RR^{\ZZ^d_h})$,
where $\Borel(\RR^{\ZZ^d_h})$ 
denotes the Borel-$\sigma$-algebra 
defined by the product topology on $\RR^{\ZZ^d_h}$.
The partition function, $\mathscr{Z}_{\Upsilon}(\beta,\omega,h)$,
is a normalization factor.
For $\omega\in\RR^{\ZZ^d_h}\setminus\big(
\bigcup_{\alpha\in\NN}\ell_{-\alpha}^2\big)$,
we set $\mu_{\beta,\Upsilon,h}(A,\omega)=0$.
A probability measure, $\mu_{\beta,h}$, on the Borel sets of
$\RR^{\ZZ^d_h}$ is called a Gibbs measure
(for the model determined by $E$), iff it satisfies 
the Dobrushin-Lanford-Ruelle equilibrium equations,
$$
\int\limits_{\RR^{\ZZ^d_h}}
\mu_{\beta,h,\Upsilon}(A,\omega)\;d\mu_{\beta,h}(\omega)
\,=\,
\mu_{\beta,h}(A)\,,\qquad
\Upsilon\subset\ZZ^d_h\;\;\textrm{finite},\;\;A\in\Borel(\RR^{\ZZ^d_h})\,.
$$
A Gibbs measure $\mu_{\beta,h}$ is called tempered iff
$\mu_{\beta,h}(\ell^2_{-\alpha})=1$, for some $\alpha>d/2$.

Under Hypothesis~\ref{hyp-UV} the results of \cite{AKRT}
are applicable and ensure, for all 
$\beta\in[1,\infty)$, $J>0$, and $h\in(0,1]$,
the existence of a convex set, $\mathscr{G}(\beta,J,h)$, of
tempered Gibbs measures. This convex set contains its extreme points,
which are called pure, tempered Gibbs measures.
We fix such a
pure, tempered Gibbs measure, $\mu_{\beta,h}$,
in the following,
let ${\sf p}$ denote the projection onto the constant
functions in $\HR^0:=L^2(\mu_{\beta,h})$, and set 
${\sf p}^\bot=\id-{\sf p}$. 
We are interested in the asymptotic behaviour, as $h\rightarrow 0$,
of the truncated two-point correlation functions,
$$
\Cor_{\beta,h}\big(\sigma_x\,;\,\sigma_y\big)\,=\,
\SPn{\sigma_x}{{\sf p}^\bot\,\sigma_y}_{\HR^0}\,,
$$
for fixed $x,y\in\ZZ^d_h$,
at very large inverse temperatures 
$\beta$. In this situation we expect the spin configurations
to be localized with high probability near the constant
configuration $(0)_{x\in\ZZ^d_h}$, 
because all on-site potentials $D(x,\cdot)$ have a
unique minimum at zero. In fact, it turns out that the
asymptotic behaviour of $\Cor_{\beta,h}$ is determined
by the formal Hessian of $E$ at zero, $E''(0)\in\LO(\ell^2(\ZZ^d_h))$,
where $E''(\sigma)$ is given by
\begin{equation}
\big(E''(\sigma)\big)_{xy}\,:=\,\partial_{\sigma_x}\partial_{\sigma_y}
E_{\{y\}}(\sigma_y|\sigma_{\{y\}^c})
\,=\,\big(E''(\sigma)\big)_{yx}\,,
\end{equation}
for $x,y\in\ZZ^d_h$ and $\sigma\in\RR^{\ZZ^d_h}$.
Here the second equality follows from
(\ref{karla5}), which we also use to derive the following
explicit formulas,
\begin{equation}\label{def-E''}
\big(E''(0)\big)_{xy}\,=\,
\left\{
\begin{array}{ll}
U_h(x)\,,&x=y\,,\\
-V\big(\xyhalf,\xyh\big)\,,&0<|x-y|\klg h\,R\,,
\\
0\,,&|x-y|>h\,R\,,
\end{array}
\right.
\end{equation}
for $x,y\in\ZZ^d_h$, where
\begin{eqnarray}
U_h(x)&:=&D_{\theta\theta}''(x,0)\,+\,J
\sum_{0<|\ell|\klg R}W_{\theta\theta}''(x+\tfrac{h\ell}{2},\ell,0)
\label{def-Uh}\,,
\\
V(x,\ell)&:=&J\,W_{\theta\theta}''(x,\ell,0)\,,\label{def-V}
\end{eqnarray}
for $x\in\RR^d$, $0<|\ell|\klg R$.
We state the precise relationship between $E''(0)$ and
$\Cor_{\beta,h}$ 
in the following theorem
which is essentially due to Bach and M{\o}ller \cite{BaMo2}.
The latter authors actually consider finite lattice 
systems uniformly in the number of spins. Using the
estimates obtained in \cite{Ma} it is, however, easy to
see that their results extend to our situation.

We remark that, due to (\ref{heribert}) and (\ref{karla8}),
we know that $E''(0)$ is continuously invertible, if
$J>0$ is sufficiently small.

\begin{theorem}\label{thm-MaBaMo}
Assume that Hypothesis~\ref{hyp-UV} 
is fulfilled and let
$\mu_{\beta,h}\in\mathscr{G}(\beta,J,h)$ be pure.
Then there exist $\beta_0\grg1$, $J_0>0$, and $C>0$,
such that, for all $\beta\grg\beta_0$, $J\in[0,J_0]$, and $h\in(0,1]$,
$$
\frac{1-C/\sqrt{\beta}}{\beta}
\:\big(E''(0)^{-1}\big)_{xy}^{1+\frac{C}{\sqrt{\beta}}}
\;\klg\;
\Cor_{\beta,h}\big(\sigma_x\,;\,\sigma_y\big)
\;\klg\;
\frac{1+C/\sqrt{\beta}}{\beta}
\:\big(E''(0)^{-1}\big)_{xy}^{1-\frac{C}{\sqrt{\beta}}}.
$$
\end{theorem}

\proof
The claim follows from the results obtained in \cite{BaMo2}
together with the estimates derived in \cite{Ma}.
We remark that all
statements and estimates of these articles when applied
to our situation
hold uniformly in $h$ since all constants
in Hypothesis~\ref{hyp-UV} are $h$-independent.
We present the details in Appendix~\ref{app-BaMo}. 
\qed

\bigskip

\noindent So, our aim will be to calculate
the leading asymptotics, as $h\rightarrow0$, of 
the matrix elements $\big(E''(0)^{-1}\big)_{xy}$,
for fixed $x,y\in\ZZ^d_h$.
The main point here is that we do not assume
the spin system to be invariant
with respect to translations on the lattice.
In the translation invariant case,
that is, if $D$ and $W$ do not depend on 
the first variable $x$,
the asymptotic behaviour of the two-point correlation functions
is well-known from
Ornstein-Zernike theory; see, e.g. \cite{P-L}.
We recall the precise result in Theorem~\ref{thm-cor-asymp-ti}.
In the translation invariant 
case 
we can actually set $h$ equal to one by a scaling
transformation and consider the large distance
asymptotics instead. 
The detailed form of the leading asymptotics
(\ref{eq-OZ-ti1}) given in Theorem~\ref{thm-cor-asymp-ti}
has been derived in 
\cite{BJS,Sj1,Sj2} for translation invariant systems
on finite discrete tori. 
As a starting point the authors used the Helffer-Sj\"ostrand
formula for the correlation which involves a certain
Witten-Laplacian associated to the lattice model.
The validity of  
(\ref{eq-OZ-ti1}) for pure, translation invariant
tempered Gibbs measures has been proven
in \cite{Ma} again using a Helffer-Sj\"ostrand formula.
In the infinite-dimensional setting of the latter paper
the Witten-Laplacian is replaced by a certain
supersymmetric Dirichlet operator.

Before we state the next theorems we have to introduce
some notation.
First, we define a Hamilton function, 
$H\in C^\infty(T^*\RR^d,\RR)$,
by 
\begin{equation}\label{def-H}
H(x,p)\,:=\,\sum_{{\ell\in\ZZ^d:\atop0<|\ell|\klg R}}
V(x,\ell)\,e^{-\SPn{\ell}{p}}
\,-\,U(x)\,,
\end{equation}
for $x\in\RR^d$, $p\in\RR_d$, where $V$ is given by (\ref{def-V})
and
\begin{equation}
U(x)\,:=\,D_{\theta\theta}''(x,0)\,+\,
\sum_{{\ell\in\ZZ^d:\atop0<|\ell|\klg R}}\label{def-U}
V(x,\ell)\,.
\end{equation}
For each $x\in\RR^d$, $H(x,\cdot)$ is strictly convex
and even and $H(x,0)<0$, provided $J>0$ is small enough.
It therefore makes sense to define the polar body
$$
\Bx_x^*\,:=\,\big\{p\in\RR_d\,\big|\;H(x,p)\,\klg\,0\,\big\}\,,
$$
which is a strictly convex set that is symmetric about the origin
and has a smooth boundary.
Moreover, we let
$$
F(x,v)\,:=\,\sup\big\{\,\SPn{p}{v}\,|\;p\in\Bx_x^*\,\big\}\,,
\qquad v\in T_x\RR^d=\RR^d\,,
$$
denote the support function of $\Bx_x^*$. 
For $v\not=0$, we have the explicit formula
$$
F(x,v)\,=\,\SPn{p(x,v)}{v}\,,
$$
where $p(x,v)\in\Figx_x:=\partial\Bx_x^*$ is the unique point,
where the exterior normal field on $\Figx_x$ points in the direction
of $v$. 
In particular, $F$ is smooth on the slit tangent bundle
$\dot{T}\RR^d=T\RR^d\setminus(\RR^d\times\{0\})$
and absolutely homogenous of degree one in $v$,
i.e. $F(x,\lambda\,v)=|\lambda|\,F(x,v)$, for
$\lambda\in\RR$, $(x,v)\in T\RR^d$.
We also introduce the positive definite
$d\times d$-matrix $G(x,v)$
with entries
$$
G(x,v)_{ij}\,:=\,F(x,v)\,F_{v^iv^j}''(x,v)\,
+\,F_{v^i}'(x,v)\,F_{v^j}'(x,v)\,,\qquad (x,v)\in\dot{T}\RR^d\,.
$$
$G_{ij}$ is homogenous of degree zero in $v$,
i.e. $G_{ij}(x,\lambda\,v)=\,G_{ij}(x,v)$, for
$\lambda\in\RR$, $(x,v)\in\dot{T}\RR^d$.
Finally, we let $H_{pp}''(x,p)^\bot$ denote the Hessian
of $H(x,\cdot)$ at $p\not=0$ restricted to the orthogonal complement
of $\nabla_pH(x,p)$ in $\RR_d$.
In the translation invariant setting of the next theorem
the functions
$H,F,p$, and $G$ are $x$-independent, whence
we drop the reference to $x$ in the notation of its statement.

\begin{theorem}\label{thm-cor-asymp-ti}
Assume that Hypothesis~\ref{hyp-UV} is fulfilled,
that $J>0$ is sufficiently small, and
assume additionally that $D(x,\theta)$
and $W(x,\ell,\theta)$ are constant in $x$.
Then, for sufficiently large $\beta$, there
is a unique tempered Gibbs measure determined
by $D$ and $W$. It is invariant with respect
to translations on the lattice and
its truncated two-point correlation functions
fulfill
\begin{eqnarray}
\label{eq-OZ-ti1}
\Cor_{\beta,h}\big(\sigma_x\,;\,\sigma_y\big)&=&
\frac{1}{\beta}\,
\frac{\big|\nabla_pH(p(x-y))\big|^{\frac{d-3}{2}}}{\sqrt{\det H_{pp}''(p(x-y))^\bot}}
\,\frac{1+\bigO(\frac{h}{|x-y|})+
\bigO(\beta^{-1/2})}{(2\pi\frac{|x-y|}{h})^{\frac{d-1}{2}}}
\\
& &\;\;\nonumber
\times\exp\Big(-\big(1+\bigO(\beta^{-1/2})\big)\,F(\tfrac{x-y}{h})\Big)
\\
&=&\nonumber\frac{1}{\beta}\,
\frac{\sqrt{\det G(x-y)}}{\SPb{p(x-y)}{\nabla_pH(p(x-y))}}\,
\frac{1+\bigO(\frac{h}{|x-y|})+
\bigO(\beta^{-1/2})}{\big(2\pi F(\tfrac{x-y}{h})\big)^{\frac{d-1}{2}}}
\\
& &\;\;\label{eq-OZ-ti2}
\times\exp\Big(-\big(1+\bigO(\beta^{-1/2})\big)\,F(\tfrac{x-y}{h})\Big)\,,
\end{eqnarray}
as $\beta\rightarrow\infty$ and $|x-y|/h\rightarrow\infty$, 
$x,y\in\ZZ^d_h$.
\end{theorem} 

\proof
Using (\ref{eq-OZ-ti1}), which is derived in the above mentioned
papers, and some elementary linear algebra
we deduce Formula (\ref{eq-OZ-ti2}) in Appendix~\ref{app-OZ-ti2}.
\qed

\bigskip

\noindent 
In the following theorem we state the main result of this article.
To formulate it, we introduce the Finsler distance associated
to $F$,
$$
\Fdist(x,y)\,=\,\inf_q\int F(q,\dot{q})\,,\qquad x,y\in\RR^d\,,
$$ 
where the infimum is taken over all piecewise smooth curves
$q:[0,\tau]\to\RR^d$ such that $\tau>0$, $q(0)=y$,
$q(\tau)=x$. It turns out that the infimum is 
always attained
by at least one (smooth) geodesic from $y$ to $x$. Since $F$
is absolutely homogenous of degree one in $v$
any reparametrization of a minimizing geodesic
yields again a minimizing geodesic and $\Fdist$ is
symmetric.
We can always reparametrize any minimizing geodesic
in such a way that we obtain the projection of a Hamiltonian
trajectory defined by $H$ which runs in the level surface
$\Figx:=H^{-1}(\{0\})$.
If there is, up to reparametrization,
a unique minimizing geodesic from $y$ to $x$
we denote its initial (resp. end)
velocity and momentum in the just mentioned special
parametrization by $v_y$ and $p_y$
(resp. $v_x$ and $p_x$).
We remark that we have $\SPn{p_x}{v_x}>0$
and $v_x=\nabla_p H(x,p_x)$
and analogous statements for $y$. 
As in Riemannian geometry one may define
the notion of a conjugate point
for $x$ and $y$ with respect to $F$; see Section~\ref{sec-weight-fct}
below.

Finally, we set $h_n:=1/2^n$, $n\in\NN$.
If $N\in\NN$ and $x\in\ZZ_{h_N}^d$,
then of course $x\in\ZZ_{h_n}^d$, for all integers $n\grg N$.

\begin{theorem}\label{thm-main}
Assume that Hypothesis~\ref{hyp-UV} is fulfilled
and that $J>0$ is sufficiently small.
Let $N\in\NN$, $x,y\in\ZZ^d_{h_N}$, and 
assume that, up to reparametrization, there exists a unique geodesic
minimizing Finslerian arc length
from $y$ to $x$ and that $x$ and $y$ are not conjugate to each
other.
(This assumption is always fulfilled, for fixed $y$,
provided $x$ is sufficiently close to $y$.)
Then 
\begin{eqnarray*}
\big(E''(0)^{-1}\big)_{xy}&=&
\frac{\sqrt[4]{\det\big(G(x,v_x)\,G(y,v_y)\big)}}{
\Delta(x,y)\,\sqrt{\SPn{p_x}{v_x}\SPn{p_y}{v_y}}}
\frac{\exp\big(-\Fdist(x,y)/h_n\big)}{
\big(2\pi\,\Fdist(x,y)/h_n\big)^{\frac{d-1}{2}}}
\\
& &
\;+\,\bigO(h_n^{\frac{d+1}{2}})\,\exp\big(-\Fdist(x,y)/h_n\big)\,,
\end{eqnarray*}
as $n\rightarrow\infty$, where
$$
\Delta(x,y)\,=\,1\,+\,\bigO(\Fdist(x,y)^{1/2})\,,
\qquad \textrm{as}\;\;x\rightarrow y\,.
$$
\end{theorem}

\proof
The theorem is proved by combining 
Propositions~\ref{prop-vp}(ii),~\ref{le-karin},~\ref{prop-asymp1} 
\& \ref{prop-ina} and Equations
(\ref{ansatz-inv1})\&(\ref{PundI}).
\qed

\begin{remark}
The function $\Delta(x,y)$ can be expressed explicitely 
in terms of transversal Jacobi fields along the 
unit speed geodesic, $q$,
from $y$ to $x$; see Proposition~\ref{prop-ina}
and the remarks preceeding it. If the geodesics emanating
from $y$ are dispersing along $q$, then
$\Delta(x,y)$ increases when $\Fdist(x,y)$ gets large.
If they are bunching together, $\Delta(x,y)$ decreases
and vanishes, if $x$ is conjugate to $y$.
If the flag curvature \cite{BCS} happens to be zero
along $q$, then $\Delta(x,y)$ is equal to one. 
\end{remark}


\section{A related $h$-pseudodifferential operator}\label{sec-psido}

\noindent The basic idea underlying our analysis is the fact that,
for every $f\in\ell^2(\ZZ^d_h)$, the expression
(recall (\ref{def-E''})-(\ref{def-V}))
\begin{equation}\label{for-E''f}
\big(E''(0)\,f\big)(x)\,=\,
U_h(x)\,f(x)\,-\,\sum_{{y\in\ZZ^d_h:\atop
0<|x-y|\klg hR}}V\big(\xyhalf,\xyh\big)\,f(y)\,,
\end{equation}
can be interpreted as the image of any $\tilde{f}\in C^\infty(\RR^d)$
with $\tilde{f}\!\!\upharpoonright_{\ZZ^d_h}=f$
under a $h$-pseudodifferential operator evaluated at 
the point $x\in\ZZ^d_h$.
Before we explain this correspondence in detail, 
we fix some notation
and recall some
general facts.

For our purposes it is sufficient to consider
only bounded symbols:
Let $\Omega$ be an open subset of
$\RR^d$, $\RR^d\times\RR_d$, or $[0,\infty)\times\RR^d\times\TT^d$
etc., 
where $\TT^d:=(\RR/2\pi\ZZ)^d$
is the $d$-dimensional torus.
Then we write 
$f\in \Symbb{\Omega}$, 
for a function
$f:\Omega\times(0,1]\to\CC$, iff $f(\cdot\,;h)\in C^\infty(\Omega)$,
for all $h\in(0,1]$, and there is some $h_0\in(0,1]$ such that,
for all multi-indices $\alpha$,
$$
\sup_{(\omega,h)\in\Omega\times(0,h_0]}|\partial_\omega^\alpha f(\omega\,;h)|\,<\,\infty\,.
$$
We recall that the Weyl quantization,
$\Op(a)$, of any symbol $a\in\Symbb{\RR^d\times\RR_d}$
is determined by the oscillatory integrals
\begin{equation}\label{def-OpW}
\Op(a)\,\tilde{f}(x)\,=\,
\int e^{i\SPn{\xi}{x-y}/h}a\big(\xyhalf\,,\,\xi\big)\,\tilde{f}(y)
\;\frac{dyd\xi}{(2\pi h)^d}\,,\qquad \tilde{f}\in \Schwartz(\RR^d)\,,
\;\;x\in\RR^d\,,
\end{equation}
where $\Schwartz(\RR^d)$ is the space of Schwartz test functions
on $\RR^d$.
The possibility to regard (\ref{for-E''f}) as the action of
a pseudodifferential operator is a consequence of the
following:
If $a\in\Symbb{\RR^d\times\RR_d}$ is a symbol which
is $(2\pi\ZZ)^d$-periodic in $\xi$, then its Weyl quantization
has the distribution kernel
\begin{equation}\label{distr-ker}
K_{a}(x,y)\,=\,(\FT_h^{-1} a)(\xyhalf,x-y)\,
=\,h^d\sum_{z\in\ZZ^d_h}\widehat{a}(\xyhalf,z)\,\delta_z(y-x)\,,
\end{equation}
where $(\FT_h\, u)(\xi)=\int e^{-i\SPn{\xi}{x}/h}\,u(x)\,dx$
is the semiclassical Fourier transform and
\begin{equation}\label{Fourier-coeff}
\widehat{a}(x,z)\,=\,
\int_{\TT^d}e^{-i\SPn{z}{\xi}/h}a(x,\xi)\:
\frac{d\xi}{(2\pi h)^d}\,,\qquad z\in\ZZ^d_h\,,
\end{equation}
are the Fourier coefficients of $a(x,\cdot)$.
So if $\tilde{f}\in \Schwartz(\RR^d)$ 
we get,
\begin{equation}\label{soeren}
\Op(a)\,\tilde{f}(x)\,=\,h^d\sum_{y\in x+\ZZ^d_h}
\widehat{a}\big(\xyhalf,\,y-x\big)\,\tilde{f}(y)\,,\qquad x\in\RR^d\,.
\end{equation}
To apply these remarks to $E''(0)$ we consider
the Fourier
series corresponding to the $V(\cdot,\ell)$,
\begin{equation}\label{Vhat}
\widetilde{V}(x,\xi)\,:=\,\sum_{{\ell\in\ZZ:\atop0<|\ell|\klg R}}
e^{i\SPn{\xi}{\ell}}V(x,\ell)
\,=\,\sum_{{\ell\in\ZZ:\atop0<|\ell|\klg R}}
V(x,\ell)\,\cos(\SPn{\xi}{\ell})\,,\qquad x\in\RR^d\,,\;\xi\in\CC_d\,.
\end{equation}
(Here the second
expression follows from the first
since $V$ is even with respect to $\ell$.)
As a trigonometric polynomial $\widetilde{V}(x,\cdot\,)$
is of course
entire and $(2\pi\ZZ)^d$-periodic with respect to real $\xi$. 
By Hypothesis~\ref{hyp-UV}, $U_h,V(\cdot,\ell)\in\Symbb{\RR^d}$,
and we see that
the symbol $U_h-\widetilde{V}$ is contained in the class
$\Symbb{\RR^d\times\RR_d}$.

\begin{proposition}\label{prop-elfriede}
For all 
$f\in\ell^2(\ZZ^d_h)$ and $\tilde{f}\in C^\infty(\RR^d)$
with $\tilde{f}\!\!\upharpoonright_{\ZZ^d_h}=f$,
\begin{equation}\label{elfriede}
\big(E''(0)\,f\big)(x)\,=\,\Op(U_h-\widetilde{V})\,\tilde{f}(x)\,,
\qquad x\in\ZZ^d_h\,.
\end{equation}
\end{proposition}

\proof
Since 
$(\widetilde{V})^{\wedge}(x,z)=h^{-d}\,V(x,z/h)=h^{-d}\,V(x,-z/h)$,
$0<|z|\klg h\,R$, and $(\widetilde{V})^{\wedge}(x,z)=0$
otherwise,
(\ref{elfriede})
is a special case of 
(\ref{soeren}) with $x\in\ZZ^d_h$, provided
$\tilde{f}\in \Schwartz(\RR^d)$.
We know, however, that $\Op(U_h-\widetilde{V})$ maps
$C^\infty(\RR^d)$ into itself, because only
finitely many Fourier coefficients of $\widetilde{V}$
are non-vanishing. Consequently, (\ref{soeren}) is
available, for every $\tilde{f}\in C^\infty(\RR^d)$, 
in this case.
\qed

\noindent Proposition~\ref{prop-elfriede} suggests to study the inverse
of $E''(0)$ by means of
the semiclassical microlocal analysis of 
the operator $\Op(U_h-\widetilde{V})$.
To work out the exponential decay of $(E''(0)^{-1})_{xy}$, we shall,
however, first conjugate the operators in (\ref{elfriede})
with suitable exponential weights and then construct
a parametrix for the conjugated operators.
Therefore, we assume that we are given some
weight function $\phi\in \Symbb{\RR^d}$
in the following and use
the symbol $e^{\phi/h}$
also to denote the diagonal multiplication operator
on $\ell^2(\ZZ^d_h)$, whose action on the canonical
orthonormal basis vectors is given
by $e^{\phi/h}e_x:=e^{\phi(x)/h}e_x$, $x\in\ZZ^d_h$.
We remark that, eventually, 
it will be necessary to consider also
conjugations with exponential weight functions
that depend additionally
on a time and a momentum parameter.

In the next lemma
we derive a classical asymptoic expansion of the Weyl symbol
corresponding to the conjugated operators.
As we shall see in the proof
this is actually possible
using only completely elementary arguments since
we are dealing with trigonometric polynomials.

\begin{lemma}\label{le-ernst1}
There is a classical Weyl symbol, $a^\phi_W\in\Symbb{\RR^d\times\RR_d}$,
such that
\begin{equation}\label{lisa4a}
e^{\phi/h}\,\Op(U_h-\widetilde{V})\,e^{-\phi/h}\,\tilde{f}(x)
\,=\,
i\,\Op(a_W^\phi)\,\tilde{f}(x)\,,
\qquad x\in\RR^d\,,
\end{equation}
and, in particular,
\begin{equation}\label{lisa4b}
e^{\phi/h}\,E''(0)\,e^{-\phi/h}\,f(x)
\,=\,
i\,\Op(a_W^\phi)\,\tilde{f}(x)\,,
\qquad x\in\ZZ^d_h\,,
\end{equation}
for all $\tilde{f}\in C^\infty(\RR^d)$
and $f\in\ell^2(\ZZ^d_h)$
such that
$\tilde{f}\!\!\upharpoonright_{\ZZ^d_h}=f$.
$a^\phi_W$ is a trigonometric polynomial in $\xi$
and
admits a classical asymptotic expansion in $\Symbb{\RR^d\times\RR_d}$,
\begin{equation}\label{lisa4c}
a^\phi_W(x,\xi)\,\asymp\,
\sum_{\nu=0}^\infty h^{2\nu}\,a_{2\nu}^\phi(x,\xi)\,,
\end{equation}
where
$a^\phi_{0},a^\phi_2,\dots$
are trigonometric polynomials in $\xi$, too.
Its principal symbol is given by
(recall (\ref{def-U}))
\begin{equation}\label{def-aphi}
a^\phi(x,\xi)\,:=\,a_0^\phi(x,\xi)\,=\,i\,\widetilde{V}(x,\xi+i\phi'(x))\,-\,
i\,U(x)\,,\qquad (x,\xi)\in\RR^d\times\RR_d\,.
\end{equation}
If $\phi\equiv\phi(t,x,\eta)\in 
C^\infty([0,\infty)\times\RR^d\times\RR_d)$
is such that 
$\phi+i\SPn{x}{\eta}$ defines an element of
$\Symbb{[0,\infty)\times\RR^d\times\TT^d}$, 
then all symbols $a^\phi_W,a^\phi_{0},a^\phi_2,\dots$
can be viewed as elements of 
$\Symbb{[0,\infty)_t\times\RR^d_x\times\RR_{d,\xi}\times\TT^d_\eta}$
and the expansion
(\ref{lisa4c}) is valid in 
$\Symbb{[0,\infty)\times\RR^d\times\RR_d\times\TT^d}$.
\end{lemma}

\proof
First, we consider the multiplication operator 
$e^{\phi/h}\,\Op(U_h)\,e^{-\phi/h}=U_h$.
Using (\ref{karla5}), which
implies $V(x,\ell)=V(x,-\ell)$, for all $x\in\RR^d$
and $0<|\ell|\klg R$, we write
$$
U_h(x)\,=\,\,D_{\theta\theta}''(x,0)\,+\,\frac{1}{2}
\sum_{0<|\ell|\klg R}\big(\,V(x+\tfrac{h\ell}{2},\ell)
\,+\,V(x-\tfrac{h\ell}{2},\ell)\,\big)\,.
$$ 
By Taylor's formula,
we find, for every $N\in\NN$, some
$\Psi_N(\cdot,\ell\,;h)\in\Symbb{\RR^d}$ such that
$$
V(x+\tfrac{h\ell}{2},\ell)
\,+\,V(x-\tfrac{h\ell}{2},\ell)
\,=\,
\sum_{\nu=0}^N\frac{2}{(2\nu)!}\,\Big(\frac{h}{2}\Big)^{2\nu}
\,\SPn{\ell}{\nabla}^{2\nu}V(x,\ell)\,+\,h^{2N+2}\,
\Psi_N(x,\ell\,;h)\,.
$$
By (\ref{Fourier-coeff})-(\ref{Vhat}) 
we further have,
for all $x\in\RR^d$, 
\begin{eqnarray}
\lefteqn{\nonumber
e^{\phi/h}\,\Op(\widetilde{V})\,e^{-\phi/h}\,\tilde{f}(x)
}
\\
&=&\label{mathilde}
h^{d}\!\!\!\!\sum_{y\in x+\ZZ^d_h}
\sum_{{\ell\in\ZZ^d:\atop0<|\ell|\klg R}}\int_{\TT^d}
e^{i\SPn{\xi}{x-y}/h+i\SPn{\xi}{\ell}}\:e^{(\phi(x)-\phi(y))/h}
\,V(\xyhalf,\ell)\,\tilde{f}(y)\:
\frac{d\xi}{(2\pi h)^d}\,.
\end{eqnarray}
Here the integrals under the double sum are non-zero
only if $y-x=h\,\ell$. If the latter condition
is fulfilled we have, however, again by 
Taylor's formula,
\begin{eqnarray}
\lefteqn{
\frac{1}{h}\:\big(\phi(x)-\phi(y)\big)
\,=\,
\frac{1}{h}\:\Big(\phi\big(\xyhalf-\tfrac{h\ell}{2}\big)
\,-\,\phi\big(\xyhalf+\tfrac{h\ell}{2}\big)\Big)\label{mathilde2}
}
\\
&=&
-\SPn{\ell}{\nabla\phi(\xyhalf)}\,-\,
\sum_{\nu=1}^N\nonumber
\frac{1}{(2\nu+1)!}\,\Big(\frac{h}{2}\Big)^{2\nu}\:
\SPn{\ell}{\nabla}^{2\nu+1}\phi(\xyhalf)
\,+\,h^{2N+2}\,\Phi_N(\xyhalf,\ell\,;h)\,,
\end{eqnarray} 
where $\Phi_N(\cdot,\ell\,;h)\in\Symbb{\RR^d}$, for every
$N\in\NN$. 
Plugging (\ref{mathilde2}) into (\ref{mathilde}),
writing out the exponential series 
\begin{equation}\label{knut88}
\exp\Big(\,-\,
\sum_{\nu=1}^N
\frac{1}{(2\nu+1)!}\,\Big(\frac{h}{2}\Big)^{2\nu}\:
\SPn{\ell}{\nabla}^{2\nu+1}\phi(\xyhalf)
\,+\,h^{2N+2}\,\Phi_N(\xyhalf,\ell\,;h)\Big)
\end{equation}
and ordering the terms
with respect to powers of $h$
we thus get,
for every $N\in\NN$,
\begin{eqnarray}
\lefteqn{\label{lisa3}
e^{\phi/h}\,\Op(U_h-\widetilde{V})\,e^{-\phi/h}\,\tilde{f}(x)
}
\\
&=&
h^{d}\!\!\!\sum_{y\in x+\ZZ^d_h}\int_{\TT^d}\nonumber
e^{i\SPn{\xi}{x-y}/h}\Big(\sum_{\nu=0}^{N-1} h^{2\nu}\,
i\,a_{2\nu}^\phi(\xyhalf,\xi)
\,+\,h^{2N}r_{2N}(\xyhalf,\xi\,;h)\Big)\,
\tilde{f}(y)\,\frac{d\xi}{(2\pi h)^d}\,,
\end{eqnarray}
where $a_0^\phi,a_{2}^\phi,\dots$ 
and $r_{2},r_4,\dots$ are Weyl symbols, 
which
are trigonometric polynomials in $\xi$.
For the principal symbol we find
$$
a_0^\phi(x,\xi)\,=\,i\sum_{{\ell\in\ZZ^d:\atop0<|\ell|\klg R}}
V(x,\ell)\,e^{i\SPn{\ell}{\xi}-\SPn{\ell}{\phi'(x)}}\;-\;i\,U(x)\,,
$$
which is (\ref{def-aphi}).
Applying
(\ref{Fourier-coeff}) and (\ref{soeren}) to (\ref{lisa3})
we obtain
(\ref{lisa4a})-(\ref{lisa4c}).

If now $\phi\equiv\phi(t,x,\eta)$
is such that $\phi+i\SPn{x}{\eta}$ is $2\pi$-periodic in $\eta$,
we observe first that $\partial_t\phi$, $\phi_\eta'$,
and $\partial_{(t,x,\eta)}^\alpha\phi$, for $|\alpha|\grg2$,
are again $2\pi$-periodic functions. For the first derivative
with respect to $x$ we get 
$\phi_x'(t,x,\eta+e)=\phi_x'(t,x,\eta)-i\,e$, for $e\in(2\pi\ZZ)^d$.
This shows that all derivatives of $\phi$ appearing in
(\ref{knut88}) are $2\pi$-periodic in $\eta$
as well as $e^{-\SPn{\ell}{\phi_x'(t,x,\eta)}}$, for $\ell\in\ZZ^d$,
$0<|\ell|\klg R$.
Therefore, all symbols $a_0^\phi,a_{2}^\phi,\dots$ 
and $r_{2},r_4,\dots$ can be viewed as elements of
$\Symbb{[0,\infty)\times\RR^d\times\RR_d\times\TT^d}$,
if this is possible for $\phi+i\SPn{x}{\eta}$,
and it is clear that the last statement holds true.
\qed

\bigskip

\noindent In order to construct
a parametrix for $\Op(a_W^\phi)$
we will solve a corresponding
heat equation by means of a WKB construction.
For this purpose we state the following lemma, 
which is a simple consequence of the method of stationary phase.
We suppose again that 
$\phi$
depends on two additional
parameters
such that 
$\phi+i\SPn{x}{\eta}\in\Symbb{[0,\infty)\times\RR^d\times\RR_d}$ 
is $2\pi$-periodic in $\eta$.
We further introduce the vector field
$$
Y(t,x,\eta)\,:=\sum_{i=1}^d\partial_{\xi^i}
a^{\phi(t,\cdot,\eta)}(x,\xi)\big|_{\xi=0}\,\partial_{x^i}\,,
$$
and set $a_1^\phi\equiv a_3^\phi\equiv\dots\equiv0$. 

\begin{lemma}\label{le-stat-phase}
Assume that $b_\nu\in \Symbb{[0,\infty)\times\RR^d\times\TT^d}$, 
for $\nu\in\NN_0$, and suppose that
there is some compact subset, 
$\mathcal{K}\Subset\RR^{d}$,
such that $\supp\big(b_\nu(t,\cdot,\eta)\big)\Subset\mathcal{K}$,
for $\nu\in\NN_0$ and $(t,\eta)\in[0,\infty)\times\TT^d$.
Let
$b(t,x,\eta\,;h)$ be a Borel resummation of the formal series
$\sum_{\nu=0}^\infty h^\nu b_\nu(t,x,\eta)$. 
Then
we find symbols
$\widetilde{r}_N\in \Symbb{[0,\infty)\times\RR^{d}\times\TT^d}$,
$N\in\NN$, with
$$
\supp\big(\widetilde{r}_N(t,\cdot,\eta)\big)\,\subset\,
\mathcal{K}+\{|x|\klg R\}\,\qquad
(t,\eta)\in[0,\infty)\times\TT^d\,,
$$
such that, for all 
$(t,x,\eta)\in[0,\infty)\times\RR^{d}\times\TT^d$
and sufficiently small $h>0$,
\begin{eqnarray*}
\lefteqn{
\big(i\,\Op(a_W^\phi)\,b\big)(t,x,\eta)
}
\\
&=& i\,a^{\phi(t,\cdot,\eta)}(x,0)\,b(t,x,\eta\,;h)\,+
\,h\,\big((Y+\tfrac{1}{2}\,\diz Y)\,b_0\big)(t,x,\eta)
\\
& &+\;\sum_{\nu=2}^{N-1}h^\nu 
\Big(\big((Y+\tfrac{1}{2}\,\diz Y)\,b_{\nu-1}\big)(t,x,\eta)\,
+\,\sum_{\vk=0}^{\nu-2}
(P_{\nu-\vk}\,b_\vk)(t,x,\eta)\Big)
\\
& &\,+\,h^N\,\widetilde{r}_N(t,x,\eta\,;h)\,,
\end{eqnarray*}
where, for $\tilde{f}\in C^\infty(\RR^d)$,
\begin{equation}\label{def-Pnu}
(P_\nu\, \tilde{f})(t,x,\eta)\,=\,i
\sum_{\vk=0}^\nu\frac{1}{i^\vk\vk!}
\Big(\sum_{i=1}^d\partial_{y^i}\partial_{\xi_i}\Big)^\vk
\big(a_{\nu-\vk}^{\phi(t,\cdot,\eta)}(\xyhalf,\xi)\,\tilde{f}(y)\big)
\big|_{\xi=0,\,y=x}\,.
\end{equation}
\end{lemma}


\section{Construction of a weight function}
\label{sec-weight-fct}

\noindent In order to study the exponential decay
of the Green kernel of $E''(0)$ we distinguish two
points $x^\star$ and $y^\star$
and design a suitable
weight function, $\vp$, which models the exponential decay
of $\big(E''(0)^{-1}\big)_{x^\star y^\star}$.
Here we can choose any pair of points $x^\star,y^\star$
satisfying a certain condition introduced below.
An appropriate weight function will essentially
be given as a solution
of 
the Hamilton-Jacobi equation
\begin{equation}\label{eq-T-U}
T(x,\vp'(x))\,-\,U(x)\;=\;0\;,
\end{equation}
where $T(x,p):=\widetilde{V}(x,ip)$.
For our purposes it is actually sufficient
to solve (\ref{eq-T-U}) only in a small
neighbourhood of a certain Finsler geodesic from
$y^\star$ to $x^\star$. 
The solution $\vp$ will then be extended in such a way
that, for all $x$
outside
that neighbourhood, it holds $T(x,\vp'(x))-U(x)<0$.
The construction of $\vp$ is presented in Subsection~\ref{ssec-vp}
below.
In Subsection~\ref{ssec-avp} we collect various properties
of the principal symbol $a^\vp$ defined by (\ref{def-aphi})
with $\phi=\vp$, which play an important role in the sequel.

\subsection{The construction of $\vp$}\label{ssec-vp}

There is no reason to restrict ourselves
to the special Hamilton function $T-U$
appearing in (\ref{eq-T-U})
for the purpose of this subsection.
In Subsection~\ref{ssec-avp} we will observe that
$T-U$ is an element of the set of
all Hamilton
functions, $H$,
that satisfy

\begin{hypothesis}\label{hyp-H}
It holds $H\in C^\infty(T^*\RR^d,\RR)$. 
For all $x\in\RR^d$, the function 
$H(x,\cdot\,):T^*_x\RR^d\rightarrow\RR$ is 
strictly convex, even, and $H(x,0)<0$.
\end{hypothesis}

\smallskip

\noindent 
We thus seek for
solutions of the Hamilton-Jacobi equation
$
H(x,\vp'(x))=0
$, where $H$ satisfies Hypothesis~\ref{hyp-H}.
The appropriate solution
will essentially be given by the Finsler distance function to some
prescribed point on the prolongation
of the geodesic from $y^\star$ to $x^\star$. 
Before we turn to the construction of $\vp$
we recall 
some required notions and facts from Finsler geometry. 
In Section~\ref{sec-7} we shall also make use of them. 
All the standard results 
collected below can be found, e.g., in \cite{BCS,GH,Li}.

We have already introduced the polar bodies
$$
\Bx_x^*\;:=\;\big\{\,p\in\RR_d=T^*_x\RR^d\,:\,H(p,x)\,\klg\,0\,\big\}\,,
\qquad x\in\RR^d\,.
$$
For every $x\in\RR^d$, $\Bx_x^*$
is strictly convex and symmetric about the origin.
Its boundary,
$$
\Figx_x\;:=\;\big\{\,p\in T^*_x\RR^d\,:\,H(p,x)\,=\,0\,\big\}\;,
$$
which is called the figuratrix at $x$,
is a smooth submanifold.
We further set $\Figx:=H^{-1}(\{0\})$.
We also recall that the support function, 
$\FS(x,\cdot):\RR^d=T_x\RR^d\rightarrow[0,\infty)$,
of $\Bx_x^*$ is given by
\begin{equation}\label{def-supp-fct}
\FS(x,v)\;:=\;\sup \big\{\,\SPn{p}{v}:\:p\in \Bx_x^*\,\big\}
\;,\qquad v\in\RR^d\;.
\end{equation}
The set
$$
\Bx_x\;:=\;\big\{v\in\RR^d=T_x\RR^d\,:\,F(x,v)\,\klg\,1\big\}
$$
is again strictly convex and symmetric about the origin,
and its smooth boundary,
$$
\Indx_x\,:=\,\big\{v\in T_x\RR^d\,:\,F(x,v)\,=\,1\big\}\,,
$$
is called the indicatrix at $x\in\RR^d$. 
For $v\not=0$, the value of the support function is given by
$\FS(x,v)=\SPn{p(x,\nv)}{v}$, where $p(x,\nv)$ is the unique
point on $\Figx_x$ at which the exterior normal field on $\Figx_x$,
i.e. the normalized gradient of $H$ with respect to $p$,
equals $\nv:=v/|v|$. 
The function $\FS$ is continuous on $T\RR^d$ and smooth on 
$\dot{T}\RR^d$.
(Here and in the following we write
$\dot{T}\RR^d$ for the tangent bundle with the zero section
deleted and use a similar notation for the cotangent bundle.)
Moreover,
$\FS(x,\cdot)$ is absolutely homogenous of degree one,
i.e. $\FS(x,\theta\,v)=|\theta|\,\FS(x,v)$,
for all $(x,v)\in T\RR^d$ and $\theta\in\RR$.
Euler's theorem thus implies 
\begin{equation}\label{eq-F'F''}
F_v'(x,v)\,=\,p(x,\nv)\quad \textrm{and}\quad
F_{vv}''(x,v)v\,=\,0\,,\quad(x,v)\in \dot{T}\RR^d\,. 
\end{equation}
The restriction of 
$F_{vv}''(x,v)$ to the orthogonal complement of $v$ is strictly
positive.
The matrix $G$ with entries
$$
G_{ij}\,:=\,\big(\tfrac{1}{2}F^2\big)_{v^iv^j}''
\,=\,F\,F_{v^iv^j}''\,+\,F_{v^i}'\,F_{v^j}'\,,\quad1\klg i,j\klg d\,,
$$
is therefore positive definite at every point of 
$\dot{T}\RR^d$. 
Altogether $F$ enjoys all properties required for a Finsler structure
on $T\RR^d$; see, e.g., \cite{BCS}.
In the calculus of variations $F$ is called an elliptic parametric
Lagrangean and $H$ is an associated Hamiltonian
in the sense of Carath\'eodory,
as $H_p'\not=0$ on $\dot{T}^*\RR^d$ and 
$H(x,F_v'(x,v))=0$, for $(x,v)\in \dot{T}\RR^d$;
see, e.g., \cite{GH}.
We also introduce

\begin{hypothesis}\label{hyp-mM}
$$
\inf\big\{\,F(x,\mathring{v})
\,\big|\,x\in\RR^d\,,\,\,\mathring{v}\in S^{d-1}\,\big\}\,>\,0
\,.
$$
\end{hypothesis}

\bigskip

\noindent 
Since $F$ is absolutely homogenous of degree one in $v$,
the map
$\Fdist:\RR^d\times\RR^d\rightarrow[0,\infty)$,
given by
\begin{equation}\label{def-Fdist}
\Fdist(x,y)\;:=\;\inf_{q}\mathscr{A}(q)\,,\qquad
\mathscr{A}(q)\,:=\,\int\limits\FS(q,\dot{q})\,,
\end{equation}
defines a metric on $\RR^d$. 
Here the infimum is taken over all piecewise smooth curves
in $\RR^d$
from $y$ to $x$.
(One could equally well take the minimum over Lipschitz continuous
curves defined on the unit interval, which would give the same
results.)
In \cite[\textsection5.3]{Li} it is shown that
\begin{equation}\label{teresa}
\widetilde{\vp}(x)\,:=\,\Fdist(x,y)\,,\qquad x\in\RR^d\,,
\end{equation}
is the (unique) viscosity solution of
$$
H(x,\widetilde{\vp}'(x))\,=\,0\,,\;\;\textrm{for}\;\;
x\in\RR^d\setminus\{y\}\,,
\qquad \widetilde{\vp}(y)\,=\,0\,.
$$
In particular, this means that $\widetilde{\vp}$ is continuous,
differentiable on $\RR^d\setminus(\{y\}\cup\Sigma)$, 
where $\Sigma$ is some closed set of finite
$(d-1)$-dimensional Hausdorff measure \cite{LiNi},
and satisfies
$H(x,\widetilde{\vp}'(x))=0$, for
$x\in\RR^d\setminus(\{y\}\cup\Sigma)$. 
Moreover, it is known that, for any function
$u:\RR^d\to\RR$ with $u(y)=0$, which is locally Lipschitz continuous
and satisfies $H(x,u'(x))\klg0$ almost everywhere,
it holds $u\klg\widetilde{\vp}$.
In Section~\ref{sec-Ham-Jac} it will become clear that
the latter property shows that $\widetilde{\vp}$ is an optimal
weight function for our purposes. Since it is not
smooth we have, however, to modify it suitably.
To this end we recall some further notions from Finsler geometry
in the following.

Extremals of the functional $\mathscr{A}$
are called geodesics.
It is well-known that under Hypothesis~\ref{hyp-mM}
the infimum in (\ref{def-Fdist})
is actually attained and that each geodesic
is smooth. The homogenity of $F$ implies that $\mathscr{A}$
is invariant under reparametrizations. 
In particular, every reparametrization of a geodesic 
is again a geodesic. Moreover, we have the following well-known
lemma relating the geodesics to trajectories
of the Hamiltonian vector field
$$
X_H\,:=\,{\nabla_pH\choose-\nabla_xH}\,.
$$
Its proof can be found, e.g., in \cite[pp. 197]{GH}.

\begin{lemma}\label{manfred}
Any geodesic defined by means of $F$ can be reparametrized in such
a way, that we obtain the
projection onto $\RR^d$ of a Hamiltonian trajectory defined by means of
$H$ and running in $\Figx\subset T^*\RR^d$. Conversely, any projection
of a Hamiltonian trajectory running in $\Figx$
is a geodesic.
\end{lemma}

\bigskip

\noindent Next, we recall that 
under Hypothesis~\ref{hyp-mM}
the exponential map determined by the Finsler
structure $F$ is defined on $T\RR^d$ by
$\exp_x(v):=q_{x,v}(1)$, for $(x,v)\in T\RR^d$, $v\not=0$,
and $\exp_x(0):=x$, for $x\in\RR^d$. 
Here $q_{x,v}$ is the unique geodesic which passes through
$x$ at time $t=0$ with velocity $v$. (It holds
$q_{x,v}(\theta\,t)=q_{x,\theta\,v}(t)$, $\theta>0$.)
If $v\in\Indx_x$, then $t\mapsto q_{x,v}(t)=\exp_x(t\,v)$
is the {\em unit speed} geodesic passing through $x$
at $t=0$ in the direction of $v$,
that is, $F\big(q_{x,v}(t),\dot{q}_{x,v}(t)\big)=1$, for all $t\in\RR$.
We remark that the projections of the Hamiltonian trajectories
onto $\RR^d$ might have different velocities.

We assume that $x,y,v\in\RR^d$, and
$x=\exp_y(v)$ in the following. If the derivative
of $\exp_y:T_y\RR^d\to\RR^d$
is singular at $v$, then $x$ is called a conjugate point
for $y$. If $x$ is the first conjugate point for $y$
along the geodesic $t\mapsto q(t):=\exp_y(tv)$, then we know that,
for all $\tau\in(0,1)$,
$q\!\!\upharpoonright_{[0,\tau]}$
is, again up to reparametrization, the unique minimizing geodesic
among all geodesics from $y$ to $x$
that run in some small neighbourhood of
$q([0,\tau])$. It does not have, however, to be globally
minimizing.
For $\tau>1$, $q\!\!\upharpoonright_{[0,\tau]}$
will definitely lose its unique minimizing property even among
nearby geodesics. The cut point of $y$ along $q$
is by definition $q(r)$, where $r$ is the supremum of all those
$\tau>0$ such that $q\!\!\upharpoonright_{[0,\tau]}$
is globally minimizing. The cut point 
of $y$ along $q$ always appears before or at the first conjugate point.
Notice that, if $x$ is simultanously the cut and first conjugate point
for $y$ along $q$, then $q\!\!\upharpoonright_{[0,1]}$
might still be globally minimizing. 
The unique minimizing property of $q$ will, however,
definitely fail at the cut point, if the latter occurs
strictly before the first conjugate point.

We can always find some open set, $\sW\subset T_y\RR^d$, 
which is star-shaped
with respect to zero, such that 
$\exp_y\!\!\upharpoonright_{\sW}\in C^1(\sW)\cap 
C^\infty(\sW\setminus\{0\})$
is bijective from $\sW$ onto $\sU:=\exp_y(\sW)$, and
it is well-known that
$$
\widetilde{\vp}(x)\,=\,\Fdist(x,y)
\,=\,F\big(y,(\exp_{y}\!\!\upharpoonright_\sW)^{-1}(x)\big)\,,
\qquad 
x\in \sU\,;
$$ see, e.g., 
\cite[Chapter~8, \textsection3.3]{GH}.
Moreover, we have 
\begin{equation}\label{clara}
\widetilde{\vp}'(x)\,=\,F_v'(y,\dot{\gamma}(\tau))\,=\,\xi(\tau)\,,
\qquad x\in\sU\setminus\{y\}\,,
\end{equation}
where $t\mapsto(\gamma(t),\xi(t))\in\Figx$ 
is the Hamiltonian trajectory
corresponding to the minimizing geodesic from $y$ to $x$
such that $\gamma(0)=y$ and $\gamma(\tau)=x$, for some $\tau>0$.

In the following we fix two
distinguished points $x^\star,y^\star\in\RR^d$,
and assume

\begin{hypothesis}\label{hyp-geo}
$x^\star\not=y^\star$.
Up to reparametrization, there is a unique minimizing
geodesic
from $y^\star$ to $x^\star$. $x^\star$ and $y^\star$
are not conjugate to each other.  
\end{hypothesis}

\bigskip

\noindent We recall that Hypothesis~\ref{hyp-geo} is always fulfilled
provided $x^\star$ and $y^\star$ are sufficiently close to each other.
Since $F$ is absolutely homogenous there exists of course,
up to reparametrization, again
only one minimizing
geodesic from $x^\star$ to $y^\star$ if Hypothesis~\ref{hyp-geo}
is satisfied.
Moreover, we can prolong the geodesic from $y^\star$ to $x^\star$
or from $x^\star$ to $y^\star$ a little bit such that
it is still minimizing.

\begin{proposition}\label{prop-vp}
Assume that $H\in C^\infty(T^*\RR^d,\RR)$
fulfills Hypotheses~\ref{hyp-H} \& \ref{hyp-mM} 
and that $x^\star,y^\star\in\RR^d$ fulfill
Hypothesis~\ref{hyp-geo}.
Then there exist a point, $y_0$, on the prolongation of the geodesic
from $x^\star$ to $y^\star$, a compact neighbourhood, $K_0$,
of the geodesic segment from $y^\star$ to $x^\star$,
some open set, $\sW\subset T_{y_0}\RR^d$, 
which is star-shaped
with respect to zero, and some function,
$\vp\in C_b^\infty(\RR^d,\RR)$, such that
the following holds true:
\begin{enumerate}
\item[(i)] For all $x\in\RR^d$,
\begin{eqnarray*}
H(x,\vp'(x))&\klg&0\,,
\\
H(x,\vp'(x))&=&0\quad\Leftrightarrow\quad x\in K_0\,.
\end{eqnarray*}
\item[(ii)] $\vp(x)-\vp(y^\star)=\Fdist(x,y^\star)$, for all
$x$ on the
geodesic segment from $y^\star$ to $x^\star$. 
\item[(iii)] 
$\exp_{y_0}\!\!\upharpoonright_{\sW}\in C^1(\sW)\cap 
C^\infty(\sW\setminus\{0\})$ is injective on $\sW$
and
$$
K_0\,\subset\,\exp_{y_0}(\sW)\setminus\{y_0\}\,.
$$
\item[(iv)] For every $x\in K_0$, there is a unique pair 
$(\tau,p_0)\in(0,\infty)\times\Figx_{y_0}$
such that the projection of 
$[0,\tau]\ni t\mapsto\exp(t\,X_H)(y_0,p_0)$ onto $\RR^d$
is a minimizing geodesic from $y_0$ to $x$.
We have
$$
\exp(\tau\,X_H)(y_0,p_0)\,=\,(x,\vp'(x))\,.
$$
\end{enumerate}
\end{proposition}

\proof
Let $I$ be an open interval and $q:I\rightarrow\RR^d$ be
a geodesic with $q(t_1)=y^\star$ and $q(t_2)=x^\star$,
for some $t_1,t_2\in I$, $t_1<t_2$. We can find some $t_0\in I$,
$t_0<t_1$, such that $q\!\!\upharpoonright_{[t_0,t_2]}$
is still, up to reparametrizations, the unique geodesic
segment from $y_0:=q(t_0)$ to $x^\star$.
We can moreover find some open set $\sW\subset T_{y_0}\RR^d$,
star-shaped with respect to zero,
such that the restriction of $\exp_{y_0}$ to $\sW$ is
bijective from $\sW$
onto $\sU:=\exp_{y_0}(\sW)$, where $q([t_0,t_2])\subset \sU$.
We have $\exp_{y_0}\in C^1(\sW)\cap C^\infty(\sW\setminus\{0\})$.
Defining $\widetilde{\vp}$ by (\ref{teresa}) with $y=y_0$
we know that
$\widetilde{\vp}(x)
=F\big(y_0,(\exp_{y_0}\!\!\upharpoonright_\sW)^{-1}(x)\big)$, 
$x\in \sU$, solves
$H(x,\vp'(x))=0$ on $\sU\setminus\{y_0\}$.

Next, we pick some open set, $\sU_\star$, and two compact
sets, $K_\star$ and $K$, with 
$$
q([t_0,t_2])\,\Subset\,
\mathring{K}_\star\,\subset\,K_\star\,\Subset\,
\mathring{K}\,\subset\,K\,\Subset\,\sU_\star\,\subset\,
\overline{\sU}_\star\Subset \sU\,.
$$
We suppose that 
the boundaries $\partial K_\star$ and $\partial K$
are smooth submanifolds.
Furthermore, we can assume that
$\exp_{y_0}^{-1}(K_\star)$, $\exp_{y_0}^{-1}(K)$, and
$\exp_{y_0}^{-1}(\sU_\star)$ are star-shaped with respect to zero.
We pick some $\widetilde{H}\in C_b^\infty(\RR^d,\RR)$ with
$\widetilde{H}\equiv0$ on $K_\star$, and $0<\widetilde{H}(x)\klg-H(x,0)$, 
for $x\in\RR^d\setminus K_\star$,
and set
\begin{equation}\label{def-Hve}
H_\ve\,:=\,H\,+\,\ve\,\widetilde{H}\,,\qquad\ve\in(0,1]\,.
\end{equation}
For every $\ve\in(0,1]$, we let $\vp_\ve$ denote the function
constructed exactly in the same way as $\widetilde{\vp}$
using the Hamiltonian $H_\ve$ instead.
We notice that $\vp_\ve\equiv\widetilde{\vp}$ on
$K_\star$, since the Finsler structures and, hence,
the geodesics emanating from $y_0$
corresponding
to $H_\ve$ and $H$ are identical on $K_\star$.

In the following we show that, for all sufficiently small
$\ve\in(0,1]$, $\vp_\ve$ is, apart from $y_0$, smooth at
every point in a neighbourhood
of $K$.
We know that all geodesics of the Hamiltonian $H$
contained in $\sU$ and
emanating from $y_0$ form a central field about $y_0$.
No conjugate or cut points of $y_0$ are present in $\sU$.
First we claim 

\begin{lemma}\label{le-geo1}
There is some $\ve_0\in(0,1]$ such that, for all $\ve\in(0,\ve_0]$,
every geodesic of the Hamiltonian $H_\ve$
emanating from $y_0$ leaves $K_\star$ and touches a
point of $\RR^d\setminus \sU_\star$ before it 
possibly enters $K_\star$ again.
\end{lemma}

\proof
Let $v\in\Indx_{y_0}$, and let $\tau>0$ 
be the first exit time of $K_\star$ for the 
geodesic $t\mapsto\exp_{y_0}(t\,v)$ defined by $H$.
Then the
geodesics $t\mapsto\exp_{y_0}(t\,v)$
and $t\mapsto\exp_{y_0}^\ve(t\,v)$
are identical on $[0,\tau]$, for $\ve\in(0,1]$.
Here we designate the exponential map defined by $H_\ve$
by a superscript $\ve$.
We also know that $t\mapsto\exp_{y_0}(t\,v)$ 
touches a point in $\RR^d\setminus \sU$, say, at time $\bar{t}>0$,
before it possibly enters $K_\star$ again.
There is an open neighbourhood, $\mathscr{I}_v\subset\Indx_{y_0}$,
of $v$ in the indicatrix at $y_0$ such that
$\exp_{y_0}(\bar{t}\,w)\in\RR^d\setminus \sU$, 
for all $w\in\mathscr{I}_v$.
Since all geodesics depend smoothly on the parameter $\ve$ and
on $w\in\mathscr{I}_v$, we find some $\ve_v\in(0,1]$ such that
$\exp_{y_0}^\ve(\bar{t}\,w)\in\RR^d\setminus \sU_\star$, 
for all $w\in\mathscr{I}_v$ and $\ve\in(0,\ve_v]$.
 The claim now 
follows from the compactness of $\Indx_{y_0}$.
For the open cover $\bigcup_{v\in\Indx_{y_0}}\mathscr{I}_v$
contains a finite subcover $\bigcup_{\iota=1}^\nu\mathscr{I}_{v_\iota}$
and we can set $\ve_0:=\min\{\ve_{v_1},\dots,\ve_{v_\nu}\}$.
\qed

\begin{lemma}\label{le-geo2}
There is some $\ve_1>0$ such that, for all $\ve\in(0,\ve_1]$,
there is no conjugate or cut point for $y_0$
contained in $K$, if the geodesics are defined 
using the Hamiltonian $H_\ve$.
\end{lemma}

\proof
The assertion about the conjugate points is clear
since they are characterized by singularities of the
exponential map, which depends smoothly on $\ve$.
Concerning the cut points we prove a slightly
stronger statement, namely that there are
$\ve_1\in(0,1]$ and $c>0$ such that,
for all $\ve\in(0,\ve_1]$, and every pair
of unit speed geodesics, $q_1^\ve$ and $q_2^\ve$,
$q_1^\ve\not=q_2^\ve$,
defined by $H_\ve$ and emanating from $y_0$, we have
$$
\Big(\exists\;t,\tilde{t}>0\,:\;q_1^\ve([0,t])\subset K \;\wedge\;
q_1^\ve(t)=q_2^\ve(\tilde{t})\Big)
\;\Longrightarrow\;\tilde{t}\grg t+c\,.
$$
Note that, for unit speed geodesics, the time parameter
equals the Finsler arc length.
We argue by contradiction and suppose that there are sequences,
$(\ve_n)$, $(c_n)$,  $(t_n)$, $(\tilde{t}_n)$,
$(v_n)$, $(\tilde{v}_n)$,
such that
$\ve_n\searrow0$, $c_n\searrow0$, as $n\rightarrow\infty$,
and
$t_n,\tilde{t}_n>0$, $v_n,\tilde{v}_n\in\Indx_{x^\star}$,
$v_n\not=\tilde{v}_n$,
for $n\in\NN$, which satisfy $\tilde{t}_n<t_n+c_n$,
$$
\exp_{y_0}^{\ve_n}(\,t_n\,v_n)\,=\,
\exp_{y_0}^{\ve_n}(\,\tilde{t}_n\,\tilde{v}_n)\,,
\quad\textrm{and}\quad \exp_{y_0}^{\ve_n}(\,s\,v_n)\in K\,,\;
s\in[0,t_n]\,,
$$
for $n\in\NN$.
By definition of $K_\star$ we know that the unit speed geodesic
with initial velocity $\tilde{v}_n$
does not stay inside $K_\star$, for all $t\in[0,\tilde{t}_n]$.
By Lemma~\ref{le-geo1} we thus find $\tau_n\in(0,\tilde{t}_n)$
such that $\exp_{y_0}^{\ve_n}(\,\tau_n\,\tilde{v}_n)\in \sU_\star^c$,
for $n\in\NN$.
By compactness and by a choice of suitable subsequences
we can assume that the sequences $(t_n)$, $(\tilde{t}_n)$,
$(v_n)$, $(\tilde{v}_n)$, and $(\tau_n)$, have a limit, as $n$
tends to infinity, which we
denote by $t,\tilde{t},v,\tilde{v}$, and $\tau$,
respectively. Then it holds
$\tilde{t}\klg t$, $\exp_{y_0}(\tau\,\tilde{v})\in \sU_\star^c$,
$$
\exp_{y_0}(\,t\,v)\,=\,
\exp_{y_0}(\,\tilde{t}\,\tilde{v})\,\in\,K\,,
\quad\textrm{and}\quad \exp_{y_0}(\,s\,v)\in K\,,\;
s\in[0,t]\,.
$$
It also follows that $v\not=\tilde{v}$ and we get a contradiction,
because $[0,t]\ni s\mapsto\exp_{y_0}(s\,v)$ is the unique
minimizing geodesic for $H$ from $y_0$ to $\exp_{y_0}(t\,v)$.
\qed

\bigskip

\noindent Returning to the proof of Proposition~\ref{prop-vp}
we notice that Lemma~\ref{le-geo2} implies that, for all
$\ve\in(0,\ve_1]$, $\vp_\ve$ is smooth in a 
(in general $\ve$-dependent) neighbourhood of $K\setminus\{y_0\}$,
since the set where $\vp_\ve$ is not smooth is closed.

Next, we pick some compact set, $K'$, satisfying
$K_\star\Subset\mathring{K}'\subset K'\Subset\mathring{K}$,
and some cut-off function
$\chi\in C^\infty(\RR^d,[0,1])$ such that
$\chi\equiv1$ on $K'$ and $\chi\equiv0$ on $K^c$.
We note that there is some
$C>0$ such that $H_\ve-H\grg C\,\ve$ on $\supp(\chi')$.
We further pick a family, $\{j_\delta\}_{\delta>0}$,
of Friedrichs mollifiers such that $\supp(j_\delta)\Subset\{|x|\klg\delta\}$ 
and observe that
\begin{equation}\label{eq-der-vp}
\big(\chi\,\vp_\ve\,+\,(1-\chi)\,j_\delta *\vp_\ve \big)'
\,=\,\chi\,\vp_\ve'\,+\,(1-\chi)\,j_\delta *\vp_\ve' 
\,+\,(\vp_\ve-j_\delta *\vp_\ve)\,\chi'
\end{equation}
holds on $\RR^d\setminus\{y_0\}$.
Here we note that the expression $(j_\delta *\vp_\ve')(x)$
makes sense, 
also for $x\in K^c$,
since $\vp_\ve$ is differentiable almost everywhere.
Since the function $\widetilde{H}$ appearing 
in the definition (\ref{def-Hve})
of
$H_\ve$  depends only on $x$, the set
$$
\Bx_x^\ve\,:=\,\big\{p\in\RR_d\,:\,H_\ve(x,p)\,\klg\,0\big\}
$$
is still convex, and by the properties of $\widetilde{H}$ it holds 
$\Bx_x^\ve\subset\Bx_x$, for all $x\in\RR^d$. This inclusion
is proper if $x\in K_\star^c$.
Moreover, we have $H_\ve(x,\vp_\ve'(x))=0$,
that is, $\vp_\ve'(x)\in\Bx_x^\ve$,
for almost every $x\in\RR^d$.
Since $j_\delta$ is
a probability density and $\Bx_x^\ve$ is convex, it holds
$(j_\delta *\vp_\ve')(x)\in\Bx_x^\ve$, for all $x\in\RR^d$.
Consequently, for every $x\in\RR^d\setminus\{y_0\}$ 
the convex combination
$\chi(x)\,\vp_\ve'(x)\,+\,(1-\chi(x))\,(j_\delta *\vp_\ve')(x)$
is contained in $\Bx_x^\ve$, too.
On the support of $\chi'$ the set $\Bx_x^\ve$ and the indicatrix
$\Indx_x$ have a strictly positive distance which is uniformly
bounded
from below by some contant $\widetilde{C}(\ve)>0$.
We also know that
the term $(\vp_\ve-j_\delta *\vp_\ve)$ converges to zero
uniformly on the compact set $\supp(\chi')$, as
$\delta$ tends to zero. 
Choosing $\delta>0$ small enough, we can therefore ensure that
the right side of (\ref{eq-der-vp}) is contained in $\Bx_x$,
for all $x\in\RR^d\setminus\{y_0\}$.

The function $\chi\,\vp_\ve+(1-\chi)\,j_{\delta(\ve)} *\vp_\ve$
is smooth everywhere except at $y_0$.
To obtain a smooth function 
we pick some $\Theta\in C^\infty(\RR,[0,\infty))$
such that $0\klg\Theta'\klg1$,
$\Theta=\const>0$ in a neighbourhood of zero,
$\Theta(t)=t$, for all
$\Fdist(y_0,y^\star)/2<t<2\Fdist(y_0,x^\star)$, and
$\Theta=\const$, on $[3\Fdist(y_0,x^\star),\infty)$.
Finally, we choose
some $\ve\in(0,\ve_1]$, and some sufficiently small
$\delta(\ve)>0$, and  set
$$
\vp\,:=\,\Theta\circ
\big(\,\chi\,\vp_\ve\,+\,(1-\chi)\,j_{\delta(\ve)} *\vp_\ve \,\big)\,.
$$
We observe that we still have $\vp'(x)\in\Bx_x$, for $x\in\RR^d$,
since $0\klg\Theta'\klg1$.
To complete the proof we note that,
by the minimizing property of the geodesic from
$y_0$ to $x^\star$,
$$
\vp(x)\,-\,\vp(y^\star)\,=\,
\Fdist(x,y_0)\,-\,\Fdist(y^\star,y_0)\,=\,
\Fdist(x,y^\star)\,,
$$
for all $x$ on the unique geodesic from $y^\star$ to $x^\star$.
Finally, (iii) is fulfilled by construction
and (iv) follows from (\ref{clara}).
\qed


\subsection{The principal symbol $a^\vp$}\label{ssec-avp}

\noindent We turn to the discussion of the special Hamilton 
function $H:=T-U$, where
\begin{equation}\label{formula-T}
T(x,p)\,=\,\widetilde{V}(x,ip)
\,=\sum_{{\ell\in\ZZ^d:\atop 0<|\ell|\klg R}}
\!\!V(x,\ell)\,\cosh\big(\SPn{\ell}{p}\big)\,,
\qquad x\in\RR^d\,,\;p\in\RR_d\,,
\end{equation}
and $U$ is given by (\ref{def-U}).
First, we observe
as in \cite[\textsection4]{Sj1} that,
for every $x\in\RR^d$, the function
$T(x,\cdot):\RR_d\rightarrow(0,\infty)$
is strictly convex and even.
Moreover, $0<T(x,0)\klg U(x)-\vr$,
for some $x$-independent $\vr>0$, provided $J>0$ is
sufficiently small. 
In other words, 
$H$ satisfies Hypothesis~\ref{hyp-H}.
Using the uniform bounds on $U(x)$ and $V(\ell,x)$
of Hypothesis~\ref{hyp-UV}
it is not difficult to see that Hypothesis~\ref{hyp-mM}
is fulfilled by $H$, too.
So, 
Proposition~\ref{prop-vp} is applicable, for each 
pair of points $x^\star,y^\star$ satisfying Hypothesis~\ref{hyp-geo}.

Since we want to use the results of Section~\ref{sec-psido}
with $\phi=\vp$, we collect some important properties of
the principal symbol 
\begin{equation}\label{def-avp}
a^\vp(x,\xi)\;=\; i\,\widetilde{V}(x,\xi+i\vp'(x))\,-\,i\,U(x), 
\qquad (x,\xi)\in\RR^d\times\RR_d\;,
\end{equation}
in the next lemma.
In its statement and henceforth we use the short-hand notation
$(a^\vp)_{\xi x}'':=d_x\nabla_\xi a^\vp$ etc. 
The Hamiltonian vector field of $\Re a^\vp$
is denoted by
$X_{\Re a^\vp}:={\nabla_\xi\Re a^\vp\choose-\nabla_x\Re a^\vp}$.

\begin{lemma}\label{le-prop's-avp}
Fix two distinguished points $x^\star,y^\star\in\RR^d$
fulfilling Hypothesis~\ref{hyp-geo}
and let $\vp$ and $K_0$ be 
as in Proposition~\ref{prop-vp}.
Then the symbol $a^\vp\in \Symbb{\RR^d\times\RR_d}$ 
has the following properties:

\smallskip

\noindent(i) For $(x,\xi)\in\RR^d\times\RR_d$,
\begin{eqnarray}\label{imaklg0}
\Im a^\vp(x,\xi)&\klg&0\,,
\\
\Im a^\vp(x,\xi)&=&0\quad\Leftrightarrow
\quad\label{ima=0}
(x,\xi)\in K_0\times(2\pi\ZZ)^d\,.
\end{eqnarray}
(ii) For $(x,\xi)\in K_0\times(2\pi\ZZ)^d$,
\begin{eqnarray}
(a^\vp)(x,\xi)&=&0\,,\label{a=0}
\\
(a^\vp)_{\xi}'(x,\xi)&=&\label{rea'not=0}
(\Re a^\vp)_{\xi}'(x,\xi)
\,=\,H_p'(x,\vp'(x))\;\not=\;0\,,
\\
(a^\vp)_x'(x,\xi)&=&0\label{a'x=0}\,,
\\
(a^\vp)_{\xi\xi}''(x,\xi)&=&-i\,H_{pp}''(x,\vp'(x))\,,
\\
(a^\vp)_{\xi x}''(x,\xi)&=&H_{px}''(x,\vp'(x))\,+\,
H_{pp}''(x,\vp'(x))\,\vp''(x)\,,
\\
(a^\vp)_{xx}''(x,\xi)&=&0\label{axx}
\,.
\end{eqnarray}
(iii) $\rho:[0,\tau]\to K_0\times(2\pi\ZZ)^d$ with $\tau>0$
is an integral curve of $X_{\Re a^\vp}$, if and only if, 
for $(x(t),\xi(t)):=\rho(t)$, the curve
$[0,\tau]\ni t\mapsto\big(x(t),\vp'(x(t))\big)$
is a piece of an integral curve of $X_H$ emanating from
some point on $\{y_0\}\times\Figx_{y_0}\subset T^*\RR^d$.

\smallskip

\noindent(iv) Let $\gamma:[0,\tau]\to K_0$ denote the geodesic
from $y^\star=\gamma(0)$ to $x^\star=\gamma(\tau)$
parametrized such that it is the projection onto $\RR^d$
of a trajectory of $X_H$. Then, for every
constant vector $e\in(2\pi\ZZ)^d$,  $(\gamma,e)$
is an integral curve of $X_{\Re a^\vp}$.
\end{lemma}

\proof
(i): Similarly to analogous considerations in \cite{Sj1}
we deduce (\ref{imaklg0}) and (\ref{ima=0})
from the representation
\begin{eqnarray}
\widetilde{V}(x,\xi+i\vp'(x))&=&\!\!\!
\sum_{{\ell\in\ZZ^d:\atop0<|\ell|\klg R}}\!\!V(x,\ell)\,\Big\{
\cosh\big(\SPn{\ell}{\vp'(x)}\big)\,
\cos\big(\SPn{\ell}{\xi}\big)\nonumber
\\
& &\quad\qquad
\,-\,i\,\sinh\big(\SPn{\ell}{\vp'(x)}\big)\,
\sin\big(\SPn{\ell}{\xi}\big)
\Big\}\,,\label{formula-a}
\end{eqnarray}
the positivity of $V$, and the inequality
$H(x,\vp'(x))=\widetilde{V}(x,i\vp'(x))-U(x)\klg0$,
$x\in\RR^d$.
In the derivation of (\ref{ima=0}) we 
make use of the assumption that $V(x,\ell)\grg1$,
if $|\ell|=1$, and the trivial fact that, 
for all $\xi \in\RR_d\setminus(2\pi\ZZ)^d$, 
we find some $k\in\ZZ^d$, $|k|=1$, such that $\cos(\SPn{k}{\xi})<1$.

(ii) follows from straightforward computations. In particular,
we obtain,
for all $x\in K_0$
and $\xi\in(2\pi\ZZ)^d$, 
\begin{equation}\label{pablo}
(\Re a^\vp)_{\xi}'(x,\xi)\,=\sum_{{\ell\in\ZZ^d:\atop0<|\ell|\klg R}}\!
V(x,\ell)\,{}^t\ell\,\sinh\big(\SPn{\ell}{\vp'(x)}\big)
\,=\,T_p'(x,\vp'(x))\,\not=\,0\,,
\end{equation}
To see that the terms in (\ref{pablo}) are in fact non-zero,
for $(x,\xi)\in K_0\times(2\pi\ZZ)^d$, we can refer
to the strict convexity of $T(x,\cdot)$, or multiply
(\ref{pablo}) with $\vp'(x)$ and
use again that $V(x,\ell)>0$, for $|\ell|=1$.

(iii) Let $\rho=(x,\xi):[0,\tau]\to K_0\times(2\pi\ZZ)^d$ 
be a trajectory of $X_{\Re a^\vp}$.
Then (\ref{rea'not=0}) implies
$\dot{x}(t)=\nabla_\xi a^\vp(\rho(t))=\nabla_p H\big(x(t),\vp'(x(t))\big)$,
for all $t\in[0,\tau]$. Moreover, differentiating the identity
$H(x,\vp'(x))=0$, $x\in K_0$, we get
$\vp''(x)\,\nabla_pH(x,\vp'(x))=-\nabla_xH(x,\vp'(x))$,
$x\in K_0$.
From this we infer 
$\frac{d}{dt}\,\vp'(x(t))\,=\,-\nabla_xH(x(t),\vp'(x(t)))$,
and
we see that $[0,\tau]\ni t\mapsto(x(t),\vp'(x(t)))$
is an integral curve of $X_H$. By Proposition~\ref{prop-vp}
we know that,
for every $x\in K_0$, $(x,\vp'(x))$ lies on an
integral curve of $X_H$ emanating from some point
on $\{y_0\}\times\Figx_{y_0}$. By the unique solvability of Hamilton's
equations we conclude
that $[0,\tau]\ni t\mapsto\big(x(t),\vp'(x(t))\big)$
is a piece of an integral curve of $X_H$ emanating from
$\{y_0\}\times\Figx_{y_0}$. 

Conversely, if $[0,\tau]\ni t\mapsto\big(x(t),\vp'(x(t))\big)$
is a piece of an integral curve of $X_H$ emanating from
$\{y_0\}\times\Figx_{y_0}$, then we use (\ref{rea'not=0})
and (\ref{a'x=0}) to check that 
$[0,\tau]\ni t\mapsto\big(x(t),e\big)$ is an integral curve
of $X_{\Re a}$, for every $e\in(2\pi\ZZ)^d$.

(iv) follows from (iii) and Proposition~\ref{prop-vp}(iv).
\qed


\section{The Hamilton-Jacobi equation}\label{sec-Ham-Jac}

In this section we take
a first step in the construction of a parametrix for 
$\Op(a^\vp_W)$,
where $\vp$ is the weight function constructed in 
Proposition~\ref{prop-vp} and $a^\vp_W$ is given by
Lemma~\ref{le-ernst1}.
Since we shall obtain the parametrix by integrating a parametrix
for the corresponding heat operator we seek
for approximate
solutions
of the time-dependent Hamilton-Jacobi equation
\begin{equation}\label{hansi}
\partial_t\psi(t,x,\eta)\,+\,a^\vp(x,\psi_x'(t,x,\eta))\,=\,0
\,,\qquad
\psi(0,x,\eta)=\SPn{x}{\eta}\,,
\end{equation}
where $t\in[0,\infty)$, $x\in\RR^d$, and $\eta\in\RR_d$.
Here  and $a^\vp$ is given by
(\ref{def-aphi}) with $\phi=\vp$.
Since the symbol $a^\vp$ is complex-valued 
we have to work with almost analytic extensions
and we can only hope to solve (\ref{hansi})
up to error terms of order $\bigO_N(|\Im\psi(t,x,\eta)|^N)$,
for every $N\in\NN$.
Approximate solutions of (\ref{hansi}) have been constructed
in \cite{MeSj2} for a class of symbols which are homogenous of degree
one and in \cite{MenSj1} for purely imaginary symbols
satisfying some further assumptions. 
Since neither of these works applies directly to our situation
we will explain in detail how the proofs of \cite{MeSj2}
can be adapted to fit with our hypotheses.
We remark that the same class of problems is adressed 
in \cite{Ku} where approximate solutions
to the heat equation are obtained in the framework
of Maslov's canonical operator theory.
Approximate solutions of complex eikonal and transport
equations are also derived in \cite{Tr2}.

This section is split in two parts:
In Subsection~\ref{subsec-est} we extend the crucial
estimates from \cite{MeSj2} on the real Hamiltonian flow associated
with $a^\vp$ to our situation. The
essential properties of the symbol $a^\vp$ required here
are
that the imaginary part of $a^\vp$ has a fixed sign
and that the derivative of $\Re a^\vp$ is non-zero
if $\Im a^\vp$ vanishes. 
Both of them are ensured by our special choice of $\vp$
as we have already observed in Lemma~\ref{le-prop's-avp}.
In Subsection~\ref{subsec-HamJac}
we construct the approximate solution $\psi$.
Some basic facts about almost analytic extensions
are collected in Appendix~\ref{subsec-aa}.


\subsection{Estimates for the Hamiltonian and contact flows}
\label{subsec-est}

We will obtain the approximate solution, $\psi$, of (\ref{hansi})
essentially
as the generating function of the flow
of a real Hamiltonian vector field associated with $a^\vp$.
In order to control the error terms
and to assure that the imaginary part of $\psi$ has
the right sign it is necessary to derive suitable
estimates on the flow of this vector field.
This is carried out in the present subsection by
adapting the arguments of \cite{MeSj2},
where the symbol is assumed to be homogenous of degree one.
The same problem has also been considered in \cite{Ku}.
We remark that the reasoning of \cite{MeSj2}
and, hence, of this subsection works globally, too.

In the following we denote the complex coordinates in 
$T^*\CC^d\cong\CC^d\times\CC_d$ again by $(x,\xi)=\rho$, so that
$\partial_{x^i}=\frac{1}{2}(\partial_{\Re x^i}-i\partial_{\Im x^i})$,
$\partial_{\overline{x}^i}
=\frac{1}{2}(\partial_{\Re x^i}+i\partial_{\Im x^i})$, etc.
We pick almost analytic extensions of $\vp$,
$U$, and $V(\cdot,\ell)$, $0<|\ell|\klg R$,
and denote them again by the same symbols. 
Plugging them into (\ref{def-avp}) we obtain
an extension of $a^\vp$ to $\CC^d\times\CC_d$
which is almost analytic with respect to the first and
analytic with respect to the second variable.
More precisely, $a^\vp$ is a trigonometric
polynomial in $\xi$ with almost analytic
coefficients.
{\bf We denote this
extension of $a^\vp$ simply by $a$ in the rest of 
Section~\ref{sec-Ham-Jac}.}

Since $a$ as well as any partial derivative of $a$
satisfies the Cauchy-Riemann differential equations
on the real domain we know that 
$a_\rho'(\rho)=a_{\Re\rho}'(\rho)$,
$a_{\rho\rho}''(\rho)=a_{\Re\rho\Re\rho}''(\rho)$,
for every real $\rho$, and, hence,
(\ref{rea'not=0})-(\ref{axx})
still hold for $(x,\xi)\in K_0\times(2\pi\ZZ)^d$,
if the derivatives are interpreted as complex derivatives.
We emphasize that (\ref{imaklg0}) and (\ref{ima=0}) 
are available only for
{\em real} $(x,\xi)$.

We are going to study the flow of the real field \cite{Sj0}
$$
\widehat{X}_{a}\,:=
\,\HVF{a}\,+\,\overline{X}_{a}\,,
$$
where
$$
\HVF{a}\;:=\;\sum_{i=1}^d\big(\,
a_{\xi_i}'\,\partial_{x^i}\,-\,
a_{x^i}'\,\partial_{\xi_i}
\,\big)\,.
$$
We verify that
\begin{equation}\label{lara}
\widehat{X}_{a}\,=\,\sum_{i=1}^d
\big\{
\Re(a_{\xi_i}')\,\partial_{\Re x^i}\,+\,
\Im(a_{\xi_i}')\,\partial_{\Im x^i}\,-\,
\Re(a_{x^i}')\,\partial_{\Re \xi_i}\,-\,
\Im(a_{x^i}')\,\partial_{\Im \xi_i}
\big\},
\end{equation}
so that the definition of $\widehat{X}_{a}$
amounts to considering $\HVF{a}$ as a vector field
on $\RR^{4d}$.
(If we took another almost analytic extension of $a$
then the components
of the corresponding real fields would be equivalent
in the sense explained in Appendix~\ref{subsec-aa}.)
We denote the flow of
$\HFA$ by
$$
\kappa_t(\rho)\,:=\,\big(Q(t,\rho),\Xi(t,\rho)\big)
\,:=\,\exp(t\HFA)(\rho)\,,
\qquad \rho\in T^*\CC^d\,.
$$ 
On account of (\ref{lara}) we see that the Hamiltonian equations
for $Q$ and $\Xi$ read
\begin{equation}\label{HEQXi}
\dot{Q}\,=\,\nabla_\xi a(Q,\Xi)\,,\qquad
\dot{\Xi}\,=\,-\nabla_xa(Q,\Xi)\,.
\end{equation}
Due to the periodicity of $a$ in $\Re\xi$ we have,
for $(t,y,\eta)\in \RR\times T^*\CC^d$
and $e\in(2\pi\ZZ)^d$,
\begin{equation}\label{per-QXi}
Q(t,y,\eta+e)\,=\,Q(t,y,\eta)\,,\qquad
\Xi(t,y,\eta+e)\,=\,\Xi(t,y,\eta)\,+\,e\,.
\end{equation} 
It will be convenient to extend $T^*\CC^d$
by an extra variable, $s$, which parametrizes the action.
We may view the resulting space $\CC\times T^*\CC^d$ 
as a contact manifold
equipped with the contact form 
$ds+\sum_{i=1}^d\xi_i\,dx^i$.
Then we may call the vector field
$$
\widehat{K}_{a}\,
:=\,\KVF{a}\,+\,\overline{K}_{a}\,,
$$
where
$$
\KVF{a}\;:=\;
-A\,\partial_s
\,+\,\HVF{a}\;,\qquad A\,:=\,\SPn{\xi}{a_{\xi}'}\,-\,a\,,
$$
the real contact field associated with $a$.
Here and in the following $\SPn{\cdot}{\cdot\!\cdot}$
denotes the bilinear extension to $\CC^d$
of the Euclidean scalar product on $\RR^d$.
The function $A$ is called the elementary action.
To study the flow of $\KFA$ we introduce the function
$\SMS:\CC\times T^*\CC^d\rightarrow\CC$, where
$$
\SMS(s,x,\xi)\;:=\;-\Im s\,-\,\SPn{\Im x}{\Re \xi}\,,
\qquad (s,x,\xi)\in\CC\times T^*\CC^d\,.
$$ 
The relationship between these objects
and the ones
considered in \cite{MeSj2}, 
where the symbol is assumed to be homogenous
of degree one, can be easily seen by means of the
the standard reduction from non-homogenous to homogenous
symbols: Denoting the momentum variable conjugate to
the extra variable $s$ by
$\sigma$, we set 
$a^\mathrm{hom}(s,x,\sigma,\xi):=
\sigma\,a(x,\frac{1}{\sigma}\,\xi)$,
for 
$(s,x,\sigma,\xi)\in T^*\CC^{d+1}$ with $\sigma\not=0$.
The results of \cite{MeSj2} apply directly to the flow
of the real field $\widehat{X}_{a^\mathrm{hom}}$, 
which is defined
analogously to $\widehat{X}_a$.
Since $\sigma$ is obviously constant along the flow lines
of $\widehat{X}_{a^\mathrm{hom}}$ 
it is easily checked that
$\widehat{X}_{a^\mathrm{hom}}\!\!\upharpoonright_{\{\sigma=1\}}$
can be identified with 
the real contact field $\widehat{K}_{a}$.
Moreover, the function $-\SPn{\Im(s,x)}{\Re(\sigma,\xi)}$
used in \cite{MeSj2} reduces to $\SMS$, for $\sigma=1$.
(A direct application of \cite{MeSj2}
would, however, give suboptimal versions of
the estimates (\ref{est-MeSj1})-(\ref{est-MeSj3})
below.)

\begin{lemma}
For all compact subsets $\Omega\Subset T^*\CC^d$,
there is some $C_{\Omega}\in(0,\infty)$,
such that, for all
$(s,\rho)\in\CC\times\Omega$, 
\begin{equation}\label{est-SMS}
\widehat{K}_{a}(\SMS)(s,\rho)\;\grg\;
-\frac{1}{2}\:\Im a(\Re \rho)\,-\,
C_{\Omega}\,|\Im\rho|^3\,.
\end{equation}
\end{lemma}

\proof
Taylor expanding at $\Re\rho$
in the third and fourth step, we get
\begin{eqnarray}
\lefteqn{\nonumber
\KFA(\Im s\,+\,\SPn{\Im x}{\Re \xi})
}
\\
&=&\nonumber
\Im\big(\,
a(\rho)\,-\,\SPn{\xi}{a'_\xi(\rho)}
\,+\,\SPn{\xi-i\,\Im\xi}{a'_\xi(\rho)}
\,\big)\,-\,\SPn{\Im x}{\Re a_x'(\rho)}
\\
&=&\nonumber
\Im a(\rho)\,-\,\SPn{\Re a_{\rho}'(\rho)}{\Im\rho}
\\
& =&\nonumber
\Im a(\Re\rho)\,+\,\,\frac{1}{2}\:
\SPb{ \Im a_{\rho\rho}''(\Re \rho)\,\Im \rho}{\Im\rho}\,+\,
\bigO\big(|\Im\rho|^3\big)
\\
& =&\nonumber
\frac{3}{4}\:\Im a(\Re\rho)\,+\,
\frac{1}{8}\:\big(\,\Im a(\Re\rho+2\,\Im\rho)
\,+\,\Im a(\Re\rho-2\,\Im\rho)\,\big)
\,+\,
\bigO\big(|\Im\rho|^3\big)
\\
&\klg&\frac{3}{4}\:\Im a(\Re\rho)\,+\,\bigO\big(|\Im\rho|^3\big)
\,.\label{est-Taylor}
\end{eqnarray}
\qed

\begin{remark}
Exactly as in \cite[\textsection1]{MeSj2},
we may check that $\SMS$ is invariantly
defined modulo $\bigO(|\Im(x,\xi)|^3)$
on compact subsets
with respect to coordinate changes $x\mapsto\tilde{x}$.
(Of course, the covariant variables $\xi$
have to be transformed accordingly.)
Then the arguments of \cite[Remark~1.8]{MeSj2} 
directly imply that Estimate (\ref{est-SMS})
does not depend on the particular coordinate system.
\end{remark}

\bigskip

\noindent
In the following we fix some
$\rho_0=(x_0,\xi_0)\in T^*\RR^d$. 
We assume
that there exists some complex neighbourhood, $\mathscr{V}\subset\CC^d$,
of $x_0$ and an almost analytic function,
$\psi_0\in C^\infty(\mathscr{V})$ such that
$\Im\psi_0\grg0$ on $\mathscr{V}\cap \RR^d$, $\Im\psi_0(x_0)=0$,
and
$\psi_0'(x_0)=\xi_0$.
By a calculation similar to (\ref{est-Taylor}) we then see that,
for every real, compact subset 
$\mathscr{V}_\RR'\Subset\mathscr{V}\cap\RR^d$, we find some
compact $\mathscr{V}'\Subset\mathscr{V}$ and some $C\in(0,\infty)$,
such that $\mathscr{V}'\cap\RR^d=\mathscr{V}_\RR'$ and
\begin{equation}\label{mildred}
\Im\psi_0(x)\,-\,\SPn{\Im x}{\Re\psi_0'(x)}\,
\grg\,-C\,|\Im x|^3\,, \qquad x\in \mathscr{V}'\,.
\end{equation}
We introduce the set
$$
\mathfrak{L}_0\,:=\,
\big\{\,(-\psi_0(x),x,\psi_0'(x))\,:\:x\in\mathscr{V}\,\big\}\,,
$$
which we may call an 
almost analytic Legendre
submanifold.
(In our applications below we will only 
consider the case where
$\psi_0(x):=\psi(0,x,\eta):=\SPn{x}{\eta}$, for some real
$\eta\in\RR_d$.)
Furthermore, we assume that there is some $\tau\grg0$
such that
$\kappa_t(\rho_0)=\exp(t\HFA)(\rho_0)$ is real, for all
$t\in[0,\tau]$.
If $\tau>0$, we additionally assume that
$\Im a(\rho_0)=0$. In this case it holds 
$(\Im a)'(\kappa_t(\rho_0))=0$, and,
hence, $\Im a(\kappa_t(\rho_0))=0$, 
for $t\in[0,\tau]$. By (\ref{rea'not=0}) we also know that 
$(\Re a)'(\kappa_t(\rho_0))\not=0$, for $t\in[0,\tau]$.
Denoting the flow of $\KFA$ by
$$
\Phi_t\,:=\,\exp(t\KFA)
$$ 
we have that
\begin{equation}\label{edgardo}
\Phi_t\big(s,\rho\big)=\big(\vs_t(s,\rho),\kappa_t(\rho)\big)
\,,\qquad
\vs_t(s,\rho)\,:=\,s\,-\int_0^tA(\kappa_r(\rho))\,dr\,.
\end{equation}
In particular, $\Phi_t\big(-\psi_0(x_0),\rho_0\big)$
is also real, for $t\in[0,\tau]$.

\begin{lemma}\label{le-Prop3.1}
In the situation described above there exist $\ve>0$, $C\in(0,\infty)$,
and some neighbourhood $\mathscr{O}\subset \CC\times T^*\CC^d$
of $(-\psi_0(x_0),\rho_0)$ such that the following inequalities
hold on $\mathfrak{L}_0\cap\mathscr{O}$, 
for all $0\klg r\klg t\klg\tau+\ve$,
\begin{eqnarray}\label{est-MeSj1}
\SMS\big(\Phi_t\big)&\grg&-\,\frac{1}{3}\,\int_0^t\Im a(\Re\kappa_u)\,du
\,-C\,\big|\Im\kappa_t \big|^3\,,
\\
\big|\Im\kappa_t\big|^2\,+\,\SMS(\Phi_t)
&\grg&
\frac{1}{C}\Big\{
\big|\Im\kappa_r\big|^2\,+\,\SMS(\Phi_r)
\,-\,\int_r^t\Im a(\Re\kappa_u)\,du
\Big\}\,,\label{est-MeSj2}
\\
\big|\Im\kappa_r \big|^2&\klg&C\,\big(\big|\Im\kappa_t \big|\,
-\,\Im \vs_t\big)
\,.\label{est-MeSj3}
\end{eqnarray}
In particular, if $(s,\rho)\in\mathfrak{L}_0\cap\mathscr{O}$
and $\kappa_t(\rho)$ 
and $\vs_t(s,\rho)$ are both 
real, for some $t\in[0,\tau+\ve]$, then
$\kappa_r(\rho)$ is real for all $r\in[0,t]$.
\end{lemma}

\proof
Taylor expanding the right side of 
$\frac{d}{dt}\:\Im \kappa_t\,=\,\Im\HFA(\kappa_t)$
at $\Re\kappa_t$ and using Duhamel's formula we obtain, exactly as
in \cite[pp. 351]{MeSj2}, the estimate
\begin{equation}\label{est-Duhamel}
|\Im\kappa_u|\,\klg\,\bigO(1)\,\Big(\,
|\Im\kappa_t|\,+\,\int_u^t|\Im a'(\Re\kappa_r)|\,dr\,\Big)\;.
\end{equation}
It holds for all $\rho$ in a compact, complex
neighbourhood of $\rho_0$ and $u,t\in[0,\tau+\ve_1]$, for some $\ve_1>0$.
Since, for $\rho$ in a compact set, 
the curves $[0,\tau+\ve_1]\ni t\mapsto\Re\kappa_t(\rho)$ stay in a
compact set, we may apply the 
standard estimate for positive functions to (\ref{est-Duhamel}),
which together with H\"{o}lder's inequality gives
\begin{equation}\label{est-Duhamel2}
|\Im\kappa_r|\,\klg\,\bigO(1)\,\Big(\,
|\Im\kappa_t|\,+\,\Big(\int_u^t-\Im a(\Re\kappa_v)\,dv\,\Big)^{1/2}\Big)\,,
\end{equation}
for $0\klg u\klg r\klg t\klg\tau+\ve_1$.
Next, we integrate the estimate (\ref{est-SMS}) for 
$\frac{d}{dt}\SMS(\Phi_t)=\KFA(\SMS)(\Phi_t)$
from $u$ to $t$, use (\ref{est-Duhamel2}) to bound $|\Im\kappa_r|$,
$r\in[u,t]$, and arrive at
\begin{eqnarray}\label{est-Duhamel3}
\SMS(\Phi_t)&\grg&\SMS(\Phi_u)
\,-\,\frac{1}{2}\,\int_u^t\Im a(\kappa_r)\,dr
\,-\,\bigO(1)\,(t-u)\,|\Im\kappa_t|^3
\,-\,\Big(\int_u^t-\Im a(\Re\kappa_r)\,dr\Big)^{3/2}.
\end{eqnarray}
For $u=0$, (\ref{mildred}) implies 
$$
\SMS\big(\Phi_0(-\psi_0(x),x,\psi_0'(x))\big)
\,\grg\,-\,\bigO(1)\,|\Im x|^3\,,
$$
for $x$ in a sufficiently small neighbourhood of $x_0$.
Here the last term can again be estimated
by means of (\ref{est-Duhamel2}).
We further notice that we can make the integral appearing
in (\ref{est-Duhamel3}) arbitrarily small by assuming
that $\rho$ is contained in a sufficiently small 
neighbourhood of $\rho_0$, and that $\ve_1>0$
is sufficiently small. For, if $t\in[0,\tau]$, the integral  
vanishes at $\rho_0$.
Setting $u=0$ we thus get (\ref{est-MeSj1}).
The proofs of (\ref{est-MeSj2}) and (\ref{est-MeSj3})
are identical to those in
\cite[pp. 355]{MeSj2}, except for Equation~(3.30)
in \cite{MeSj2}, where $-\Im\vs_t$ has to be added on the
right side.
\qed


\subsection{Approximate solution of the Hamilton-Jacobi equation}
\label{subsec-HamJac}

\noindent The aim of this subsection is to construct
an approximate solution, $\psi$, of (\ref{hansi}).
We will define $\psi$ by a formula well-known from
classical mechanics and proceed along traditional
arguments to check that $\psi$ has the required
properties. In doing so we use Lemma~\ref{le-Prop3.1}
to controll the error terms and to ensure that $\Im\psi$
is positive.
Since Lemma~\ref{le-Prop3.1} is applicable near
real integral curves
of $\HFA$, we start by considering the latter more closely. 
All the time we keep on
using the notation of Proposition~\ref{prop-vp}
with $H=T-U$.

\begin{lemma}\label{le-HFA-XH}
Let $\tau>0$ and $\rho:[0,\tau]\to T^*\RR^d$
be a real integral curve of $\HFA$ with
$\rho(0)\in\{\Im a=0\}$.
Then $\rho([0,\tau])\subset K_0\times(2\pi\ZZ)^d$.
If we write $\rho(t)=(x(t),\xi(t))$, then 
$[0,\tau]\ni t\mapsto(x(t),\vp'(x(t)))$ is a piece
of an integral
curve of $X_H$ emanating from some point
on $\{y_0\}\times\Figx_{y_0}\subset T^*\RR^d$.
Conversely, every point
of $K_0$ lies on the projection of an integral curve
of $X_H$ emanating from some point on 
$\{y_0\}\times\Figx_{y_0}$ and
the piece of that projection inside $K_0$ is 
the projection of a real integral
curve of $\HFA$ with momentum $e\in(2\pi\ZZ)^d$.
\end{lemma}

\proof
$\rho$ being a real integral curve of $\HFA$ means 
$\Im{\nabla_\xi a\choose -\nabla_x a}(\rho(t))=\frac{d}{dt}\,\Im\rho(t)=0$.
Since $a$ fulfills the Cauchy-Riemann differential equations
on the real domain, it follows that
$(\Im a)_{\Re \rho}'(\rho(t))=\Im(a_\rho')(\rho(t))=0$.
Hence, the derivative of $(\Im a)\!\!\upharpoonright_{T^*\RR^d}$ 
vanishes along $\rho$
and $\Im a(\rho(0))=0$ thus implies $\rho([0,\tau])\subset\{\Im a=0\}$.
Using (\ref{ima=0}) we conclude
$\rho([0,\tau])\subset K_0\times(2\pi\ZZ)^d$.
The remaining assertions follows from 
Proposition~\ref{prop-vp} and Lemma~\ref{le-prop's-avp}(iii).
\qed

\bigskip 

\noindent 
Of course, there might be real integral curves of
$\HFA$ defined on a non-trivial interval
running in a region where $\Im a<0$.
These do, however, not play any important role
since the imaginary part of the action is strictly
decreasing along them.

The approximate solution
$\psi$ will turn out to be ``almost'' a generating
function
of the canonical relations
$$
C_t\,=\,\big\{\,
(x,\xi,y,\eta)\in T^*\CC^d\times T^*\CC^d\::\;
(x,\xi)=\kappa_t(y,\eta)\,
\big\}\;,\qquad t\grg0\,.
$$
We recall our notation $\kappa_t=\exp(t\,\HFA)$.
For every $t\grg0$, the relation $C_t$
is $2\pi$-periodic in the sense that
$(x,\xi,y,\eta)\in C_t$
implies $(x,\xi+e,y,\eta+e)\in C_t$, for every $e\in(2\pi\ZZ)^d$.
To study $C_t$ we denote the bilinear extension of the
canonical symplectic form on $T^*\RR^d$ to $T^*\CC^d$
by $\widetilde{\sigma}$, i.e.
$$
\widetilde{\sigma}\big(\vartheta\,,\,\vartheta'\big)
\,=\,{}^t\vartheta\matr{0&-\id\\ \id&0}\vartheta'\,,
\qquad \vartheta,\vartheta'\in TT^*\CC^d\,.
$$
Furthermore, we
write
$$
\widetilde{\sigma}\ominus\widetilde{\sigma}(\vt,\vt')\,
:=
\,\widetilde{\sigma}\big((\vt_x,\vt_\xi)\,,\,(\vt_x',\vt_\xi')\big)
\,-\,\widetilde{\sigma}\big((\vt_y,\vt_\eta)\,,\,(\vt_y',\vt_\eta')\big)\,,
$$
where $\vt=(\vt_x,\vt_\xi,\vt_y,\vt_\eta)\in T(T^*\CC^d)\times T(T^*\CC^d)$
and similarly for $\vt'$.

\begin{lemma}\label{le-klaus}
Let $\tau>0$ and
$\rho:[0,\tau]\rightarrow K_0\times(2\pi\ZZ)^d$ be
a real integral curve of $\HFA$. Then $C_t$
is strictly positive along $\rho$, that is,
for all $t\in[0,\tau]$, 
\begin{equation}\label{klaus}
\frac{1}{2i}\:\widetilde{\sigma}\ominus\widetilde{\sigma}
(\vt,\overline{\vt})\,\grg\,0\,,
\qquad \vt\in T_{(\rho(t),\rho(0))}C_t\,,
\end{equation}
with equality if and only if 
$\vt$ is contained in the complexification
of the tangent space at $(\rho(t),\rho(0))$
of $C_t\cap\RR^d$. The latter
condition holds, if and only if
$\vt=(\vt_x,0,\vt_y,0)$.
\end{lemma}

\proof
Along the real curve
$\rho(t)=:(x(t),e)\in K_0\times(2\pi\ZZ)^d$ 
the fundamental matrix of $a$ is given by
\begin{equation}\label{eq-fund-matr}
\FM_{a}(x(t),e)\,=
\,\matr{a_{\xi x}''&a_{\xi\xi}''\\-a_{xx}''&-a_{x\xi}''}(x(t),e)
\,=\,
\matr{\bbB(t)&-i\,\bbA(t)\\0&-{}^t\bbB(t)}\,,
\end{equation}
where 
\begin{equation}
\bbA(t)\,:=\,H_{pp}''\big(x(t),\vp'(x(t))\big)\,,
\qquad \bbB(t)\,:=\,H_{px}''\big(x(t),\vp'(x(t))\big)\,
+\,\bbA(t)\,\vp''(x(t))\,,
\end{equation}
for $t\in[0,\tau]$.
It holds $\vt=(\theta(t),\theta(0))
=\big(\theta_x(t),\theta_{\xi}(t),\theta_x(0),\theta_\xi(0)\big)$, 
where 
$\theta(t)=\big(\theta_x(t),\theta_{\xi}(t)\big)
=\kappa_t'(\rho(0))\,\theta(0)$,
$t\in[0,\tau]$, for some $\theta(0)\in T_{\rho(0)}T^*\CC^d$,
and we know that $\theta(t)$ satisfies 
$\frac{d}{dt}\,\theta(t)=\FM_{a}(\rho(t))\,\theta(t)$, for 
$t\in[0,\tau]$. We thus obtain
\begin{eqnarray*}
\frac{d}{dt}\:\frac{1}{2i}\:\SYC{\theta(t)}{\overline{\theta(t)}}
&=&
\SPb{\Re \theta_\xi(t)}{\bbA(t)\,\Re \theta_\xi(t)}
\,+\,
\SPb{\Im \theta_\xi(t)}{\bbA(t)\,\Im \theta_\xi(t)}\,,
\end{eqnarray*}
for 
$t\in[0,\tau]$.
Integrating the previous identity from $0$ to $\tau$
and using the fact that $\bbA(t)$ is positive definite,
for all $t\in[0,\tau]$, we see that assertion of the lemma
holds true. In fact, the structure of $\FM_{a}(x(t),e)$
shows that we have equality in (\ref{klaus}) if and only
if $\theta_\xi(0)=0$. 
\qed

\bigskip

\noindent In the following we set
\begin{eqnarray*}
\widetilde{\mathscr{D}}&:=&
\big\{\,
(t,y,e)\in[0,\infty)\times K_0\times(2\pi\ZZ)^d\,:\;
\;\;\kappa_{t'}(y,e)\in K_0\times\{e\}\,,\;t'\in[0,t]\,
\big\}\,,
\\
\mathscr{D}&:=&
\big\{\,
(t,x,e)\in[0,\infty)\times K_0\times(2\pi\ZZ)^d\,:\;
\;\;\kappa_{-t'}(x,e)\in K_0\times\{e\}\,,\;t'\in[0,t]\,
\big\}\,,
\\
\widetilde{\mathscr{E}}&:=&
\big(\{0\}\times\RR^d\times\RR_d\big)\cup\widetilde{\mathscr{D}}\,,
\qquad 
\mathscr{E}\,:=\,
\big(\{0\}\times\RR^d\times\RR_d\big)\cup\mathscr{D}\,.
\end{eqnarray*}
By Proposition~\ref{prop-vp} and Lemma~\ref{le-HFA-XH} it is clear
that there is some $t_0\in(0,\infty)$ such that
$\widetilde{\mathscr{D}},\mathscr{D}
\subset[0,t_0]\times K_0\times(2\pi\ZZ)^d$ and,
moreover,
$$
\forall\:(t,y,e)\in\widetilde{\mathscr{D}}\::\quad
\dot{Q}(t,y,e)\,=\,\nabla_p H\big(Q(t,y,e),\vp'(Q(t,y,e))\big)\,.
$$

\begin{lemma}\label{le-proj} 
There is some neighbourhood, $\mathscr{G}$, of $\mathscr{E}$
in $[0,\infty)\times\CC^d\times\CC_d$ with the following properties:
There are two smooth
functions, $k$ and $g$, defined on $\mathscr{G}$, such that,
for every $(t,x,\xi,y,\eta)\in[0,\infty)\times T^*\CC^d\times T^*\CC^d$,
$$
(x,\xi,y,\eta)\in C_t\quad\Leftrightarrow\quad
y\,=\,k(t,x,\eta)\;\;\wedge\;\;\xi\,=\,g(t,x,\eta)\,.
$$
$\mathscr{G}$ is $2\pi$-periodic in $\Re\eta$ and,
for all $(t,x,\eta)\in\mathscr{G}$ 
and $e\in(2\pi\ZZ)^d$, 
\begin{equation}\label{gk-per}
g(t,x,\eta+e)\,=\,g(t,x,\eta)+e\,,\qquad
k(t,x,\eta+e)\,=\,k(t,x,\eta)\,.
\end{equation}
Moreover, 
\begin{equation}\label{gaufD}
\forall\;(t,x,e)\in\mathscr{D}\::\qquad g(t,x,e)\,=\,e\,,\quad
k(t,x,e)\,=\,\pi\,\exp\big(-t\,X_{H}\big)(x,\vp'(x))\,.
\end{equation}
\end{lemma}

\proof
It suffices to show that the
projection
$C_t\ni(x,\xi,y,\eta)\mapsto(x,\eta)$ is bijective 
in a neighbourhood of each point on $C_t$ with 
$(t,x,\eta)\in\mathscr{E}$.
For $t=0$, this is trivial by definition of $C_0$.
So, it remains to treat the case $(t,x,\eta)\in\mathscr{D}$, $t>0$.

Let $t>0$ and let 
$\rho:[0,t]\rightarrow K_0\times(2\pi\ZZ)^d$ be
a real integral curve of $\HFA$.
Let
$\vt=(\vt_x,\vt_\xi,\vt_y,\vt_\eta)\in T_{(\rho(t),\rho(0))}C_{t}$ 
be such that
$\vt_x=0$ and $\vt_\eta=0$. 
It clearly holds 
$\widetilde{\sigma}\ominus\widetilde{\sigma}(\vt,\overline{\vt})=0$.
Using the same notation as in the proof of
Lemma~\ref{le-klaus} we thus have that $\theta_\xi(r)=0$, for
all $r\in[0,t]$. 
In particular, $\vt_\xi=0$. 
Then it is also clear that $\vt_y=0$, because
$\theta_x(r)$ solves the initial value problem
$\frac{d}{dr}\theta_x(r)=\bbB(r)\,\theta_x(r)$, $\theta_x(t)=0$.
Consequently,
the projection $C_t\ni(x,\xi,y,\eta)\mapsto(x,\eta)$
is bijective in a neighbourhood of 
$(\rho(t),\rho(0))\in C_{t}$. 

Finally, (\ref{gk-per}) follows from the periodicity of $C_t$
and (\ref{gaufD}) from Lemma~\ref{le-HFA-XH} and the definition
of $\mathscr{D}$.
\qed

\bigskip

\noindent In the remaining part of this section we show that
an approximate solution of (\ref{hansi}) is given by
\begin{equation}\label{def-psi}
\psi(t,x,\eta)\,:=\,
\SPn{k(t,x,\eta)}{\eta}
\,+\,\int_0^tA(\kappa_r(k(t,x,\eta),\eta))\,dr\,,
\qquad (t,x,\eta)\in\mathscr{G}\,.
\end{equation}
We observe that $\psi(0,x,\eta)=\SPn{x}{\eta}$ and 
a direct computation using (\ref{HEQXi}), (\ref{per-QXi}), and
(\ref{gk-per}) shows, for $(t,x,\eta)\in\mathscr{G}$ 
and $e\in(2\pi\ZZ)^d$,
\begin{equation}\label{per-psi}
\psi(t,x,\eta+e)\,=\,\psi(t,x,\eta)\,+\,\SPn{\eta}{e}\,,
\end{equation}
so that
the difference $\psi-\SPn{x}{\eta}$ is $(2\pi\ZZ)^d$-periodic
in $\Re\eta$.
By definition of $\mathscr{E}$ and (\ref{ima=0}) and (\ref{rea'not=0})
we also know that
\begin{equation}\label{psiaufE}
\forall\;(t,x,\eta)\in\mathscr{E}\;:\qquad
\Im\psi(t,x,\eta)\,=\,0\,.
\end{equation}
To study $\psi$ further we introduce, for $(t,y,\eta)\in\RR\times T^*\CC^d$,
\begin{eqnarray*}
Z(t,y,\eta)&:=&-\vs_t(-\SPn{y}{\eta},y,\eta)
\,=\,\SPn{y}{\eta}
\,+\,\int_0^tA(\kappa_r(y,\eta))\,dr\,,
\\
\Theta(t,y,\eta)&:=&(t,\kappa_t(y,\eta))\,,
\\
\wtG(t,y,\eta)&:=&
\big|\Im(Q,\Xi)(t,y,\eta)\big|^2\,+\,\SMS(-Z,Q,\Xi)(t,y,\eta)\,.
\end{eqnarray*}

\begin{lemma}\label{est-MeSj3-QXi}
There is a small neighbourhood, $\widetilde{\mathscr{N}}$, 
of $\widetilde{\mathscr{E}}$ in 
$[0,\infty)\times\CC^d\times\RR_d$, which is $(2\pi\ZZ)^d$-periodic
in the last variable, such that,
for all $(t,y,\eta)\in\widetilde{\mathscr{N}}$ and
$r\in[0,t]$,
\begin{eqnarray}
\Im Z(t,y,\eta)
&\grg&\SPn{\Im Q(t,y,\eta)}{\Re \Xi(t,y,\eta)}\label{eq-est-MeSj1-QXi}
\,-\,\frac{1}{3}\int_0^t\Im a(\Re\kappa_u(y,\eta))\,du
\\
& &\quad\qquad
\,-\,\bigO(1)\,
\big|\Im(Q,\Xi)(t,y,\eta)\big|^3\,,\nonumber
\\
\frac{1}{2}\:\big|\Im(Q,\Xi)(r,y,\eta)\big|^2
&\klg&
\wtG(r,y,\eta)\,\klg\,\bigO(1)\,\wtG(t,y,\eta)
\,,\label{eq-est-MeSj3-QXi}
\end{eqnarray}
where the $\bigO$-symbols are uniform on compact subsets of 
$\widetilde{\mathscr{N}}$.
\end{lemma}

\proof
For each fixed $\eta\in\RR_d$,
we apply Lemma~\ref{le-Prop3.1}
with $\psi_0(y):=\SPn{y}{\eta}$, $y\in\CC^d$, 
and $\rho_0\in\RR^d\times\{\eta\}$.
If $\rho_0\in K_0\times(2\pi\ZZ)^d$, we choose
$\tau:=\max\{t>0:\,\kappa_t(\rho_0)\in K_0\times(2\pi\ZZ)^d\}$.
If $\rho_0\notin K_0\times(2\pi\ZZ)^d$, we set $\tau=0$. 
So, if $\tau>0$,
then Lemma~\ref{le-HFA-XH} ensures that $\kappa_t(\rho_0)$
is real, for $t\in[0,\tau]$, 
and by (\ref{ima=0}) we know that $\Im a(\rho_0)=0$.
Then (\ref{eq-est-MeSj1-QXi}) follows from (\ref{est-MeSj1}).
The first inequality in (\ref{eq-est-MeSj3-QXi})
follows from (\ref{imaklg0}), (\ref{eq-est-MeSj1-QXi}), 
and the fact that we can assume $\big|\Im(Q,\Xi)(r,y,\eta)\big|$
to be arbitrarily small by restricting our attention to
some sufficiently small neighbourhood of $\widetilde{\mathscr{E}}$,
where $\big|\Im(Q,\Xi)(r,y,\eta)\big|$ vanishes.
The second inequality in (\ref{eq-est-MeSj3-QXi})
follows from (\ref{imaklg0}) and (\ref{est-MeSj2}).
\qed

\bigskip

\noindent In the next lemma we compare the differential of $Z$
with the pull-back under $\Theta$ of the Cartan form,
$$
\CF\,:=\,\xi\,dx\,-\,a(x,\xi)\,dt\,,
$$
considered as a form on $\RR\times T^*\CC^d$.
Following a standard proof known from classical mechanics
(see, e.g., \cite[pp.~479]{GH})
we only have to keep track of the error terms.

\begin{lemma}\label{le-Cartan}
(i) On every compact subset $\Omega\Subset\RR\times T^*\CC^d$
such that $|t|\klg t_0$ on $\Omega$,
\begin{equation}\label{Cartan1}
dZ\,=\,
-\Theta^*\CF\,+\,y\,d\eta\,+\,
\bigO_N\big(\max_{|\tilde{t}|\klg t_0}
\wtG(\tilde{t},y,\eta)^N\big)\,.
\end{equation}
(ii) Let $\widetilde{\mathscr{N}}$ be as in Lemma~\ref{est-MeSj3-QXi}.
Then
$$
dZ\,=\,
-\Theta^*\CF\,+\,y\,d\eta\,+\,
t\,\bigO_N\big(\wtG^N\big)
$$
on $\widetilde{\mathscr{N}}$,
where the $\bigO_N$-symbols are uniform on compact subsets
of $\widetilde{\mathscr{N}}$.
\end{lemma}

\proof
We set $\lambda:=dZ+\Theta^*\CF$ and use
\begin{equation}\label{erikah}
\dot{Z}\,=\,\SPn{\Xi}{a_\xi'(Q,\Xi)}\,-\,a(Q,\Xi)
\,=\,\SPn{\Xi}{\dot{Q}}\,-\,a(Q,\Xi)
\end{equation}
to obtain
$$
\lambda\,=\,
(Z_{y}'-\SPn{\Xi}{Q_{y}'})\,dy\,+\,
(Z_{\overline{y}}'-\SPn{\Xi}{Q_{\overline{y}}'})\,d\overline{y}\,+\,
(Z_{\eta}'-\SPn{\Xi}{Q_{\eta}'})\,d\eta\,+\,
(Z_{\overline{\eta}}'-\SPn{\Xi}{Q_{\overline{\eta}}'})\,d\overline{\eta}
\,.
$$
Now, let $\alpha$ denote any of the variables
$y^i,\overline{y}^i,\eta_i,\overline{\eta}_i$, $1\klg i\klg d$,
and set $\lambda_\alpha:=(Z_{\alpha}'-\SPn{\Xi}{Q_{\alpha}'})$.
Using successively (\ref{erikah}), the Hamiltonian equations
(\ref{HEQXi})
and the almost analyticity of $a$
we find
\begin{eqnarray*}
\dot{\lambda}_\alpha&=&
\partial_\alpha\big(\SPn{\Xi}{\dot{Q}}\,-\,a(Q,\Xi)\big)
\,-\,\SPn{\dot{\Xi}}{Q_\alpha'}\,-\,
\SPn{\Xi}{\dot{Q}_\alpha'}
\\
&=&
\SPn{\Xi_\alpha'}{a_\xi'(Q,\Xi)}\,-\,\partial_\alpha \big(a(Q,\xi)\big)
\,+\,\SPn{a_x'(Q,\Xi)}{Q_\alpha'}
\\
&=&
-a_{\overline{x}}'(Q,\Xi)\,\overline{Q_{\overline{\alpha}}'}
\,-\,a_{\overline{\xi}}'(Q,\Xi)\,\overline{\Xi_{\overline{\alpha}}'}
\\
&=&\bigO_N\big(|\Im(Q,\Xi)|^N\big)\,.
\end{eqnarray*}
Due to the initial conditions
$(Z,Q,\Xi)|_{t=0}=(\SPn{y}{\eta},y,\eta)$ we have
$$
Z_y'|_{t=0}=\eta\,,\quad 
Z_\eta'|_{t=0}=y\,,\quad
Z_{\overline{y}}'|_{t=0}=Z_{\overline{\eta}}'|_{t=0}=0\,,
\quad
Q_y'|_{t=0}\,=\,\id\,,\quad
Q_\eta'|_{t=0}=
Q_{\overline{y}}'|_{t=0}=Q_{\overline{\eta}}'|_{t=0}=0
\,.
$$
We conclude
$$
\lambda(t,y,\eta)\,=\,y\,d\eta
\,+\,\int_0^t\bigO_N\big(|\Im(Q,\Xi)(r,y,\eta)|^N\big)\,dr
\,,
$$
which implies (i). (ii) follows from 
Lemma~\ref{est-MeSj3-QXi}. (Note that (\ref{eq-est-MeSj3-QXi})
is available only for {\em real} $\eta$.)
\qed

\bigskip

\noindent
In what follows we set
$$
K(t,x,\eta)\,:=\,\big(t,k(t,x,\eta),\eta\big)\,,
$$
for all $(t,x,\eta)\in\mathscr{G}$, so that
\begin{equation}\label{sascha}
K^*Z\,=\,Z\circ K\,=\,\psi\,,\quad
K^*Q\,=\,Q\circ K\,=\,x\,,\quad
K^*\Xi\,=\,\Xi\circ K\,=\,g\,.
\end{equation}
Furthermore, we write
$\mathscr{N}:=K^{-1}(\widetilde{\mathscr{N}})$ and
$$
\Gamma\,:=\,K^*\wtG\,=\,\wtG\circ K
\,=\,\big|\Im(x,g)\big|^2\,-\,\SPn{\Im x}{\Re g}\,+\,\psi\,.
$$
The set $\mathscr{N}\subset\mathscr{G}$ is
a neighbourhood of 
$\mathscr{E}$ in $[0,\infty)\times\CC^d\times\RR_d$,
which is $2\pi$-periodic in $\eta$.

\begin{proposition}\label{prop-psi-1}
Let $\psi$ be given by (\ref{def-psi}).
Then
\begin{eqnarray}
\partial_{(t,x,\overline{x},\Re\eta)}^\alpha
\big(\,\partial_t\psi\,+\,a(x,\psi_x')\,\big)
&=&
\bigO_{N,\alpha}(\Gamma^N)\,,
\label{t-dep-Ham-Jac-eq-alpha-1}
\\
\partial_{(t,x,\overline{x},\Re\eta)}^\alpha(\psi_x'\,-\,g)
&=&
\bigO_{N,\alpha}(\Gamma^N)\,,
\label{t-dep-Ham-Jac-eq-alpha-2}
\\
\partial_{(t,x,\overline{x},\Re\eta)}^\alpha(\psi_\eta'\,-\,k)
&=&
\bigO_{N,\alpha}(\Gamma^N)
\,,\label{t-dep-Ham-Jac-eq-alpha-3}
\\
\partial_{(t,x,\overline{x},\Re\eta)}^\alpha\psi_{\overline{x}}'
&=&
\bigO_{N,\alpha}(\Gamma^N)
\,,\label{t-dep-Ham-Jac-eq-alpha-4}
\\
\partial_{(t,x,\overline{x},\Re\eta)}^\alpha
\psi_{\overline{\eta}}'
&=&
\bigO_{N,\alpha}(\Gamma^N)
\,,\label{t-dep-Ham-Jac-eq-alpha-5}
\end{eqnarray}
on $\mathscr{N}$,
for $N\in\NN$ and every multi-index $\alpha\in\NN_0^{3d+1}$.
All $\bigO_{N,\alpha}$-symbols are uniform on compact subsets
of $\mathscr{N}$. 
\end{proposition}

\proof
On account of Lemma~\ref{le-Cartan}(ii) we obtain
\begin{eqnarray*}
d\psi&=&d(K^*Z)\;=\;K^*(dZ)
\\
&=&
K^*\Big(\Theta^*\omega\,+\,y\,d\eta\,+\,t\,\bigO_N
\big(\wtG^N\big)\Big)
\\
&=&
g\,dx\,-\,a(x,g)\,dt\,+\,k\,d\eta
\,+\,t\,\bigO_N
\big(\Gamma^N\big)
\end{eqnarray*}
on $\mathscr{N}$.
Using Lemma~\ref{tech-le-MeSj}
we infer that 
(\ref{t-dep-Ham-Jac-eq-alpha-1})-(\ref{t-dep-Ham-Jac-eq-alpha-5}) 
hold true. 
\qed

\bigskip

\noindent The next corollary summarizes the properties of $\psi$
on the real domain, where the weight $\Gamma$ can actually
be replaced
by $\Im\psi$:

\begin{corollary}\label{le-psi}
(i) There is some neighbourhood, $\mathscr{M}_\RR$, 
of $\mathscr{E}$ in $[0,\infty)\times\RR^d\times\RR_d$,
which is $2\pi$-periodic in the last variable, such that,
for all $(t,x,\eta)\in\mathscr{M}_\RR$,
\begin{eqnarray}\label{psigrgg}
\Im\psi(t,x,\eta)&\grg&\frac{1}{\bigO(1)}|\:\Im g(t,x,\eta)|^2\,,
\\
\Im\psi(t,x,\eta)&=&0\quad\Leftrightarrow\quad
(t,x,\eta)\in\mathscr{E}\,.\label{psiaufE2}
\end{eqnarray}
(Here the $\bigO$-symbol is again uniform on compact subsets
of $\mathscr{M}_\RR$.) Consequently,
\begin{equation}\label{psigrgGamma}
\Im\psi\,\grg\;\frac{1}{\bigO(1)}\:\Gamma\qquad
\textrm{on}\;\;\mathscr{M}_\RR\,,
\end{equation}
so that
(\ref{t-dep-Ham-Jac-eq-alpha-1})-(\ref{t-dep-Ham-Jac-eq-alpha-5}) 
hold true on $\mathscr{M}_\RR$ with the right sides replaced
by $\bigO_N((\Im\psi)^N)$. In particular, 
\begin{equation}\label{t-dep-Ham-Jac-eq}
\partial_{(t,x,\overline{x},\Re\eta)}^\alpha
\big(\partial_t\psi\,+\,a(x,\psi_x')\big)\,=\,
\bigO_{N,\alpha}\big((\Im \psi)^N\big)
\qquad\textrm{on}\;\;\mathscr{M}_\RR\,.
\end{equation}
(ii) For all $(t,x,e)\in\mathscr{D}$,
\begin{equation}\label{lotta99}
\partial_t\psi(t,x,e)\,=\,0\,,\qquad \nabla_\eta\psi(t,x,e)\,=\,
\pi\exp\big(-t\,X_{H}\big)(x,\vp'(x))\,,\qquad
\psi_x'(t,x,e)\,=\,e
\,. 
\end{equation}
\end{corollary}

\proof
(i) We observe that (\ref{imaklg0}),
(\ref{eq-est-MeSj1-QXi}), and (\ref{sascha}) imply
$$
\Im \psi-\SPn{\Im x}{\Re g}
\,\grg\,-\bigO(1)\,
\big|\Im (x,g)\big|^3\qquad\textrm{on}\;\;\mathscr{N}\,.
$$
Let $(t_0,x_0,\eta_0)\in\mathscr{E}$. 
From (\ref{gaufD}) we know that $\Im g$
vanishes on $\mathscr{E}$ and we can thus find
some compact, real neighbourhood, $K'$, of $(t_0,x_0,\eta_0)$
in $\mathscr{N}\cap([0,\infty)\times\RR^d\times\RR_d)$,
such that $x+\ve\Im g(t,x,\eta)\in\mathscr{N}$, for
all $(t,x,\eta)\in K'$ and every $\ve\in[0,1]$. Consequently,
there exist constants
$C,C'\in(0,\infty)$ such that, for all $(t,x,\eta)\in K'$,
$$
\Im\psi(t,x-\ve\,\Im g(t,x,\eta),\eta)
\,\grg\,
-C\,\big|\Im g\big(t,x-\ve\,\Im g(t,x,\eta),\eta\big)\big|^3
\,\grg\,-C'\,\big|\Im g(t,x,\eta)\big|^3
\,.
$$
Taylor expanding the left hand side with respect to $x$
and using (\ref{t-dep-Ham-Jac-eq-alpha-2})
and (\ref{t-dep-Ham-Jac-eq-alpha-4})
we obtain
\begin{eqnarray*}
\Im\psi&\grg&\ve\,|\Im g|^2\,-\,
C''
\big(|\Im g|^3\,+\,\ve^2\,|\Im g|^2\big)
\,-\,\ve\,C_{N_0}\,\big|\Im \psi\big|^{N_0}\qquad
\textrm{on}\;\;K'\,,
\end{eqnarray*}
for some ${N_0}\in\NN$, ${N_0}\grg2$, and $C'',C_{N_0}\in(0,\infty)$.
Now, we choose $\ve\in(0,\frac{1}{2C''})$ such that
$\ve\,C_{N_0}\,|\Im\psi|^{{N_0}-1}<1/2$ on $K'$.
By possibly restricting our attention to an even smaller real
compact neighbourhood, $K''\Subset K'$, of $(t_0,x_0,\eta_0)$,
we may further ensure that
$C''|\Im g|<\ve/4$ on $K''$. Then we readily obtain (\ref{psigrgg})
on $K''$.

Concerning (\ref{psiaufE2}), it only remains to prove the
implication ``$\Rightarrow$'' because of (\ref{psiaufE}).
So, suppose that $(t,x,\eta)\in\mathscr{M}_\RR$ 
and $\Im\psi(t,x,\eta)=0$. Then (\ref{psigrgg})
implies that $\Im g(t,x,\eta)=0$
and (\ref{eq-est-MeSj3-QXi}) and (\ref{sascha})
show that $(y,\eta):=(k(t,x,\eta),\eta)$
and $(x,g(t,x,\eta))$ are connected by a purely real integral curve of
$\widehat{X}_a$.
Hence, if we had $\Im a(y,\eta)<0$,
then we also had $\Im\psi(t,x,\eta)>0$ by 
(\ref{imaklg0}) and (\ref{eq-est-MeSj1-QXi}).
So, we must have $\Im a(y,\eta)=0$ and Lemma~\ref{le-HFA-XH}
implies that $t=0$
or $(t,x,\eta)\in\mathscr{D}$.

(ii) follows from (\ref{a=0}), (\ref{gaufD}), and
(\ref{t-dep-Ham-Jac-eq-alpha-1}).
\qed

\bigskip

\noindent
To solve the transport equations in the next section
we need another consequence of Proposition~\ref{prop-psi-1}:

\begin{corollary}\label{overlinebla}
Locally on $\mathscr{N}$,
\begin{equation}\label{kovx}
k_{(\overline{x},\overline{\eta})}'\,=\,\bigO_N\big(\Gamma^N\big)\,,
\qquad
g_{(\overline{x},\overline{\eta})}'\,=\,\bigO_N\big(\Gamma^N\big)\,,
\end{equation}
and locally
on some small neighbourhood, 
$\widetilde{\mathscr{N}}'\subset\widetilde{\mathscr{N}}$,
of $\widetilde{\mathscr{E}}$
in $[0,\infty)\times\CC^d\times\RR_d$
\begin{equation}\label{Qovy}
Q_{(\overline{y},\overline{\eta})}'\,=\,\bigO_N\big(\wtG^N\big)\,,
\qquad
\Xi_{(\overline{y},\overline{\eta})}'\,=\,
\bigO_N\big(\wtG^N\big)\,,
\end{equation}

\end{corollary}

\proof
(\ref{kovx}) follows immediatly from
(\ref{t-dep-Ham-Jac-eq-alpha-2})-(\ref{t-dep-Ham-Jac-eq-alpha-5}).
The first equality in
(\ref{Qovy}) is easily derived from (\ref{kovx})
by differentiating the identity
$k(t,Q(t,y,\eta),\eta)=y$ with respect to
$y$ and $(\overline{y},\overline{\eta})$.
To obtain the second we differentiate 
$g(t,Q(t,y,\eta),\eta)=\Xi(t,y,\eta)$
with respect to $(\overline{y},\overline{\eta})$
and use the first and again (\ref{kovx}).
\qed


\section{The transport equations}
\label{sec-heat}

In this section we carry out the next step
in the construction of a parametrix for
the heat operator $h\partial_t+\Op(a^\vp_W)$.
We seek for a parametrix, $\mathcal{Q}$,
of the form
\begin{equation}\label{ansatz-u}
\mathcal{Q}f(t,x)\,=\,
\int e^{i\psi(t,x,\eta)/h-i\SPn{y}{\eta}/h}b(t,x,\eta\,;h)
\,f(y)\:\frac{dyd\eta}{(2\pi h)^d}\,,
\end{equation}
where $\psi$ is given by (\ref{def-psi}).
For the amplitude $b$ we use the ansatz
\begin{equation}\label{b-asymp}
b(t,x,\eta\,;h)\,\asymp\,\sum_{\nu=0}^\infty h^\nu b_\nu(t,x,\eta)\,.
\end{equation}
To find the sequence of transport equations that determines
$b_0,b_1,\dots$ 
we ignore for the moment that
$\psi$ is defined only in a small neighbourhood
of 
$\mathscr{E}$ and argue formally.
(We shall perform everything properly in Proposition~\ref{prop-56}
where the results of this and the previous section
are put together.)
Observing that
$$
e^{-i\psi/h}\,\Op(a_W^\vp)\,e^{i\psi/h}\,=\,
e^{-i\psi/h+\vp/h}\,\Op(i\,\widetilde{V}-i\,U_h)\,
e^{-\vp/h+i\psi/h}\,=\,
\Op(a_W^{\vp-i\psi})
$$
and using Lemma~\ref{le-stat-phase} with
$\phi=\vp-i\psi$ we see that
the equation 
$$
(h\partial_t+\Op(a^\vp_W))\,e^{i\psi/h}\,b\,=\,0
$$
leads to the following
sequence of transport equations:
\begin{eqnarray*}
(\mathrm{T}_0)
\quad(\partial_t+Y+\tfrac{1}{2}\,\diz Y)\,b_0&=&
\bigO_N(\Gamma^N)\,,
\quad N\in\NN\,,
\\
(\mathrm{T}_1)
\quad(\partial_t+Y+\tfrac{1}{2}\,\diz Y)\,b_1&=&P_{2}\,b_0\,+
\,\bigO_N(\Gamma^N)
\,,\quad N\in\NN\,,
\\
&\vdots&
\\
(\mathrm{T}_\nu)
\quad(\partial_t+Y+\tfrac{1}{2}\,\diz Y)\,b_{\nu}&=&
\sum_{\vk=0}^{\nu-1}
P_{\nu+1-\vk}\,b_\vk\,+\,\bigO_N(\Gamma^N)\,,\quad N\in\NN\,,
\\
&\vdots&
\end{eqnarray*}
Here we have 
\begin{equation}\label{def-Y}
Y(t,x,\eta)\,=\,\sum_{i=1}^d\partial_{\xi_i}a^{\vp-i\psi(t,\cdot,\eta)}
(x,\xi)\big|_{\xi=0}\,
\partial_{x^i}
\,=\,\sum_{i=1}^d\partial_{\xi_i}a^\vp(x,\psi_x'(t,x,\eta))\,
\partial_{x^i}
\end{equation}
and the operators $P_2,P_3,\dots$ are given by (\ref{def-Pnu})
with $\phi=\vp-i\psi$. 
By (\ref{per-psi}) $\vp-i\psi-i\SPn{x}{\eta}$ depends 
$(2\pi\ZZ)^d$-periodically on $\eta$ and, hence,
the same holds for $Y$ and $P_\nu$,
so that $\eta$ can be viewed as an element of $\TT^d$
in the following.
As initial conditions we choose
\begin{eqnarray}\label{eq-ab-T_0}
b_0(0,x,\eta)&:=&\chi(x)\,,
\\ \label{eq-ab-T_nu}
b_\nu(0,x,\eta)&:=&0\,,\quad\nu\in\NN\,,
\end{eqnarray}
where $\eta\in\TT^d$ and $\chi\in C_0^\infty(\RR^d,[0,1])$ 
is supported in some small neighbourhood of
$K_0$ and fulfills $\chi\equiv1$
in another small neighbourhood of
$K_0$. In the following
we also assume 
that $\chi$ is extended almost analytically 
such that the extension, $\widetilde{\chi}$, 
is compactly supported in some
tiny complex neighbourhood of $\supp(\chi)$.

\begin{lemma}\label{le-winfried}
The transport equations $(T_0), (T_1),\dots$ with initial conditions
(\ref{eq-ab-T_0}) and (\ref{eq-ab-T_nu}) are solvable
in a small 
neighbourhood, 
$\mathscr{N}'\subset\mathscr{N}$,
of 
$\mathscr{E}$ in $[0,\infty)\times\CC^d\times\RR_d$, 
which is $2\pi$-periodic in $\eta$.
More generally, 
\begin{equation}\label{karlgustav}
\partial_{(t,x,\overline{x},\eta)}^\alpha
\Big((\partial_t+Y+\tfrac{1}{2}\,\diz Y)\,b_{\nu}\,-\,
\sum_{\vk=0}^{\nu-1}
P_{\nu+1-\vk}\,b_\vk\,\Big)
\,=\,
\left\{
\begin{array}{ll}
\bigO_{N,\alpha}(\Gamma^N)\,,&\textrm{on}\;\mathscr{N}'\,,
\\
\bigO_{N,\alpha}((\Im\psi)^N)\,,&\textrm{on}
\;\mathscr{N}'\cap\mathscr{M}_\RR\,,
\end{array}
\right.
\end{equation}
for every $N\in\NN$ and
any multiindex $\alpha\in\NN_0^{3d+1}$, where
the $\bigO_{N,\alpha}$-symbols are uniform
on compact subsets of $\mathscr{N}'$. 
The solutions $b_\nu$, $\nu\in\NN_0$, 
are $2\pi$-periodic
in $\eta$.
We have
$\supp\big(b_\nu(t,\cdot,\eta)\big)
\subset Q\big(t,\supp(\widetilde{\chi}),\eta\big)$,
for all $(t,\eta)$ and $\nu\in\NN_0$.
\end{lemma}

\proof
The proof is essentially standard; see, e.g., \cite[pp. 578]{Tr2}. 
We have, however,
to keep  track of the error terms. 
We argue by induction and consider the equation
$$
(\partial_t\,+\,Z)\,b
\,=\,f\,b\,+\,w\,,\qquad b|_{t=0}\,=\,c
$$
where 
$$
Z(t,x)\,:=\,\sum_{i=1}^d \partial_{\xi_i}a^\vp(x,g(t,x,\eta))\,
\partial_{x^i}\,.
$$
Moreover, $f:=-\tfrac{1}{2}\,\diz Y$ 
and $w$ is either $0$, in which case we set $c=\widetilde{\chi}$,
or $w=\sum_{\vk=0}^{\nu-1}
P_{\nu+1-\vk}\,b_\vk$, where we suppose that $b_0,\dots,b_{\nu-1}$
have already been constructed. 
In the latter case
$c=0$.
We know that
$$
f_{\overline{x}}'\,=\,\bigO_N\big(\Gamma^N\big)\qquad
\textrm{on}\;\;\mathscr{N}
$$
and as an induction hypothesis we assume that
$$
(b_0)_{\overline{x}}'\,,\dots,\,(b_{\nu-1})_{\overline{x}}'
\,=\,\bigO_N\big(\Gamma^N\big)\qquad
\textrm{on}\;\;\mathscr{N}':=K^{-1}(\widetilde{\mathscr{N}}')\,,
$$
which implies
$$
w_{\overline{x}}'
\,=\,\bigO_N\big(\Gamma^N\big)\qquad
\textrm{on}\;\;\mathscr{N}'\,.
$$
Here $\widetilde{\mathscr{N}}'$ is the neighbourhood of $\mathscr{E}$
appearing in Corollary~\ref{overlinebla}.
We observe that 
$Q$ is the flow of
$\widehat{Z}:=Z+\overline{Z}$. 
We first solve the following initial value problem
for an ordinary differential equation,
\begin{equation}\label{eq-ODE}
\partial_t \tilde{b}(t,y,\eta)\,=\,f(t,Q(t,y,\eta),\eta)\,\tilde{b}(t,y,\eta)
\,+\,w(t,Q(t,y,\eta),\eta)\,,\quad
\tilde{b}(0,y,\eta)\,=\,c(y)\,,
\end{equation}
and observe that 
$$
b(t,x,\eta)\,:=\,\tilde{b}(t,k(t,x,\eta),\eta)
$$ 
fulfills
\begin{equation}\label{petra0}
(\partial_tb\,+\,\widehat{Z}\,b)(t,Q(t,y,\eta),\eta)
\,=\,\frac{d}{dt}\:b(t,Q(t,y,\eta),\eta)\,=\,
f(t,Q(t,y,\eta),\eta)\,\tilde{b}(t,y,\eta)
\,+\,w(t,Q(t,y,\eta),\eta)
\end{equation}
on $\widetilde{\mathscr{N}}'$
because of $k(t,Q(t,y,\eta),\eta)=y$.
Next, we consider the error term
\begin{equation}\label{rudi}
b_{\overline{x}}'(t,x,\eta)\,=\,\tilde{b}_y(t,k(t,x,\eta),\eta)\,
k_{\overline{x}}'(t,x,\eta)
\,+\,\tilde{b}_{\overline{y}}(t,k(t,x,\eta),\eta)
\,\overline{k_{x}'(t,x,\eta)}
\,.
\end{equation}
Here the error $k_{\overline{x}}'$ can be controlled using (\ref{kovx}).
To investigate $\tilde{b}_{\overline{y}}$ we 
differentiate (\ref{eq-ODE}) with respect to $\overline{y}$
and obtain
\begin{equation}\label{petra2}
\partial_t\tilde{b}_{\overline{y}}'(t,y,\eta)\,-\,
f(t,Q(t,y,\eta),\eta)\,\tilde{b}_{\overline{y}}'(t,y,\eta)
\,=\,\bigO_N\big(\wtG(t,y,\eta)^N\big)
\qquad\textrm{on}\;\;\widetilde{\mathscr{N}}'
\,,
\end{equation}
if we take (\ref{Qovy}) into account.
It also holds $\tilde{b}_{\overline{y}}'(0,y,\eta)
=c_{\overline{y}}'(y)=\bigO_N(|\Im y|^N)$, so
we infer from (\ref{eq-est-MeSj3-QXi})
and (\ref{petra2}) together with 
Gronwall's lemma that
$$
\tilde{b}_{\overline{y}}'(t,k(t,x,\eta),\eta)\,=\,
\bigO_N\big(\Gamma(t,x,\eta)^N\big)\,.
$$
In view of (\ref{kovx}) we obtain
$$
b_{\overline{x}}'(t,x,\eta)\,=\,\bigO_N(\Gamma(t,x,\eta)^N)\,,\qquad
(t,x,\eta)\in\mathscr{N}'\,.
$$
Together with (\ref{petra0}) this shows that
\begin{equation}\label{sieglinde}
\partial_tb\,+\,Z\,b
\,=\,
f\,b
\,+\,w\,+\,\bigO_N(\Gamma^N)\qquad\textrm{on}\;\;\mathscr{N}'\,.
\end{equation}
Finally,
we observe that the components
of $Y$ and $Z$ satisfy 
$Y_i-Z_i=\bigO_N(\Gamma^N)$,
because of (\ref{t-dep-Ham-Jac-eq-alpha-2}).
The vector field $Z$ may hence be replaced by $Y$
in (\ref{sieglinde}).

Finally, it is clear that $b$
is $2\pi$-periodic in $\eta$. 
In fact, the maps $f$, $w$, $Q$, and $k$ 
used to construct $b$ only depend
on the equivalence class of $\eta$ in $\TT^d$. 
Moreover, it is easy to see
that the inclusion
$\supp(w(t,\cdot,\eta))\subset Q(t,\supp(\widetilde{\chi}),\eta)$
implies that the support of $b(t,\cdot,\eta)$
is again contained in
$Q(t,\supp(\widetilde{\chi}),\eta)$.
So, we also obtain the last statement of the lemma.
\qed

\bigskip

\noindent 
Next, we consider $b_0(\tau,x,0)$ more closely, for 
$(\tau,x,0)\in \mathring{\mathscr{D}}$. By definition
of $\mathscr{D}$ we know
that
there is some $y\in K_0$ 
such that $Q(\tau,y,0)=x$ and
$Q(t,y,0)\in K_0$, for all $t\in[0,\tau]$.
By (\ref{rea'not=0}), (\ref{HEQXi}), (\ref{lotta99}), and (\ref{def-Y})
we further know that, for $t\in[0,\tau]$,
\begin{equation}\label{YYYY}
Y(t,Q(t,y,0),0)\,=\,
\nabla_p H\big(Q(t,y,0),\vp'(Q(t,y,0))\big)\,=\,\dot{Q}(t,y,0)\,.
\end{equation}
In particular, $Q(t,y,0)$ and $Y(t,Q(t,y,0),0)$
are real.

Henceforth we denote the determinant
of a real square matrix, $M$, by $|M|$.

\begin{lemma}\label{le-b_0}
Assume that $(\tau,x)\in \mathring{\mathscr{D}}$. 
Then the solution of $(\mathrm{T}_0)$ with initial condition
(\ref{eq-ab-T_0}) satisfies
$$
b_0(\tau,x,0)\,=\,
\big|\,Q_y'(\tau,y,0)\,\big|^{-1/2}
\,\chi(y)\,,\qquad \textrm{where}\quad
y\,=\,\pi_x\exp(-\tau\,X_H)(x,\vp'(x))\,.
$$
\end{lemma}

\proof
The proof is well-known. We recall it for the sake of completeness.
From the proof of Lemma~\ref{le-winfried} and (\ref{YYYY})
we infer that, for
$(t,y,0)\in\widetilde{\mathscr{D}}$,
$$
\frac{d}{dt}\:b_0(t,Q(t,y,0),0)\,=\,
-\frac{1}{2}\:\big(b_0\,\diz Y\big)(t,Q(t,y,0),0)\,.
$$
The Liouville formula shows that
$$
\frac{d}{dt}\:\big|\,Q_y'(t,y,0)\,\big|^{-1/2}
\,=\,-\frac{1}{2}\:\big|\,Q_y'(t,y,0)\,\big|^{-1/2}
\,\big(\diz Y\big)(t,Q(t,y,0),0)\,,
$$
and it follows that
$$
b_0(t,Q(t,y,0),0)
\,=\,\big|\,Q_y'(t,y,0)\,\big|^{-1/2}\,\chi(y)\,.
$$
\qed

\bigskip

\noindent In the next proposition
we draw the main conclusion of the results
attained in Sections~\ref{sec-Ham-Jac} and~\ref{sec-heat} so
far.

Up to now the restrictions of the coefficients $b_\nu(t,x,\eta)$,
$\nu\in\NN_0$, to the real domain
are defined in some $2\pi$-periodic neighbourhood, 
$\mathscr{N}'\cap\mathscr{M}_{\RR}$,
of $\mathscr{E}$ which is contained in the domain
of $\psi$. In what follows
we consider them as functions on 
$[0,\infty)\times\RR^d\times\TT^d$
by simply setting $b_\nu$ equal to zero
outside $\mathscr{N}'\cap\mathscr{M}_{\RR}$.
We pick some smooth cut-off
function, $\vr\in C_b^\infty([0,\infty)\times\RR^d\times\TT^d)$,
such that $\vr\equiv1$ in a neighbourhood of $\mathscr{E}$
and $\supp(\vr)\subset\mathscr{N}'\cap\mathscr{M}_{\RR}$,
and let $b$
denote a Borel resummation
in $\Symbb{[0,\infty)\times\RR^d\times\TT^d}$ of the formal
series on the right side of
$$
b(t,x,\eta\,;h)\,\asymp\,
\sum_{\nu=0}^\infty h^\nu\,b_\nu(t,x,\eta)\,\vr(t,x,\eta)\,.
$$
By Proposition~\ref{le-winfried} we then know
that
there is some $t_0>0$ and some compact set $\mathscr{K}\Subset\RR^d$
such that
$\supp(b)\subset[0,t_0]\times\mathscr{K}\times\TT^d$.

We may assume that we find another compactly supported
cut-off function,
$\tilde{\vr}\in C^\infty_0([0,\infty)\times\RR^d\times\TT^d,[0,1])$,
such that $\tilde{\vr}\equiv1$ on $\supp(b)$
and $\supp(\tilde{\vr})\subset\mathscr{M}_{\RR}$.
Extending $\psi$ again by zero outside $\mathscr{M}_\RR$
we define
$$
\widetilde{\psi}(t,x,\eta)
\,:=\,\tilde{\vr}(t,x,\eta)\,\psi(t,x,\eta)\,+\,(1-\tilde{\vr}(t,x,\eta))
\,\big(\SPn{x}{\eta}\,+\,i\,\big)\,,
$$
for $t\grg0$, $x\in\RR^d$, and $\eta\in\RR_d$.
$\widetilde{\psi}$ is a smooth function 
such that $\Im\widetilde{\psi}>0$ on 
$([0,\infty)\times\RR^d\times\RR_d)\setminus\mathscr{E}$
and such that $\widetilde{\psi}(t,x,\eta)-\SPn{x}{\eta}$
can be viewed as an element of $\Symbb{[0,\infty)\times\RR^d\times\TT^d}$.

\begin{proposition}\label{prop-56}
For every $N\in\NN$, there is some
compactly supported
$\check{r}_N\in\Symbb{[0,\infty)\times\RR^d\times\TT^d}$
with 
$\supp\big(\check{r}_N(t,\cdot,\eta\,;h)\big)\subset
\mathscr{K}+\{|x|\klg R\}$ such that,
for sufficiently small $h>0$,
$$
\big(\,h\,\partial_t\,+\,\Op(a_W^\vp)\,\big)\,\big(
e^{i\psi/h}\,b\big)
\,=\,
h^N\,e^{i\widetilde{\psi}/h}\,\check{r}_N\,.
$$
\end{proposition}

\proof
By definition of $a_W^\phi$ with $\phi=\vp$
and $\phi=\vp-i\widetilde{\psi}$, respectively,
\begin{equation}\label{knut99}
e^{-i\widetilde{\psi}/h}\,\Op(a^\vp_W)\,e^{i\psi/h}\,b
\,=\,
e^{-i\widetilde{\psi}/h}\,\Op(a^\vp_W)\,e^{i\widetilde{\psi}/h}\,b
\,=\,
\Op(a^{\vp-i\widetilde{\psi}}_W)\,b\,.
\end{equation}
By our choice of $\widetilde{\psi}$
the phase $\vp-i\widetilde{\psi}+i\SPn{x}{\eta}$
defines an element of
$\Symbb{[0,\infty)\times\RR^d\times\TT^d}$.
We may thus apply
Lemma~\ref{le-stat-phase} with $\phi=\vp-i\widetilde{\psi}$
to the last term in (\ref{knut99}) and
obtain the following asymptotic expansion
in $\Symbb{[0,\infty)\times\RR^d\times\TT^d}$,
\begin{eqnarray*}
\lefteqn{
e^{-i\widetilde{\psi}(t,x,\eta)/h}\,
\big((h\,\partial_t\,+\,\Op(a^W))\,
e^{i\psi/h}\,b\big)(t,x,\eta)
}
\\
&\asymp&
i\,\big(\partial_t\psi(t,x,\eta)\,+\,a^\vp(x,\psi_x'(t,x,\eta))\big)
\,b(t,x,\eta\,;h)
\,+\,
h\,\big((\partial_t\,+\,Y\,+\,\tfrac{1}{2}\,\diz Y)\,b_0\,\vr\big)(t,x,\eta)
\\
& &
+\;\sum_{\nu=2}^{\infty}h^\nu 
\Big(\big((\partial_t\,+\,
Y+\tfrac{1}{2}\,\diz Y)\,b_{\nu-1}\,\vr\big)(t,x,\eta)\,
+\,\sum_{\vk=0}^{\nu-2}
(P_{\nu-\vk}\,b_\vk\,\vr)(t,x,\eta)\Big)\,,
\end{eqnarray*}
where $Y$ and $P_\nu$ are given by (\ref{def-Y}) and (\ref{def-Pnu})
with $\phi=\vp-i\psi$. (Here we again use that $\psi=\widetilde{\psi}$
on
$\supp(b)$.)
We observe that
\begin{eqnarray*}
(\partial_t\,+\,Y\,+\,\tfrac{1}{2}\,\diz Y)\,(b_\nu\,\vr)
&=&
\big((\partial_t\,+\,Y\,+\,\tfrac{1}{2}\,\diz Y)\,b_\nu\big)\,\vr\,+\,m_\nu\,,
\qquad \nu=0,1,2,\dots
\\
P_{\nu-\vk}\,(b_\vk\,\vr)
&=&
(P_{\nu-\vk}\,b_\vk)\,\vr\,+\,\widetilde{m}_{\vk,\nu}\,,\qquad\;
0\klg\vk\klg\nu\,,\;\nu=2,3,4,\dots,
\end{eqnarray*}
where $m_\nu,\widetilde{m}_{\vk,\nu}
\in C_0^\infty([0,\infty)\times\RR^d\times\TT^d)$
are supported in $\supp(|\vr'|)\subset\{\Im\psi>0\}$, which implies
$$
e^{i\widetilde{\psi}/h}\,m_\nu\,=\,\bigO(h^\infty)\qquad
\textrm{and}
\qquad
e^{i\widetilde{\psi}/h}\,\widetilde{m}_{\vk,\nu}\,=\,\bigO(h^\infty)
\qquad\textrm{in}\;\;
\Symbb{\RR\times\RR^d\times\TT^d}\,.
$$
We next use (\ref{t-dep-Ham-Jac-eq}) and the fact that $b_0,b_1,\dots$
solve the transport equations $(T_0),(T_1),\dots$.
The error terms in (\ref{t-dep-Ham-Jac-eq}) and (\ref{karlgustav})
lead to contributions of order $\bigO(h^\infty)$
in $\Symbb{\RR\times\RR^d\times\TT^d}$, too,
because of 
the elementary estimates
$$
e^{-\Im \psi/h}\,(\Im \psi)^N\,\klg\, N!\,h^N\,, \qquad N\in\NN\,.
$$
\qed


\section{The construction of $E''(0)^{-1}$}
\label{sec-inv}

\noindent Proposition~\ref{prop-56} supplies the piece of the
parametrix for $E''(0)$ corresponding to the set $K_0\times(2\pi\ZZ)^d$
as we shall see in the proof of Proposition~\ref{le-karin} below.
Outside $K_0\times(2\pi\ZZ)^d$ the Weyl symbol $a^\vp_W$ is elliptic
and we can use standard methods of pseudodifferential calculus
to construct the missing piece of the parametrix. Here are
the details:

If 
$a^\vp_W$ were elliptic,
we had a well-known asymptotic expansion, 
$\breve{q}(x,\xi)\asymp\sum_{\nu=0}^\infty h^\nu\,q_\nu(x,\xi)$, 
for the Weyl symbol of 
the inverse operator $\big(\Op(a^\vp_W)\big)^{-1}$. 
We can of course always write down this expansion formally
and determine $q_\nu$ at all points $(x,\xi)$
where $a^\vp_W$ is invertible. 
We proceed in this way and pick some 
$\widetilde{\chi}\in C_0^\infty(\RR^{d},[0,1])$ 
such that
$\widetilde{\chi}\equiv1$ in a small neighbourhood
of $K_0$ and $\supp(\widetilde{\chi})\subset \{\chi=1\}$.
We recall that $\chi$ has been introduced in (\ref{eq-ab-T_0}).
Then we define the symbol 
$\tilde{q}\in \Symbb{\RR^{d}\times\RR_d}$ 
to be a Borel
resummation of
$$
\sum_{\nu=0}^\infty h^\nu\,q_\nu(x,\xi)\,(1-\widetilde{\chi}(x))\,,
$$
where now each term is a well-defined function.
We let $\#_h^W$ denote the composition of Weyl symbols, i.e.
\begin{eqnarray}
\lefteqn{
(a\#_h^W c)(x,\xi)\:=\:\nonumber
e^{ih[D_\eta D_x-D_y D_\xi]}\, a(y,\eta)\,c(x,\xi)
\,\big|_{y=x,\,\eta=\xi}
}
\\
&=&\label{def-sharp}
\int e^{i\SPn{\xi-\eta}{v-x}/h+i\SPn{\xi-\zeta}{x-u}/h}
\,a(\tfrac{x+u}{2},\eta)\,c(\tfrac{x+v}{2},\zeta)\:
\frac{dudvd\eta d\zeta}{(2\pi h)^{2d}}\,,
\end{eqnarray}
for $a,c\in\Symbb{\RR^d\times\RR_d}$; see, e.g., 
\cite[\textsection2.7]{Mar}.
Then we get,
by definition of $q_\nu$, $\nu\in\NN_0$,
\begin{eqnarray}
a^\vp_W\#_h^W\tilde{q}(x,\xi)&\asymp&\nonumber
\sum_{\alpha,\beta\in\NN_0^d}
\frac{h^{|\alpha+\beta|}(-1)^{|\alpha|}}{(2i)^{|\alpha+\beta|}\alpha!\beta!}
\,\big(\partial_x^\alpha\partial_\xi^\beta a^\vp_W(x,\xi)\big)
\,\big(\partial_\xi^\alpha\partial_x^\beta \tilde{q}(x,\xi)\big)
\\
&=&\label{egon}
1-\widetilde{\chi}(x)\,+\,\sum_{\nu=0}^\infty h^\nu \,
\breve{r}_\nu(x,\xi)\,,
\end{eqnarray}
where each error term $\breve{r}_\nu$, $\nu\in\NN_0$, contains some
partial derivative of $\widetilde{\chi}$, which shows
that $\supp \breve{r}_\nu\subset \{\chi=1\}$, for all $\nu\in\NN_0$. 
We now set $q:=\tilde{q}\#_h^W(1-\chi)$ and infer from
(\ref{egon}) that
$$
a^\vp_W\#_h^W q\,\asymp\,1-\chi
$$
in $\Symbb{\RR^d\times\RR_d}$. 
We observe that the symbol $q$ is again
$2\pi$-periodic in $\xi$. In fact, if we suppose that $a$ and $c$
in (\ref{def-sharp}) are $2\pi$-periodic in 
the momentum variables, then we see that
$(a\#_h^W c)(x,\xi+2\pi\,\ell)=(a\#_h^W c)(x,\xi)$, $\ell\in\ZZ^d$,
by a translation of the variables $\eta$ and $\zeta$
in the oscillatory integral (\ref{def-sharp}).
We let $K_q(x,y)$ denote the distribution kernel  
of $\Op(q)$.
The periodicity of $q$ implies that 
$$
K_q(x,y)\,=\,h^d\sum_{z\in\ZZ^d_h}\widehat{q}(\xyhalf,z)\,\delta_z(y-x)\,,
\qquad x,y\in\ZZ^d_h\,,
$$
where the Fourier coefficients, $\widehat{q}$,
are defined as in (\ref{Fourier-coeff}).

Recalling that $b$ vanishes, for large $t$, 
we set  
\begin{eqnarray}
I(x,y)&:=&\int\limits_0^{\infty}\int\limits_{\TT^d} 
e^{i\psi(t,x,\eta)/h-i\SPn{y}{\eta}/h}\,
b(t,x,\eta\,;h)\,\frac{d\eta\,dt}{(2\pi)^{d}\,h}
\,,\label{ansatz-inv}
\end{eqnarray}
for all $x,y\in\ZZ^d_h$, 
and try 
the operator $\mathcal{P}$ given by 
\begin{equation}\label{ansatz-inv1}
\mathcal{P}(x,y)\,:=\,I(x,y)\,+\,h^d\,\widehat{q}(\xyhalf,y-x)\,,\qquad
x,y\in\ZZ^d_h\,,
\end{equation}
as an approximate inverse for
$e^{\vp/h}E''(0) e^{-\vp/h}$.

\begin{proposition}\label{le-karin}
It holds $\mathcal{P}\in\LO(\ell^2(\ZZ^d_h))$ and
$$
e^{\vp/h}E''(0) e^{-\vp/h}\,\mathcal{P}\,=\,\id\,+\,\mathcal{R}\,,
$$
where $\mathcal{R}\in\LO(\ell^2(\ZZ^d_h))$ such that 
$\|\mathcal{R}\|=\bigO_N(h^N)$, for every
$N\in\NN$. In particular,
$$
\big(E''(0)\big)^{-1}\,=\,e^{-\vp/h}\,\mathcal{P}\,\Big(\sum_{m=0}^\infty 
(-1)^m\mathcal{R}^m
\Big)\,e^{\vp/h}\,.
$$
\end{proposition}

\proof
Applying successively
Lemma~\ref{le-ernst1}, Proposition~\ref{prop-56},
and the fact that $b$ vanishes, for large $t$,
we get, for all $x,y\in\ZZ^d_h$,
\begin{eqnarray}
\lefteqn{\nonumber
\big(e^{\vp/h}E''(0) e^{-\vp/h}\circ I\big)(x,y)
}
\\
&=&\nonumber
\int\limits_{\TT^d}\int\limits_0^\infty 
\big(\, e^{\vp/h}E''(0) e^{-\vp/h}\,e^{i\psi(t,\cdot,\eta)/h}
\,
b(t,\cdot,\eta\,;h)\big)(x)\,e^{-i\SPn{y}{\eta}/h}\,
\frac{dt\,d\eta}{(2\pi )^{d}\,h}
\\
&=&\nonumber
\int\limits_{\TT^d}\int\limits_0^\infty 
\big(\,\Op(a_W^\vp)\, e^{i\psi(t,\cdot,\eta)/h}
\,
b(t,\cdot,\eta\,;h)\big)(x)\,e^{-i\SPn{y}{\eta}/h}\,
\frac{dt\,d\eta}{(2\pi )^{d}\,h}
\\
&=&\nonumber
-\int\limits_{\TT^d}\int\limits_0^\infty \partial_t\big(
e^{i\psi(t,x,\eta)/h-i\SPn{y}{\eta}/h}\,
b(t,x,\eta\,;h)
\big)\,\frac{dt\,d\eta}{(2\pi )^{d}}\,+\,\mathcal{R}_1(x,y)
\\
&=&\label{susi0}
\int\limits_{\TT^d} e^{i\SPn{x-y}{\eta}/h}\,\chi(x)\,
\frac{d\eta}{(2\pi )^{d}}\,+\,\mathcal{R}_1(x,y)
\;=\;
\chi(x)\,\delta_{xy}\,+\,\mathcal{R}_1(x,y)
\,,
\end{eqnarray}
where, for $N\in\NN$ and $x,y\in\ZZ^d_h$,
\begin{equation}
\mathcal{R}_1(x,y)\,:=\,h^N\label{kuno1}
\int\limits_0^\infty\int\limits_{\TT^d}
e^{i\widetilde{\psi}(t,x,\eta)/h-i\SPn{y}{\eta}/h}
\,\check{r}_N(t,x,\eta\,;h)
\,\frac{d\eta \,dt}{(2\pi )^{d}h}\,.
\end{equation}
Since the integrants in (\ref{kuno1}) vanish, if $t$ 
is sufficiently large and $\check{r}_N(t,\cdot,\eta\,;h)$
is supported in $\mathscr{K}+\{|x|\klg R\}$, we have 
$$
\max\big\{\,
\mathcal{R}_1(x,y)\,:\:\dist(x,{\mathscr{K}})\klg R\,,\;y\in\ZZ^d_h
\,\big\}\,=\,\bigO_N(h^N)\,,\quad N\in\NN\,.
$$ 
Moreover, since $|\widetilde{\psi}_\eta'(t,x,\eta)-x|$
is uniformly bounded on compact sets, we have 
$|\widetilde{\psi}_\eta'(t,x,\eta)-y|\grg|x-y|/2$ provided $|x-y|$
is sufficiently large.
Integrating by parts in (\ref{kuno1}) by means of the operator
$$
\frac{h}{i}
\frac{\SPb{\overline{\widetilde{\psi}_\eta'(t,x,\eta)-y}}{\nabla_\eta}}{
\big|\,\widetilde{\psi}_\eta'(t,x,\eta)-y\,\big|^2}
$$
we therefore find that
\begin{equation}\label{susi00}
|\mathcal{R}_1(x,y)|\,=\, 
\bigO\big(h^N\langle\,x-y\,\rangle^{-N}\big)\,,\qquad N\in\NN\,.
\end{equation} 
If we form an operator, $\mathcal{R}_1$, whose matrix element
at $(x,y)\in\ZZ^d_h\times\ZZ^d_h$ is given by 
$\mathcal{R}_1(x,y)$, we thus obtain an element
in $\LO(\ell^1(\ZZ^d_h))\cap\LO(\ell^\infty(\ZZ^d_h))$. 
By interpolation we then also have $\mathcal{R}_1\in\LO(\ell^2(\ZZ^d_h))$.

Next, we notice that
$$
\Op(1-\chi+r)\,=
\Op(a^\vp_W)\circ \Op(q)\,=\,\mathrm{Op}_h(a^\vp_W\#_h q)\,,
$$
where $r:=a^\vp_W\#_h^Wq-(1-\chi)$ is of order $\bigO(h^\infty)$
in $\Symbb{\RR^d\times\RR_d}$,
$$
a^\vp_W\#_h q(x,z,\eta)\,:=\,
\int e^{i\SPn{\xi-\eta}{x-y}/h}
a^\vp_W(\xyhalf,\xi)\,q(\tfrac{y+z}{2},\eta)\:\frac{dyd\xi}{(2\pi h)^d}\,,
$$
and $\mathrm{Op}_h(a^\vp_W\#_h q)$ denotes the 
$h$-pseudodifferential operator defined by the symbol
$a^\vp_W\#_h q\in\Symbb{\RR^{2d}\times\RR_d}$.
The symbols
$r$ and $a^\vp_W\#_h q$
are both $2\pi$-periodic in the momentum variable.
Moreover, 
$$
\big(e^{\vp/h}E''(0)e^{-\vp/h}\,\widehat{q}(\tfrac{\cdot\,+y}{2},y-\cdot)
\big)(x)\,=\,(a^\vp_W\#_h q)^{\wedge}(x,y,y-x)\,,
$$
for $x,y\in\ZZ^d_h$, where 
the Fourier coefficients,
$(a^\vp_W\#_h q)^{\wedge}(x,y,z)$, $z\in\ZZ^d_h$, are given
by a formula similar to (\ref{Fourier-coeff}).
Since any pseudodifferential operator 
determines uniquely its distribution kernel, we obtain
\begin{eqnarray}
\lefteqn{\nonumber
\big(e^{\vp/h}E''(0)e^{-\vp/h}\,h^d\,
\widehat{q}(\tfrac{\cdot\,+y}{2},y-\cdot)
\big)(x)
}
\\
&=&\nonumber
\int\limits_{\TT^d}e^{i\SPn{\eta}{x-y}/h}
\big(1-\chi(\tfrac{x+y}{2})+r(\tfrac{x+y}{2},\eta)\big)
\:\frac{d\eta}{(2\pi )^d}
\\
&=&\big(1\,-\,\chi(x)\big)\,\delta_{xy}
\,+\,h^d\,\widehat{r}(\tfrac{x+y}{2},y-x)
\,.
\end{eqnarray}
Setting 
\begin{equation}\label{susi2}
(\mathcal{R}_2\,f)(x)\,:=\,h^d
\sum_{y\in\ZZ^d}\widehat{r}(\tfrac{x+y}{2},y-x)\,f(y)\,,
\qquad x\in\ZZ^d_h\,,\;f\in\ell^p(\ZZ^d_h)\,,
\end{equation}
where $p\in\{1,\infty\}$,
we again have
$\mathcal{R}_2\in\LO(\ell^1(\ZZ^d_h))\cap\LO(\ell^\infty(\ZZ^d_h))$,
because
$$
\big|\,\widehat{r}(\xyhalf,y-x)\,\big|
\,=\,\bigO\big(h^N\,\SL x-y\SR^{-N}\big)\,,\qquad
N\in\NN\,,
$$
which is easily verified.
By interpolation we see that
$\|\,\mathcal{R}_2\,\|_{\LO(\ell^2(\ZZ^d_h))}=\bigO(h^\infty)$.
In view of (\ref{susi0})-(\ref{susi2})
the proof is complete
with $\mathcal{R}:=\mathcal{R}_1+\mathcal{R}_2$. 
\qed


\section{Calculation of the leading asymptotics}
\label{sec-7}

In this section we calculate the leading asymptotics of
$\mathcal{P}(x^\star,y^\star)$, where
$x^\star$ and $y^\star$ are our two distinguished points
which satisfy Hypothesis~\ref{hyp-geo}.
Integrating by parts repeatedly in the integral
defining the Fourier coefficient
$\widehat{q}(\tfrac{x^\star+y^\star}{2},y^\star-x^\star)$
we see that
\begin{equation}\label{PundI}
\mathcal{P}(x^\star,y^\star)\,=\,I(x^\star,y^\star)\,+\,\bigO(h^\infty)
\end{equation}
and it remains to determine the 
leading asymtotics of $I(x^\star,y^\star)$.
To this end we recall that $H=T-U$ and 
that $X_H$ denotes the Hamiltonian vector field
of $H$.
$\pi:T^*\RR^d\to\RR^d$ is the canonical projection
in the following.
We let $\rho:\RR\rightarrow T^*\RR^d$
denote the unique integral curve of $X_H$ running 
in $\Figx=\{H=0\}$
such that $\gamma:=\pi\rho\!\!\upharpoonright_{[0,\tau]}$
is a minimizing geodesic from
$\gamma(0)=y^\star$ to $\gamma(\tau)=x^\star$, for some suitable $\tau>0$.
Recalling the notation introduced in the paragraph 
preceeding Theorem~\ref{thm-main} we have
\begin{eqnarray}
v_{y^\star}&=&\dot{\gamma}(0)\,=\,-\dot{k}(\tau,x^\star,0)\,,\label{gamma0}
\\
v_{x^\star}&=&\dot{\gamma}(\tau)\,=\,\dot{Q}(\tau,y^\star,0)\,.
\label{gammatau}
\end{eqnarray}
Furthermore, we let 
$X(t),P(t)\in \LO(\RR^d)$, $t\in\RR$, denote
the solution of the initial value problem
\begin{equation}\label{eq-Jac-H}
\frac{d}{dt}\,{X(t)\choose P(t)}\,=\,
\matr{H_{px}''(\rho(t))&H_{pp}''(\rho(t))
\\-H_{xx}''(\rho(t))&-H_{xp}''(\rho(t))}
{X(t)\choose P(t)}\,,\qquad
{X(0)\choose P(0)}\,=\,{0\choose\id}\,.
\end{equation}

\begin{proposition}\label{prop-asymp1}
The following formula holds, as $h\searrow0$:
\begin{eqnarray}
\nonumber
I(x^\star,y^\star)&=&
\int\limits_0^\infty\int\limits_{\TT^d} 
e^{i\psi(t,x^\star,\eta)/h-i\SPn{y^\star}{\eta}/h}\,
b(t,x^\star,\eta\,;h)\,\frac{d\eta\,dt}{(2\pi)^{d}\,h}
\\
&=&\label{karin}
\Big(\frac{h}{2\pi}\Big)^{\frac{d-1}{2}}
\,
\Det{0&-{}^tv_{y^\star}\\v_{x^\star}&X(\tau)}^{\frac{1}{2}}
\,+\,\bigO(h^{\frac{d+1}{2}})
\end{eqnarray}
\end{proposition}

\proof
We apply the method of complex stationary phase
\cite[\textsection2]{MeSj1} 
(with respect to $t$ and $\eta$) to the integral
(\ref{karin}). 
The critical points of the phase are given
by
\begin{eqnarray}
0&=&\partial_t\psi(t,x^\star,\eta)\,,
\label{eq-cp1}
\\
\label{eq-cp2}
0&=&\nabla_\eta\psi(t,x^\star,\eta)\,-\,y^\star\,.
\end{eqnarray}
Of course, a critical point gives a non-vanishing contribution
to the asymptotics of (\ref{karin}) only if it is
contained in the set $\{\Im\psi=0\}$.
By Corollary~\ref{le-psi}
we know that we must have $t=0$
or $(t,x^\star,\eta)\in\mathscr{D}$
in this case. At $t=0$ we have, however,
$\nabla_\eta\psi(0,x^\star,\eta)
=x^\star\not=y^\star$. So, (\ref{lotta99})
implies
that $(\tau,0)$ is the only critical point of the phase in (\ref{karin})
where $\Im\psi$ vanishes.
Again by (\ref{lotta99}),
$$
\partial_t^2\psi(\tau,x^\star,0)\,=\,0\,.
$$
To study the remaining second derivatives of the phase we 
differentiate the identity
$Q(t,k(t,x^\star,\eta),\eta)=x^\star$
with resect to $t$ and $\eta$ and obtain
\begin{eqnarray*}
Q_y'(t,k(t,x^\star,\eta),\eta)\,\dot{k}(t,x^\star,\eta)
\,+\,Q_{\overline{y}}'(t,k(t,x^\star,\eta),\eta)\,
\overline{\dot{k}(t,x^\star,\eta)}
&=&
-\dot{Q}(t,k(t,x^\star,\eta),\eta)\,,
\\
Q_y'(t,k(t,x^\star,\eta),\eta)\,k_\eta'(t,x^\star,\eta)
\,+\,
Q_{\overline{y}}'(t,k(t,x^\star,\eta),\eta)\,
\overline{k_\eta'(t,x^\star,\eta)}
&=&
-Q_\eta'(t,k(t,x^\star,\eta),\eta)\,.
\end{eqnarray*}
Evaluated at $(t,\eta)=(\tau,0)$ this yields
\begin{eqnarray}
Q_y'(\tau,y^\star,0)\,v_{y^\star}
&=&
v_{x^\star}\,,\label{christel1}
\\
Q_y'(\tau,y^\star,0)\,
\tfrac{1}{i}\psi_{\eta\eta}''(\tau,x,0)
&=&
i\,Q_\eta'(\tau,y^\star,0)\label{christel2}
\,,
\end{eqnarray}
if we take (\ref{t-dep-Ham-Jac-eq-alpha-3}), (\ref{Qovy}), (\ref{gamma0}),
and (\ref{gammatau}) into account.
We consider  
$iQ_\eta'(\tau,y^\star,0)$ more closely
in the following.
To this end, we recall that the derivative of the flow
of $\widehat{X}_{a^\vp}$
along $(\gamma(t),0)$, $\kappa_t'(y^\star,0)$, $t\in[0,\tau]$,
is a solution of the Jacobi equation
\begin{equation}
\frac{d}{dt}\,\kappa_t'(y^\star,0)
\,=\,\FM_a(\gamma(t),0)\,\kappa_t'(y^\star,0)\,,\label{eq-Jac}
\end{equation}
where $\FM_a(\gamma(t),0)$ is given by (\ref{eq-fund-matr})
with $x(t)=\gamma(t)$ and $e=0$.
Since we want to determine 
$iQ_\eta'(\tau,y^\star,0)=i\,\pi_x (\kappa_\tau)_\eta'(y^\star,0)$, 
we consider the solution of (\ref{eq-Jac})
with initial condition 
$i(\kappa_0)_\eta'(y^\star,0)\,=\,(0,i\id)$, which
we denote by $(\widetilde{X}(t),\widetilde{P}(t))$.
It follows that $\Im\widetilde{X}\equiv0$ and 
$\Re\widetilde{P}\equiv0$ and that 
$(\Re\widetilde{X},\Im\widetilde{P})$ is a solution
of the Jacobi equation
\begin{equation}\label{eq-Jacmatr-wtH}
\frac{d}{dt}{\Re\widetilde{X}(t)\choose\Im\widetilde{P}(t)}
=
\matr{\bbB(t)&
\bbA(t)\\
0&-{}^t\bbB(t)}{\Re\widetilde{X}(t)\choose\Im\widetilde{P}(t)}.
\end{equation}
The matrix in (\ref{eq-Jacmatr-wtH}) is the fundamental matrix at 
$(x,p)=(\gamma(t),0)$
of the Hamiltonian
$H_\vp(x,p):=H(x,p+\vp'(x))=(H\circ\Xi)(x,p)$,
where $\Xi(x,p):=(x,p+\vp'(x))$, $(x,p)\in T^*\RR^d$, 
defines a symplectomorphism.
We further recall that $H_{\vp,xx}''(x,0)=0$,
for $x\in K_0$, by construction of $\vp$.
Using these remarks we check that $\Re\widetilde{X}=X$,
where $(X,P)$ is a solution of (\ref{eq-Jac-H}).
In fact, the Jacobi equation (\ref{eq-Jacmatr-wtH})
is solved by $(\Xi')^{-1}(\gamma)(X,P)=(X,-\vp''(\gamma)X+P)$.

In summary, we see that $iQ_\eta'(,\tau,y^\star,0)=X(\tau)$.
In particular, $iQ_\eta'(,\tau,y^\star,0)$ is real and so is 
$\tfrac{1}{i}\psi_{\eta\eta}''(\tau,x,0)$ by (\ref{christel2}),
because $Q_y'(\tau,y^\star,0)$ is a real
invertible matrix.

By Lemma~\ref{le-b_0} we know that
$b_0(\tau,x^\star,0)^{-1}=|Q_y'(\tau,y^\star,0)|^{1/2}$.
Using also (\ref{christel1}) and (\ref{christel2})
we thus get
$$
\frac{1}{b_0(\tau,x^\star,0)}
\,
\Det{0&\tfrac{1}{i}\partial_t\psi_\eta'\\
\tfrac{1}{i}\partial_t\nabla_\eta\psi&\tfrac{1}{i}\psi_{\eta\eta}''
}^{\frac{1}{2}}
\,=\,
\Det{1&0\\
0&\tilde{\kappa}_\tau'(y^\star)
}^{\frac{1}{2}}
\Det{0&-{}^t\dot{\gamma}(0)\\
\dot{\gamma}(0)&\tfrac{1}{i}\psi_{\eta\eta}''
}^{\frac{1}{2}}
\,=\,
\Det{0&-{}^tv_{y^\star}\\ v_{x^\star}&
X(\tau)}^{\frac{1}{2}}\!\!,
$$
where all derivatives of $\psi$ are 
evaluated at $(\tau,x^\star,0)$.
From Proposition~\ref{prop-ina} below it
follows in particular, that
the last determinant in the previous
equation is non-zero.
Therefore, the left hand side is non-zero, too.
(To show that $\psi_{\eta\eta}''(\tau,x^\star,0)$
is non-singular we could also appeal to the strict positivity
of $C_\tau$ at $(x^\star,0,y^\star,0)$.)
We may thus split the integral (\ref{karin}) into
pieces by means of suitable cut-off functions
and
apply the stationary (complex) phase formula
\cite[\textsection2]{MeSj1}
to a sufficiently small piece of 
(\ref{karin}) near the critical point $(\tau,0)$.

Finally, we remark that, for all $\delta,T>0$,
$\inf\{\Im\psi(t,x^\star,\eta):\,
t\in[\delta,T],\,\eta\in\TT^d,\,|\eta|\grg\delta\}>0$
and the corresponding piece of the integral (\ref{karin}) gives
a contribution of order $\bigO(h^\infty)$.
At $t=0$ we have
$\nabla_\eta(\psi(0,x^\star,\eta)-\SPn{y^\star}{\eta})
=x^\star-y^\star\not=0$, $\eta\in\TT^d$, and, for
$\eta=0$ and some $\delta',\delta''>0$,
$$
\big|\nabla_\eta|_{\eta=0}(\psi(t,x^\star,\eta)-\SPn{y^\star}{\eta})\big|
=|k(t,x^\star,0)-y^\star|\grg\delta'\,,\quad
t\in[\,\delta,\tau-\delta''\,]\cup[\,\tau+\delta'',T\,]\,, 
$$
Therefore, the remaining pieces of the integral (\ref{karin})
can be treated by means of integration by parts 
in $\eta$ and give contributions
of order $\bigO(h^\infty)$, too.
\qed

\bigskip

\noindent It is of course simple to check that
(\ref{karin}) agrees with the already known formula
(\ref{eq-OZ-ti1}) in the translation invariant case.
In the next proposition we 
obtain a generalization of (\ref{eq-OZ-ti2}).

To prove it we introduce special orthonormal bases
with respect to the Finsler structure \cite[pp. 31]{BCS}
in the following.

We denote the quadratic form
associated with $G(z,v_z)$ by $g_{z}$, for $z=x^\star,y^\star$,
and 
pick a basis,
$\langle\,b_1,\dots,b_d\,\rangle$, 
which is orthonormal with respect to $g_{y^\star}$
such that $b_d=F(y^\star,v_{y^\star})^{-1}v_{y^\star}$.
Similarly, we pick another basis,
$\langle\,c_1,\dots,c_d\,\rangle$, 
which is orthonormal with respect to $g_{x^\star}$
and where $c_d=F(x^\star,v_{x^\star})^{-1}v_{x^\star}$.
We denote the dual basis vectors by $b_i^*$, $c_i^*$,
$1\klg i\klg d$, so that
$b_i^*(b_j)=\sum_{k=1}^d(b_i^*)_k\,b_j^k=\delta_{ij}$, etc.

For $1\klg i\klg d-1$, we further introduce a Jacobi field,
$J_i$, which is by definition the solution
of the Jacobi equation
$$
D_{\mathscr{T}}D_{\mathscr{T}}J_i\,=\,R(J_i,{\mathscr{T}}){\mathscr{T}}\,,
\qquad J_i(0)\,=\,b_i\,.
$$
Here ${\mathscr{T}}$ is the velocity vector field of the unit speed geodesic
from $y^\star$ to $x^\star$, $D_{\mathscr{T}}$ 
denotes covariant differentiation
in the direction ${\mathscr{T}}$ with reference vector 
${\mathscr{T}}$, and
$$
R({\mathscr{T}},J_i){\mathscr{T}}\,=\,
\sum_{k,\ell,m,n=1}^d({\mathscr{T}}^kR_k{}^{\ell}{}_{mn}{\mathscr{T}}^n)
J_i^m\,\partial_{x^\ell}\,,
$$
where $R_k{}^{\ell}{}_{mn}$ are the components of the
$hh$-curvature tensor of the Chern connection defined by $F$
\cite{BCS}.
Since we only need a well-known Taylor expansion of 
mutual scalar products of the $J_i$ and the formula
$$
(\exp_{y})'(r\,b_d)\,b_i\,=\,\frac{1}{r}\:J_i(r)\,,
$$
where
$$
r\,:=\,\Fdist(x^\star,y^\star)\,,
$$
we do not explain these notions further.

\begin{proposition}\label{prop-ina}
The following identity holds:
\begin{eqnarray*}
\Det{0&-{}^tv_{y^\star}\\v_{x^\star}&X(\tau)}^{\frac{1}{2}}
&=&
\frac{1}{
\sqrt{\SPn{p_{x^\star}}{v_{x^\star}}\SPn{p_{y^\star}}{v_{y^\star}}}}
\cdot
\frac{\Det{G(x^\star,v_{x^\star})\,G({y^\star},v_{y^\star})}^{1/4}}{
\Det{\big(\,g_{x^\star}(J_i(r),J_j(r))
\,\big)_{i,j=1}^{d-1}}^{1/4}}\,,
\end{eqnarray*}
where
$$
\Det{\big(g_{x^\star}(J_i(r),J_j(r))
\big)_{i,j=1}^{d-1}}^{1/4}
\,=\,\Fdist(x^\star,y^\star)^{\frac{d-1}{2}}\,+\,
\bigO\big(\Fdist(x^\star,y^\star)^{\frac{d}{2}}\big)\,,
$$
as $x^\star\rightarrow y^\star$.
\end{proposition}

\proof
To simplify our notation we denote the distinguished points
$x^\star$ and $y^\star$ simply by $x$ and $y$
in this proof.
We let $B$ and $C$ denote the matrices
whose $i$-th row is $b_i^*$ and $c_i^*$, respectively.
From $c_d^*=F'_v(x,v_x)=p_x$ we infer that
$C\,v_x={}^t(0\dots0\,F(x,v_x))=\SPn{p_x}{v_x}{}^t(0\dots0\,1)$
and similarly ${}^tv_y\,{}^tB=\SPn{p_y}{v_y}(0\dots0\,1)$.
We thus get 
$$
\Det{1&0\\0&C}\Det{0&-{}^tv_y\\v_x&X(\tau)}\Det{1&0\\0&{}^tB}
=\SPn{p_x}{v_x}\SPn{p_y}{v_y}
\Det{0&0\;\,\cdots\;\,0\;\,\,-\!\!1\\\begin{array}{c}0\\\vdots\\0\\1
\end{array}&C\,X(\tau)\,{}^tB}\,.
$$
Here the determinant on the right side clearly equals
$$
\det\Big(\big(c_i^*\,X(\tau)\,{}^tb_j^*\big)_{1\klg i,j\klg d-1}\Big)\,.
$$
The vectors ${}^tb_j^*$, $1\klg j\klg d-1$, are orthogonal
to $b_d\in\Indx_{y}$ with respect to the
Euclidean scalar product and, hence, span the tangent space
of the figuratrix at the momentum which is conjugate
to $b_d$, i.e. at $p_y$. It therefore suffices to consider
$$
X(\tau)\!\!\upharpoonright_{T_{p_y}\Figx_{y}}
\,=\,(x_\tau\!\!\upharpoonright_{\Figx_{y}})'(p_y)\,,
\qquad
\textrm{where}\qquad
x_\tau(p)\,:=\,\pi\exp(\tau X_H)(y,p)\,.
$$ 
Let $\mathcal{L}:\Figx_{z}\rightarrow\Indx_{z}$ denote
Legendre transformation, so that $\mathcal{L}$
maps $p\in\Figx_{z}$ to its conjugate direction $v\in\Indx_{z}$,
for $z\in\RR^d$. It holds $\mathcal{L}(p_x)=c_d\parallel v_x$
and $\mathcal{L}(p_y)=b_d\parallel v_y$.
Moreover, we have \cite[p. 410]{BCS}
\begin{equation}\label{kunigunde0}
\mathcal{L}'(p_y)\,=\,
G(y,\mathcal{L}(p_y))^{-1}\!\!\upharpoonright_{T_{p_y}\Figx_{y}}
\,=\,G(y,v_y)^{-1}\!\!\upharpoonright_{T_{p_y}\Figx_{y}}.
\end{equation}
We recall that $G$ is homogenous of degree zero in $v$.
Besides we know that
$$
\big((\exp_{y})'(\,r\mathcal{L}(p_y)\big)\,\mathcal{L}(p_y)\,=\,
\frac{d}{dt}\Big|_{t=r}\exp_y(\,t\mathcal{L}(p_y))\;\parallel\;v_x\,,
$$
where $r=\Fdist(x,y)$. The image of $\mathcal{L}(p_y)$ under 
$(\exp_{y})'(\,r\mathcal{L}(p_y))$ is therefore
$g_{x}$-orthogonal to $c_i$, $1\klg i\klg d-1$.
(This 
is precisely the statement of the Gau{\ss} lemma
in Finsler geometry \cite[\textsection6.1.A]{BCS},
as the vectors $c_i$, $1\klg i\klg d-1$, span
$T_{\mathcal{L}(p_x)}\Indx_{x}$.)
In particular, 
\begin{equation}\label{kunigunde1}
c_i^*\,
(\exp_{y})'(r\mathcal{L}(p_y))\,\Big[
\mathcal{L}(p_y)\otimes\big(\Fdist(x_\tau(\cdot),y)\big)'\Big]\,=\,0\,.
\end{equation}
We shall use the following identity,
\begin{equation}\label{kunigunde2}
x_\tau(p)\,=\,\exp_{y}\big(\,\Fdist(x_\tau(p),y)\,\mathcal{L}(p)\big)\,,
\qquad p\in\Figx_{y}\,.
\end{equation}
Differentiating (\ref{kunigunde2}) and using (\ref{kunigunde0}),
(\ref{kunigunde1}),
and 
$c_i^*(v_x)=0$, 
$1\klg i\klg d-1$, we obtain
\begin{eqnarray*}
c_i^*\,(x_\tau\!\!\upharpoonright_{\Figx_{y}})'(p_y)\,{}^tb_j^*
&=&
r\,c_i^*\,\big((\exp_{y})'(r\,\mathcal{L}(p_y)\big)\,
G(y,v_y)^{-1}\,{}^tb_j^*
\\
&=&
r\,c_i^*\,\big((\exp_{y})'(r\,b_d)\big)\,
b_j\,,
\end{eqnarray*}
for $1\klg i,j\klg d-1$. Let $M$ be the matrix with
entries $c_i^*\,\big((\exp_{y})'(r\,b_d)\big)\,b_j$,
$1\klg i,j\klg d-1$. It holds
$\sum_{i=1}^{d-1}{^t}c_i^*\,c_i^*
=G(x,v_x)\!\!\upharpoonright_{T_{\mathcal{L}(p_x)}\Indx_{x}}$,
whence we get
\begin{eqnarray*}
\big({}^tM\,M\big)_{ij}
&=&g_{x}\big((\exp_{y})'(r\,b_d)\,b_i\,,
\,(\exp_{y})'(r\,b_d)\,b_j\big)
\\
&=&\frac{1}{r^2}\,g_{x}\big(J_i(r)\,,\,J_j(r)\big)
\\
&=&\delta_{ij}\,-\,\frac{r^2}{3}\:g_{y}
\big(R(b_i,b_d)b_d\,,\,b_j\big)\,+\,\bigO(r^3)
\,.
\end{eqnarray*}
Here the well-known Taylor expansion in the last line
is derived using the formulas of \cite[\textsection5.5]{BCS}.
We conclude the proof of the proposition by noticing
that $(\det B)^2=\det(^tB\,B)=\det G(y,v_y)$
and $\det C=(\det G(x,v_x))^{1/2}$.
\qed


\begin{appendix}

\section{Applicability of a result by Bach and M{\o}ller}\label{app-BaMo}

In this appendix we show that we can apply the results
of \cite{BaMo2} in our situation
and thus obtain a proof of Theorem~\ref{thm-MaBaMo}. 
The starting point
will be a Helffer-Sj\"ostrand formula established
in \cite{Ma}. It involves the inverse of
a certain 
operator which is introduced in the following.
First, we recall that we have fixed some pure, tempered
Gibbs measure, $\mu_{\beta,h}$, and have set
$\HR^0=L^2(\mu_{\beta,h})$. We further set 
$\HR^1:=L^2(\mu_{\beta,h})\widehat{\otimes}\ell^2(\ZZ^d_h)$,
where $\widehat{\otimes}$ deontes the (completed)
tensor product of Hilbert spaces,
and 
\begin{eqnarray*}
\Omega^0\,:=\,
\big\{\,
f:\RR^{\ZZ^d_h}\to\CC\,:\;\exists\;n\in\NN\,,\,
x_1,\dots,x_n\in\ZZ^d_h\,,\,f_n\in C_b^\infty(\RR^n)
\\
\forall\;\sigma\in\RR^{\ZZ^d_h}\;:\;\;
f(\sigma)\,=\,f_n(\sigma_{x_1},\dots,\sigma_{x_n})
\,\big\}\,.
\end{eqnarray*}
Then $\Omega^0$ is a dense subspace
of $\HR^0$ and the algebraic tensor product
$\Omega^1:=\Omega^0\otimes\langle\,e_x:\,x\in\ZZ^d_h\,\rangle$
is a dense subspace of $\HR^1$. 
Here $\langle\,e_x:\,x\in\ZZ^d_h\,\rangle$ is the linear hull
of the canonical orthonormal basis of $\ell^2(\ZZ^d_h)$.
We next define two operators, $\mathscr{L}^{(i)}_{\beta,h}$,
$i=1,2$,
on $\HR^i$ with domain 
$\mathrm{dom}(\mathscr{L}^{(i)}_{\beta,h}):=\Omega^i$
by
$$
\mathscr{L}^{(0)}_{\beta,h}\,f(\sigma)\,:=\,
\sum_{x\in\ZZ^d_h}\big(-\beta^{-1}\partial_{\sigma_x}^2f(\sigma)
\,+\,\partial_{\sigma_x}E_{\{x\}}(\sigma_{x}|\sigma_{\{x\}^c})
\,\partial_{\sigma_x}f(\sigma)\big)\,,
$$
for $\sigma\in\RR^{\ZZ^d_h}$ and $f\in\Omega^0$, and
$$
\mathscr{L}^{(1)}_{\beta,h}\,:=\,\mathscr{L}^{(0)}_{\beta,h}\otimes\id
\,+\,E''\,.
$$
We recall that the matrix element of $E''$
at $\sigma\in\RR^{\ZZ^d_h}$ is defined by (\ref{def-E''}).
Then $\mathscr{L}^{(i)}_{\beta,h}$
is essentially selfadjoint on $\HR^i$,
for $i=0,1$ \cite{Ma}.
We denote its unique selfadjoint extension by 
$L^{(i)}_{\beta,h}$.
We recall a result from \cite{Ma} which permits
to locate spectral gaps at the bottom of the
spectrum of $L^{(i)}_{\beta,h}$ at large inverse
temperatures. 
To this end, we set
\begin{eqnarray*}
m_\star&:=&\inf_{x\in\RR^d}D_{\theta\theta}''(x,0)
\,-\,J\sum_{|\ell|\klg R}
\sup_{x\in\RR^d}W_{\theta\theta}''(x,\ell,0)\,,
\\
m^\star&:=&\sup_{x\in\RR^d}D_{\theta\theta}''(x,0)
\,+\,2\,J\sum_{|\ell|\klg R}
\sup_{x\in\RR^d}W_{\theta\theta}''(x,\ell,0)\,,
\end{eqnarray*}
and assume that there is some $n\in\NN$ such that
\begin{equation}\label{ass-mstar}
(n+1)\,m_\star\,-\,n\,m^\star\,>\,0\,.
\end{equation}
Note that this is always true, for fixed $n$, provided $J>0$
is sufficiently small.
Moreover, we set, for $\ve>0$,
$$
\mathcal{I}(n,\ve)\,:=\,
\bigcup_{\nu=1}^n(\nu\,m_\star-\ve\,,\,\nu\,m^\star+\ve)\,.
$$
 
\begin{theorem}\label{thm-Ma}
For every $\ve>0$, there is some $\beta_0\grg1$
such that, for all $\beta\grg\beta_0$ and $h\in(0,1]$, 
\begin{eqnarray*}
\spec\big(L^{(0)}_{\beta,h}\big)\cap(-\infty\,,\,(n+1)\,m_\star-\ve\,]
&
\subset&
\mathcal{I}(n,\ve)\cup\{0\}\,,
\\
\spec\big(L^{(1)}_{\beta,h}\big)\cap(-\infty\,,\,(n+1)\,m_\star-\ve\,]
&
\subset&
\mathcal{I}(n,\ve)\,.
\end{eqnarray*}
\end{theorem}

\bigskip

\noindent Next, we introduce an effective Hamiltonian, or, 
Feshbach operator, $\Fesh_{\beta,h}$, associated with $L^{(1)}_{\beta,h}$.
To this end 
we denote the projection on $\HR^0$ onto the 
subspace of constant functions by ${\sf p}$, and set
$P:={\sf p}\otimes\id$. 
We write ${\sf p}^\bot=\id-{\sf p}$ and $P^\bot={\sf p}^\bot\otimes\id$.
In view of
Theorem~\ref{thm-Ma} $\Fesh_{\beta,h}$ is then well-defined by
\begin{eqnarray}
\Fesh_{\beta,h}&:=&P\,L^{(1)}_{\beta,h}\,P
\,-\,P\,L^{(1)}_{\beta,h}\,P^\bot\big(\nonumber
P^\bot\,L^{(1)}_{\beta,h}\,P^\bot\big)^{-1}P^\bot\,L^{(1)}_{\beta,h}\,
P
\\
&=&P\,E''\,P
\,-\,P\,E''\,P^\bot\big(
P^\bot\,L^{(1)}_{\beta,h}\,P^\bot\big)^{-1}P^\bot\,E''\,
P\label{norbert}
\,.
\end{eqnarray}
Now we are in a position to write down the Helffer-Sj\"ostrand
formula for our lattice spin model.
Its validity is also proved in \cite{Ma}.
We note that it only holds if we assume
$\mu_{\beta,h}$ to be pure. The crucial point is that
$\ker L_{\beta.h}^{(0)}=\CC\,1$
if and only if $\mu_{\beta,h}$ is pure.

\begin{theorem}
There is some $\beta_0\grg1$
such that, for all $\beta\grg\beta_0$, $h\in(0,1]$, 
and $x,y\in\ZZ^d_h$, 
\begin{equation}
\Cor_{\beta,h}\big(\sigma_x\,;\,\sigma_y\big)\,=\,\frac{1}{\beta}\:
\SPb{e_x}{\big(L^{(1)}_{\beta,h}\big)^{-1}\,e_y}_{\HR^1}\,=\,
\frac{1}{\beta}\:
\SPb{e_x}{(\Fesh_{\beta,h})^{-1}e_y}_{\HR^1}\,,
\end{equation}
where $e_z\equiv1\otimes e_z$, for $z\in\ZZ^d_h$.
\end{theorem}

\bigskip

\noindent To obtain a proof of Theorem~\ref{thm-MaBaMo}
we only have to verify certain conditions that allow to apply
\cite[Theorems~3.3 \& 3.7]{BaMo2} to our situation.
In fact, it suffices to derive a certain estimate
involving the operators $L_{\beta,h}^{(i)}$
which is done in the next lemma.
The main idea behind all this is that, due to the small temperature
localization at the global minimum $0$ of the spin system,
we expect the Feshbach operator $\Fesh_{\beta,h}$ to be well-approximated
by $E''(0)$; see (\ref{norbert}).
Before we come to that we have, however, to introduce
some further notation.

We let ${\sf D}_E\in\LO(\ell^2(\ZZ_h^d))$ denote the diagonal part
of $E''(0)$, that is,
${\sf D}_Ee_x:=E''(0)_{xx}\,e_x$, for $x\in\ZZ^d_h$.
Furthermore, we set 
\begin{eqnarray*}
{\sf T}&:=&{\sf D}^{-1/2}_E\big({\sf D}_E
-J^{-1}\, E''(0) \big){\sf D}_E^{-1/2}\,,
\\
{\sf V}&:=&\big(\id\otimes{\sf D}^{-1/2}_E\big)\,\big(E''-
\id\otimes E''(0)\big)\,\big(\id\otimes{\sf D}^{-1/2}_E\big)\,,
\\
{\sf B}_0&:=&L_\beta^{(0)}\otimes{\sf D}^{-1}_E\,+\,\id\,,
\\
{\sf B}&:=&L_\beta^{(0)}\otimes{\sf D}_E^{-1}\,+\,
\id\otimes\big( {\sf D}^{-1/2}_EE''(0){\sf D}^{-1/2}_E\big)
\;=\;{\sf B}_0\,-\,J\,\id\otimes {\sf T}\,.
\end{eqnarray*}
Identifying $\ell^2(\ZZ^d_h)$ with 
$\Ran P=\CC 1\otimes\ell^2(\ZZ^d_h)$
we have
$$
\big(\id\otimes{\sf D}^{-1/2}_E\big)\,\Fesh_{\beta,h}\,
\big(\id\otimes{\sf D}^{-1/2}_E\big)
\,=\,\id\,-\,J\,{\sf T}\,+\,\beta^{-1/2}\,{\sf Y}\,,
$$
where
$$
{\sf Y}\,:=\,\beta^{1/2}P{\sf V}P\,-\,
\beta^{1/2}P{\sf V}P^\bot\big(P^\bot( {\sf B}+{\sf V}) P^\bot\big)^{-1}P^\bot {\sf V} P\,.
$$

\begin{lemma}\label{le-app-BaMo}
There exist $\beta_0\grg1$, $J_0>0$,
and $\vt\in(0,1)$ such that, 
for all $\beta\grg\beta_0$, $J\in(0,J_0]$, and
$x,y\in\ZZ^d_h$,
$$
|{\sf Y}_{xy}|\,\klg\,\big\{(\id-\vt\,J\, {\sf T})^{-1}\big\}_{xy}\,.
$$
\end{lemma}

\proof
Combining the estimate succeeding Equation (53) of \cite{Ma}
and Lemma~7.4 of \cite{Ma} we see that there is some constant
$C\in(0,\infty)$ such that, for all sufficiently small
$J>0$ and all $x,y\in\ZZ^d_h$, 
\begin{equation}\label{lisa1}
\big(E_{xx}''(0)\,E_{yy}''(0)\big)^{-1/2}\,
\big|E_{xy}''-E_{xy}''(0)\big|\,\klg\,
\frac{C}{\beta^{1/2}}\:(\delta_{xy}+J\,{\sf T}_{xy})\,(L_\beta^{(0)}+1)
\end{equation}
in the sense of quadratic forms on 
$\Omega^0$.
Here we also use that 
we have a strictly positive uniform lower bound for
$E_{xx}''(0)$, $x\in\ZZ^d_h$,
provided $J>0$ is sufficiently small.
Moreover, we use that there is some
$c>0$ such that, for all $x,y\in\ZZ^d_h$,
we have the bound $\widetilde{J}_{xy}\klg c\,J\,{\sf T}_{xy}$,
where the $\widetilde{J}_{xy}\grg0$ are the constants appearing
in \cite[\textsection7]{Ma}.
The form bound (\ref{lisa1}) clearly implies that
$$
\big|\SPb{e_x}{P{\sf V}P\,e_y}_{\HR^1}\big|\,=\,
\frac{\big|\SPb{1}{(E_{xy}''-E_{xy}''(0))\,1}_{\HR^0}\big|}{\sqrt{E_{xx}''(0)
\,E_{yy}''(0)}}
\,\klg\,\frac{C}{\beta^{1/2}}\:(\delta_{xy}+J\,{\sf T}_{xy})\,,
$$
because $L_\beta^{(0)}\,1=0$. 
Since 
$$
\delta_{xy}+J\,{\sf T}_{xy}\,<\,\Big\{\sum_{n=0}^\infty(J\,{\sf T})^n
\Big\}_{xy}\,=\,\big\{(\id-J\,{\sf T})^{-1}\big\}_{xy}\,,
$$
due to the fact that ${\sf T}_{xy}\grg0$
with strict inequality at least for $|x-y|/h=1$, there is some
$\tilde{\vt}\in(0,1)$ such that
\begin{equation}\label{volker1}
\beta^{1/2}\big|\SPb{e_x}{P{\sf V}P\,e_y}_{\HR^1}\big|\,\klg\,
C\,\big\{(\id-\tilde{\vt}\,J\,{\sf T})^{-1}\big\}_{xy}\,.
\end{equation}
We 
set $\overline{{\sf B}}_0:=P^\bot{\sf B}_0P^\bot$,
$\overline{{\sf B}}:=P^\bot{\sf B}P^\bot$,
and $\overline{{\sf V}}:=P^\bot{\sf V}P^\bot $ in the following.
Since the numbers $E_{xx}''(0)$ are bounded from above uniformly
in $x\in\ZZ^d_h$, 
it also follows from (\ref{lisa1}) that
there is some $\widetilde{C}\in(0,\infty)$
such that
\begin{equation}\label{volker2}
\big\|\,\{{{\sf B}}_0^{-1/2}\,{{\sf V}}\,
{{\sf B}}_0^{-1/2}\}_{xy}\,\big\|\,+\,
\big\|\,\{\overline{{\sf B}}_0^{-1/2}\,\overline{{\sf V}}\,
\overline{{\sf B}}_0^{-1/2}\}_{xy}\,\big\|
\,\klg\,\frac{\widetilde{C}}{\beta^{1/2}}\:(\delta_{xy}+J\,{\sf T}_{xy})
\end{equation}
for all $x,y\in\ZZ^d_h$.
Now let $\ve\in(0,m^\star-m_\star)$.
By Theorem~\ref{thm-MaBaMo}(i) we know that,
for sufficiently large $\beta\grg1$, 
$$
\overline{{\sf B}}_0\;\grg\;\frac{m_\star-\ve}{m^\star}\;+\,1\;=:\;
\frac{1}{\vt'}\,,
$$
where $\vt'\in(0,1)$.
This implies, for all
$x,y\in\ZZ^d_h$, 
$$
\big\|\,\{\overline{{\sf B}}_0^{-1/2}\,
J\,{\sf T}\,\overline{{\sf B}}_0^{-1/2}\}_{xy}\,\big\|
\,\klg\,J\,{\sf T}_{xy}\,\|\overline{{\sf B}}_0^{-1}\|
\,\klg\,\vt'\,J\,{\sf T}_{xy}\,, 
$$
which permits to get
\begin{equation}\label{volker3}
\big\|\,\{\overline{{\sf B}}_0^{1/2}\,
\overline{{\sf B}}^{-1}\,\overline{{\sf B}}_0^{1/2}\}_{xy}\,\big\|
\,\klg\,
\Big\|\,\Big\{
\sum_{n=0}^\infty\big(\overline{{\sf B}}_0^{-1/2}\,
J\,{\sf T}\,\overline{{\sf B}}_0^{-1/2}\big)^n
\Big\}_{xy}\,\Big\|
\,\klg\,
\{(\id-\vt'\,J\,{\sf T})^{-1}\}_{xy}\,.
\end{equation}
Using (\ref{volker2}) and (\ref{volker3}) we can now copy the
proof of \cite[Theorem~4.6]{BaMo2} to get
$$
\beta^{1/2}\big|\SPb{e_x}{P{\sf V}P^\bot\big(P^\bot( {\sf B}+{\sf V}) P^\bot\big)^{-1}P^\bot {\sf V} P\,e_y}\big|
\,\klg\,
C'\,\{(\id-\vt'\,J\,{\sf T})^{-1}\}_{xy}\,,
$$
for some $C'\in(0,\infty)$. The previous estimate
together with (\ref{volker1}) implies the assertion
of the lemma.
\qed

\bigskip

\noindent The proof of Theorem~\ref{thm-MaBaMo}(iii)
now follows from Lemma~\ref{le-app-BaMo}
together with \cite[Theorems~3.3 \& 3.7]{BaMo2}.


\section{Some elementary linear algebra}\label{app-OZ-ti2}

In this appendix we prove that Formula (\ref{eq-OZ-ti1})
of Theorem~\ref{thm-cor-asymp-ti} yields Formula (\ref{eq-OZ-ti2}).

We let $N:\Figx\rightarrow S^{d-1}$ denote the exterior normal
field on $\Figx$. (Note that, of course, the figuratrix does not depend
on $x$ in the translation invariant case.)
The derivative of its inverse is given by 
$(N^{-1})'(\mathring{v})
=|H_p'(p(\mathring{v}))|\,
\big(H_{pp}''(p(\mathring{v}))^\bot\big)^{-1}$,
for $\mathring{v}\in S^{d-1}$. On the other hand we have
$F(v)=\SPn{v}{N^{-1}(\mathring{v})}$ and direct calculations show
that the Hessian of $F$ at $v$ restricted to the orthogonal
complement of $v$ in $\RR^d$ is given by
$F_{vv}''(v)^\bot=\frac{1}{|v|}\,(N^{-1})'(\mathring{v})$.
(Here and in the following we always assume $v\in\RR^d\setminus\{0\}$
and set $\mathring{v}:=v/|v|$.)
This implies
\begin{equation}\label{maria1}
\big|H_{pp}''(p(\mathring{v}))^\bot\big|^{-1}
\,=\,\frac{|v|^{d-1}}{|H_p'(p(\mathring{v}))|^{d-1}}
\big|F_{vv}''(v)^\bot\big|
\,=\,-\frac{|v|^{d-3}}{|H_p'(p(\mathring{v}))|^{d-1}}
\Det{0&{}^tv\\v&F_{vv}''(v)}.
\end{equation}
Next, we pick a basis, 
$\langle\,b_1,\dots,b_d\rangle$, which is orthonormal
with respect to
$G(v)=\frac{1}{2}\,(F^2)_{v^iv^j}''(v)$
and such that
$b_d=F(v)^{-1}v$. We denote its dual basis by 
$\langle\,b_1^*,\dots,b_d^*\rangle$. Furthermore, we denote
the matrix whose $i$-th row is $b_i^*$ by $B$ and the matrix
whose $i$-th column is $b_i$ by $Q$, so that $B\,Q=\id_d$.
Then 
\cite[\textsection2.2]{BCS}
$$
F_{vv}''(v)={}^tB\matr{\tfrac{1}{F(v)}\,\id_{d-1}&0\\0&0}B\,, \quad
|Q|^{-2}=|G(v)|\,, \quad ({}^tQv)_d=\,\frac{|v|^2}{F(v)}\,.
$$
Consequently, we get
\begin{equation}\label{maria2}
\Det{0&{}^tv\\v&F_{vv}''(v)}\,=\,
|Q|^{-2}\Det{0&{}^tv\,Q\\{}^tQ\,v&\begin{array}{cc}\tfrac{1}{F(v)}\,\id_{d-1}&0
\\0&0\end{array}}
\,=\,
-|G(v)|\frac{|v|^4}{F(v)^{d+1}}\,.
\end{equation}
From (\ref{maria1}), (\ref{maria2}) and the homogenity of $F$ we infer that
$$
\frac{|H_p'(p(v))|^{d-3}}{\big|H_{pp}''(p(\mathring{v}))^\bot\big|
\,|v|^{d-1}}
\,=\,\frac{1}{|H_p'(p(v))|^2F(\mathring{v})^2}\,\frac{|G(v)|}{F(v)^{d-1}}
\,.
$$
Finally, using the fact that $N^{-1}(\mathring{v})=p(v)$ and 
$\mathring{v}\parallel \nabla_pH(p(v))$ we see that
$$
|H_p'(p(v))|^2F(\mathring{v})^2\,=\,\SPb{\nabla_pH(p(v))}{p(v)}^2\,.
$$
This shows that (\ref{eq-OZ-ti1}) and (\ref{eq-OZ-ti2})
are equivalent.


\section{Basic facts about almost analytic extensions}
\label{subsec-aa}

\noindent We recall some notions developed in \cite{MeSj1}:
$\Omega$ always denotes some open subset of $\CC^n$
in the following and $\Omega_{\RR}:=\Omega\cap\RR^n$. 
Two functions,
$g_1,g_2\in C^\infty(\Omega,\CC)$, are called
equivalent, in symbols $g_1\sim g_2$, iff, for every compact
$\Omega'\Subset\Omega$ and $N\in\NN$, there exists a constant
$C_{N,\Omega'}\in(0,\infty)$ such that
\begin{displaymath}
|g_1(z)-g_2(z)|\,\klg\,C_{N,\Omega'}\,|\Im z|^N\,,\qquad
z\in \Omega'\,.
\end{displaymath}
We recall a technical lemma from \cite{MeSj1}
which is used very often in the main text.

\begin{lemma}\label{tech-le-MeSj}
Let $\mathcal{G}\subset\RR^n$ be open, $f\in C_b^\infty(\mathcal{G})$,
and $g:\mathcal{G}\rightarrow\RR$ 
be locally Lipschitz continuous,
i.e. for all compact $\mathcal{G}'\Subset\mathcal{G}$
there is some $C_{\mathcal{G}'}\in(0,\infty)$ 
such that $|g(x)-g(y)|\klg C_{\mathcal{G}'}|x-y|$, $x,y\in\mathcal{G}'$.
Assume that $\supp(g)=\mathcal{G}$ and that,
for all compact $\mathcal{G}'\Subset\mathcal{G}$
and $N\in\NN$, there is some $C_{N,\mathcal{G}'}\in(0,\infty)$
such that
$$
|f(x)|\,\klg\,C_{N,\mathcal{G}'}\,|g(x)|^N\,,\qquad x\in\mathcal{G}'\,.
$$
Then, for all compact $\mathcal{G}'\Subset\mathcal{G}$,
$N\in\NN$, and $\alpha\in\NN_0^n$, $|\alpha|<N$, there exists
some $C_{N,\mathcal{G}',\alpha}\in(0,\infty)$ such that
$$
|\partial_x^\alpha f(x)|\,
\klg\,C_{N,\mathcal{G}',\alpha}\,|g(x)|^{N-|\alpha|}\,,\qquad x\in
\mathcal{G}'\,.
$$
\end{lemma}

\bigskip

\noindent Let $g_1,g_2\in C^\infty(\Omega,\CC)$ be equivalent.
By the previous lemma we find, for every compact
$\Omega'\Subset\Omega$, $N\in\NN$, and
all multi-indices $\alpha,\beta\in\NN_0^n$,
some $C_{N,\Omega',\alpha,\beta}\in(0,\infty)$ such that
$$
\big|\,\partial_{\Re z}^\alpha
\partial_{\Im z}^\beta\,\big(g_1(z)-g_2(z)\big)\,\big|\,
\klg\,C_{N,\Omega',\alpha,\beta}\,|\Im z|^N\,,\qquad
z\in \Omega'\,.
$$
A function $\tilde{f}\in C^\infty(\Omega,\CC)$
is called almost analytic, iff $\partial_{\bar{z}_i}\tilde{f}\sim0$,
for $i=1,\dots,n$.
We denote the set of all almost analytic functions on
$\Omega$ by $C^\aaa(\Omega,\CC)$.
Furthermore, let 
$f\in C^\infty(\Omega_{\RR},\CC)$. Then 
$\tilde{f}\in C^\infty(\Omega,\CC)$
is called an almost analytic extension of $f$, iff
$\tilde{f}\in C^\aaa(\Omega,\CC)$ and
$\tilde{f}\!\!\upharpoonright_{\Omega_{\RR}}=f$.
We denote the set of all almost analytic extensions of
$f$ by $C^\aaa[f]$.
For instance, every $f\in C_b^\infty(\RR^n,\CC)$ possesses
almost analytic extensions and
$\tilde{f}_1,\tilde{f}_2\in C^\aaa[f]$
fulfill $f_1\sim f_2$.

Conversely, if $\tilde{f}\in C^\aaa[f]$ 
and $g\sim 0$, then again $\tilde{f}+g\in C^\aaa[f]$,
so that $C^\aaa[f]$ is an equivalence class modulo $\sim$.
For every $\delta>0$, there is some
$\tilde{f}\in C^\aaa[f]$ with support contained in 
$\{z\in\Omega:\,|\Im z|\klg\delta\}$. Moreover, 
for $\tilde{f}\in C^\aaa[f]$,
for all multi-indices
$\alpha,\beta\in\NN_0^n$, and $i=1,\dots,n$,
it holds 
$\partial_{\Re z}^\alpha\partial_{\Im z}^\beta\partial_{\overline{z}_i}
\tilde{f}\sim0$ by Lemma~\ref{tech-le-MeSj}.
In particular, it follows that any partial derivative
of
$\tilde{f}\in C^\aaa(\Omega,\CC)$ is again almost analytic.


\end{appendix}


\noindent{\bf Acknowledgement.}
{I am gratefully indepted to Johannes Sj\"ostrand 
for proposing
to study the problems addressed in this article
and for many helpful discussions.
This work was supported by the IHP network HPRN-CT-2002-00277
from the European Union.}


\end{document}